\documentclass[twocolumn]{aastex63_rt42}

\usepackage{graphicx}
\usepackage{epstopdf}
\usepackage{amsmath}
\usepackage{psfig}
\usepackage{multirow}
\usepackage{color}
\usepackage{icomma}
\usepackage{bbding}
\usepackage{pifont}

\def\keV{\textrm{keV}}
\def\GHz{\textrm{GHz}}

\def\yr{\textrm{yr}}

\def\Msun{\textrm{M}_{\odot}}
\def\Lsun{\textrm{L}_{\odot}}
\def\Rsun{\textrm{R}_{\odot}}

\def\gcm3{\textrm{g}\,\textrm{cm}^{-3}}
\def\rhoUnits{\textrm{g}\,\textrm{cm}^{-3}}

\def\km{\textrm{km}}

\def\AU{\textrm{AU}}
\def\pc{\textrm{pc}}
\def\kpc{\textrm{kpc}}
\def\Mpc{\textrm{Mpc}}

\def\NPlanet{N_{\rm planet}}

\def\REarth{\textrm{R}_{\oplus}}

\def\albedoGV{\alpha_G^V}
\def\albedoG{\alpha_G}

\def\Mass{\textrm{M}}
\def\Mmoon{\textrm{M}_{\rm moon}}
\def\MHostPlanet{\textrm{M}_{\rm host}}
\def\MHostStar{\textrm{M}_{\rm host}}
\def\MHostGalaxy{\textrm{M}_{\rm host}}
\def\MstarHostGalaxy{\textrm{M}_{\star_{\rm host}}}
\def\Msecondary{\textrm{M}_2}
\def\MJupiter{\textrm{M}_\textrm{J}}
\def\MEarth{\textrm{M}_{\oplus}}
\def\Mstar{\textrm{M}_{\star}}
\def\Mgas{\textrm{M}_{\rm gas}}
\def\MBH{\textrm{M}_{\rm SMBH}}

\def\Nstar{\textrm{N}_{\star}}
\def\Ntier{\textrm{N}_{\rm tier}}

\def\Lstar{\textrm{L}_{\star}}
\def\LHostStar{\textrm{L}_{\rm host}}
\def\THostStar{\textrm{T}_{\rm host}}

\def\tAge{t}
\def\sec{\textrm{s}}
\def\hr{\textrm{hr}}
\def\yr{\textrm{yr}}

\def\Gyr{\textrm{Gyr}}

\def\GGauss{\textrm{GG}}

\def\PGauss{\textrm{PG}}

\def\kms{\textrm{km}\,\textrm{s}^{-1}}
\def\Kelv{\textrm{K}}

\def\Hz{\textrm{Hz}}

\def\Luminosity{\textrm{L}}
\def\LHa{\textrm{L}_{{\rm H}\alpha}}
\def\LIR{\textrm{L}_{\rm IR}}
\def\LVband{\textrm{L}_{\rm V}}
\newcommand{\LuminosityAt}[1]{\textrm{L}_{#1}}
\def\Lspin{\textrm{L}_{\rm rot}}
\def\EIRP{\textrm{EIRP}}
\def\EIRPOH{\textrm{EIRP}_{\rm OH}}
\def\EIRPWater{\textrm{EIRP}_{\rm H_2O}}

\def\Radius{\textrm{R}}
\def\Reff{\textrm{R}_{\rm eff}}
\def\Rmax{\textrm{R}_{\rm max}}
\def\RHostPlanet{\textrm{R}_p}

\def\Temperature{\textrm{T}}
\def\Teff{\textrm{T}_{\rm eff}}
\def\TICM{\textrm{T}_{\rm ICM}}

\def\Period{\textrm{P}}
\def\MOID{\textrm{MOID}}
\def\PSMBH{P_{\rm SMBH}}
\def\qSMBH{q_{\rm SMBH}}
\def\Pspin{\textrm{P}_{\rm rot}}
\def\Magnification{{\cal M}}

\def\Richness{{\cal R}}

\def\vTrans{v_{\rm trans}}
\def\muVband{\mu_V}

\def\PB{\textrm{PB}}

\def\arcsec{^{\prime\prime}}
\def\arcmin{^{\prime}}

\def\omicron{o}

\def\GHAT{\^G}

\def\ga{\gtrsim}
\def\la{\lesssim}

\def\endash{\text{--}}

\def\SurveyCountLower{count}
\def\SurveyCountTitle{Count}
\def\NEntries{\edit1{963}}
\def\NDistinct{\edit1{816}}
\def\ExoticaCatalog{\emph{Exotica Catalog}}
\def\ExoticaCatalogUpper{\emph{EXOTICA CATALOG}}
\def\CatalogVersion{\edit1{20E}}

\def\edit1{}
\def\editbfOne{}
\def\editbfTwo{}

\shorttitle{Breakthrough Listen Exotica Catalog}
\shortauthors{Lacki et al.}

\begin{document}

\title{One of Everything: The Breakthrough Listen \ExoticaCatalog{}}

\newcommand{\UCB}{Department of Astronomy,  University of California Berkeley, Berkeley CA 94720}
\newcommand{\SSL}{Space Sciences Laboratory, University of California, Berkeley, Berkeley CA 94720}
\newcommand{\RAL}{Radio Astronomy Laboratory, University of California, Berkeley, CA 94720, USA}
\newcommand{\SWIN}{Centre for Astrophysics \& Supercomputing, Swinburne University of Technology, Hawthorn, VIC 3122, Australia}
\newcommand{\GBT}{Green Bank Observatory,  West Virginia, 24944, USA}
\newcommand{\OXF}{Astronomy Department, University of Oxford, Keble Rd, Oxford, OX13RH, United Kingdom}
\newcommand{\NIJ}{Department of Astrophysics/IMAPP,Radboud University, Nijmegen, Netherlands}
\newcommand{\ATNF}{Australia Telescope National Facility, CSIRO, PO Box 76, Epping, NSW 1710, Australia}
\newcommand{\HOU}{Hellenic Open University, School of Science \& Technology, Parodos Aristotelous, Perivola Patron, Greece}
\newcommand{\USQ}{University of Southern Queensland, Toowoomba, QLD 4350, Australia}
\newcommand{\SETI}{SETI Institute, Mountain View, California}
\newcommand{\KZA}{University of Malta, Institute of Space Sciences and Astronomy}
\newcommand{\UOM}{University of Manchester, Department of Physics and Astronomy}
\newcommand{\PWJD}{The Breakthrough Initiatives, NASA Research Park, Bld. 18, Moffett Field, CA, 94035, USA}
\newcommand{\ICRAR}{International Centre for Radio Astronomy Research, Curtin University, Bentley WA 6102, Australia}

\correspondingauthor{Brian C. Lacki}
\email{astrobrianlacki@gmail.com}

\author[0000-0003-1515-4857]{Brian C. Lacki}
\affiliation{Breakthrough Listen, \UCB}

\author[0000-0002-7461-107X]{Bryan Brzycki}
\affiliation{\UCB}

\author[0000-0003-4823-129X]{Steve Croft}
\affiliation{\UCB, \SETI}

\author[0000-0002-8071-6011]{Daniel Czech}
\affiliation{\UCB}

\author[0000-0003-3197-2294]{David DeBoer}
\affiliation{\UCB}

\author{Julia DeMarines}
\affiliation{\UCB}

\author[0000-0002-8604-106X]{Vishal Gajjar}
\affiliation{\UCB}

\author[0000-0002-0531-1073]{Howard Isaacson}
\affiliation{\UCB}
\affiliation{\USQ}

\author{Matt Lebofsky}
\affiliation{\UCB}

\author{David H.\ E.\ MacMahon}
\affiliation{\RAL}

\author[0000-0003-2783-1608]{Danny C.\ Price}
\affiliation{\UCB}
\affiliation{\ICRAR}

\author[0000-0001-7057-4999]{Sofia Z. Sheikh}
\affiliation{\UCB}

\author[0000-0003-2828-7720]{Andrew P.\ V.\ Siemion}
\affiliation{\UCB}
\affiliation{\SETI}
\affiliation{\UOM}
\affiliation{\KZA}

\author{Jamie Drew}
\affiliation{\PWJD}

\author{S. Pete Worden}
\affiliation{\PWJD}

\begin{abstract}
We present Breakthrough Listen's ``Exotica'' Catalog as the centerpiece of our efforts to expand the diversity of targets surveyed in the Search for Extraterrestrial Intelligence (SETI).  As motivation, we introduce the concept of survey breadth, the diversity of objects observed during a program.  Several reasons for pursuing a broad program are given, including increasing the chance of a positive result in SETI, commensal astrophysics, and characterizing systematics.  The \ExoticaCatalog{} is \edit1{a} \NEntries{} entry collection of \NDistinct{} distinct targets intended to include ``one of everything'' in astronomy.  It contains four samples: the Prototype sample, with an archetype of every known major type of non-transient celestial object; the Superlative sample of objects with the most extreme properties; the Anomaly sample of enigmatic targets that are in some way unexplained; and the Control sample with sources not expected to produce positive results.  As far as we are aware, this is the first object list in recent times with the purpose of spanning the breadth of astrophysics.  We share it with the community in hopes that it can guide treasury surveys and as a general reference work.  Accompanying the catalog is extensive discussion of classification of objects and a new classification system for anomalies.  \edit1{Extensive notes on the objects in the catalog are available online.} We discuss how we intend to proceed with observations in the catalog, contrast it with our extant Exotica efforts, and suggest similar tactics may be applied to other programs.
\end{abstract}


\section{Introduction}

Breakthrough Listen is a ten year program to conduct the deepest surveys for extraterrestrial intelligence (ETI) in the radio and optical domains \citep{Worden17}.  The core of the program is a deep search for artificial radio emission from over a thousand nearby stars and galaxies (\citealt{Isaacson17}, hereafter I17; see also \citealt{Enriquez17,Price20} for results), and commensal studies of a million more stars in the Galaxy \citep{Worden17}.  It joins other programs in the Search for Extraterrestrial Intelligence (SETI), most of which have also focused on nearby stars \citep{Tarter01}.  But where should we look for ETIs?  Indeed, how should we look for new phenomena of any kind?

Serendipity is a key ingredient in the discovery of most new types of phenomena and extraordinary new objects \citep{Harwit81,Dick13,Wilkinson16}.  From Ceres\footnote{A planet between Mars and Jupiter was ``predicted'' by the Titius-Bode Law.  Interestingly, in September 1800, a group of astronomers colloquially known as the ``Cosmic Police'' chose twenty-four astronomers to search for this planet.  Giuseppe Piazzi was among the twenty-four selected, but did not know this when he discovered Ceres serendipitously during the construction of a star catalog in January 1801 \citep{Cunningham11}.} to pulsars, from the cosmic microwave background (CMB) to gamma-ray bursts (GRBs), the majority of unknown phenomena have been found by observers that were not explicitly looking for them.\footnote{For discussion of the discovery of Ceres, \citealt{Cunningham11}; the discovery of pulsars, reported in \citealt{Hewish68}, is recounted in \citealt{BellBurnell77}; the CMB is reported as unexpected noise in \citealt{Penzias65}; \citealt{Klebesadel73} presents the discovery of GRBs by the \emph{Vela} satellites, designed to watch for for nuclear weapon tests in violation of treaty \citep[see also][]{Klebesadel12}.}  Historically, theory has rarely driven these findings.\footnote{Among the exceptions are the discovery of radio emission from interstellar HI \citep{Ewen51} and molecules \citep{Weinreb63}, small Kuiper belt objects \citep{Jewitt93}, and binary black hole mergers \citep{Abbott16}.  The CMB was almost found by a dedicated experiment \citep{Dicke65}, but \citet{Penzias65} discovered it instead before the results came in.  The discovery of Neptune -- not a new type of object but certainly significant -- was driven by theoretical calculations of its perturbations on Uranus \citep{Galle1846,Airy1846}.}  Instead, they frequently come about by new regions of parameter space being opened by new instruments and telescopes \citep{Harwit81}.  

Other discoveries -- like the moons of Mars or Cepheid variables in external galaxies -- were delayed because no thorough observations were carried out on the targets \citep{Hall1878,Dick13}.  The pattern persists to this day.  Because ultracompact dwarf galaxies have characteristics that fall in the cracks between other galaxies and globular clusters, they were only recognized recently despite being easily visible on images for decades (\editbfOne{\citealt{Phillipps01}}; \citealt{Sandoval15}).  Of relevance to SETI, hot Jupiters were speculated about in the 1950s \citep{Struve52}, but they were not discovered until 1995 in part because no one systematically looked for them \citep[for further context, see][]{Mayor12,Walker12,Cenadelli15}.  This may have delayed by years the understanding that exoplanets are not extremely rare, one of the factors in the widely-used Drake Equation in SETI relating the number of ETIs to evolutionary probabilities and their lifespan \citep{Drake62}.

Despite searches spanning several decades, no compelling evidence for ETIs has been found by the SETI community to date \citep[e.g.,][]{Horowitz93,Griffith15,Pinchuk19,Lipman19,Price20,Sheikh20}.  The continuing lack of a discovery among SETI efforts looking for various technosignatures is sometimes called the Great Silence \citep{Brin83}.  If at least some ETIs are willing and capable of expanding across interstellar space, a bolder interpretation of the null results is popularly referred to as the Fermi Paradox, the unexpected lack of any obvious technosignatures in the Solar System \citep{Cirkovic09}.\footnote{The accuracy of the name ``Fermi Paradox'' is disputed by \citet{Gray15}; \citet{Cirkovic18-Book} on the other hand applies the term to all of the Great Silence.}  Although the simplest resolution may be that we are alone in the local Universe \citep{Hart75,Wesson90}, and others question whether we should expect to have detected technosignatures yet \citep{Tarter01,Wright18-Volume}, many have suggested that ETIs are actually abundant but we are simply looking in the wrong places for them \citep[e.g.,][]{Corbet97,Cirkovic06-Cold,Davies10,DiStefano16,Benford19,Gertz19}.  It is very difficult to detect a society of similar power and technology as our own through the traditional methods of narrowband radio searches unless it makes intentional broadcasts \citep{Forgan11}.  But like hot Jupiters, might there be easy discoveries in SETI that we keep missing because we keep looking in the wrong ways or at the wrong places?

Considerations like these in astrophysics have inspired efforts to accelerate serendipity, by expanding the region of parameter space explored by instruments \citep[c.f.,][]{Harwit81,Djorgovski01,Cordes06,Djorgovski13}.\footnote{\edit1{The discovery of gamma-ray bursts arguably falls into this category \citep[c.f.,][]{Trimble04}: as \citet{Klebesadel12} recalls, they were discovered by a systematic search for unknown gamma-ray transients, although the motivation was to determine backgrounds for the \emph{Vela} satellites' primary goal (see Motivation III in Section~\ref{sec:CatalogMotivations}).}} \edit1{\citet{Zwicky57} advocated a philosophy of ``morphological astronomy'' in which essentially all possible phenomena are considered, and more specifically searching unexplored regions of parameter space to counter biases.}\footnote{\edit1{Morphological astronomy induced Zwicky to search for compact galaxies (as catalogued in \citealt{Zwicky71}), including the first recognized Blue Compact Dwarfs \citep{Sargent70}.}} This approach has been highlighted in SETI to gauge the progress of the search (\citealt{Wright18-Volume,Davenport19}; see also \citealt{Sheikh19-Merit}).\footnote{
We should be careful not to equate parameter space volume to survey value or the probability of discovery, however.  The volume depends on parameterization (for example, vastly different volumes are found when substituting wavelength for frequency).  The more general notion of \emph{measure} on parameter space is more appropriate; these include Bayesian probability distributions \citep[c.f.,][]{Lacki16-LogLog}.  A suitable measure avoids the apparent problem that current SETI efforts are worth $\sim 10^{-20}$ the value of an ETI discovery noted by \citealt{Wright18-Volume}; current and past SETI efforts do have significant value.} Breakthrough Listen harnesses expanding capabilities in several dimensions.  In radio, Breakthrough Listen has developed a unique backend, already implemented on the Green Bank Telescope \citep{MacMahon18} and the CSIRO Parkes telescope \citep{Price18-Parkes}, and more are being installed on MeerKAT \citep[an array described in][]{Jonas09}.  These allow for an unprecedented frequency coverage at high spectral and temporal resolution.  In optical, Breakthrough Listen continues to use the Automated Planet Finder (APF; \citealt{Vogt14}) for high spectral resolution observations of stars in hopes of spotting laser emission \citep[e.g.,][]{Lipman19}, and we have partnered with the VERITAS gamma-ray telescope (Very Energetic Radiation Imaging Telescope Array System; \citealt{Weekes02}) for its sensitivity to extremely short optical pulses \citep{Abeysekara16}.  But we can also consider exploring observational parameter space too, by expanding our strategies for where and when to look.

This paper presents our motivations and initial selection and strategies for exotic targets.  The centerpiece is a broad catalog of targets, most \edit1{of them} unlike those previously covered by SETI, that we intend to observe over the coming years.  Although I17 already included a broad range of stellar and galaxy types, the \ExoticaCatalog{}'s aim is to include ``one of everything'' to ensure that we are not missing some obvious technosignatures.  Also among the targets are extreme examples of cosmic phenomena, to cover the full range of environments, and mysterious anomalies that might yield interesting results if examined closely.  In addition, we describe our other efforts to expand SETI in unconventional directions and to enhance our primary scientific results with campaigns observing selected classes of interesting targets.  In these efforts, we seek evidence for both ETIs and new astrophysical phenomena.

We present this catalog in hopes that it aids in other searches for unexpected phenomena.  A program of observing as wide a range of targets as possible does not need to be restricted to SETI, or to radio and optical wavelengths.  Any new facility across the spectrum might benefit by doing a treasury survey using a catalog based off or inspired by this one.  \edit1{We also hope that the catalog is useful as a reference for educational purposes or for early researchers by providing a convenient summary to what is currently known to be ``out there'' with references.  Online notes provide further details on the entries, target selection, and further references.}

The paper has the following structure.  We discuss basic concepts motivating exotica observations and the Catalog in Section~\ref{sec:Concepts}.  A brief overview of the division of the Catalog into four samples is presented in Section~\ref{sec:CatalogDivisions}.  The next four sections each describe the construction and principles behind each of these samples: the Prototype sample in Section~\ref{sec:Prototypes}, the Superlative sample in Section~\ref{sec:Superlatives}, the Anomaly sample in Section~\ref{sec:Anomalies}, and the Control sample in Section~\ref{sec:Controls}.  We discuss the properties of the \ExoticaCatalog{} and its planned supplementary materials in Section~\ref{sec:Discussion}.  Section~\ref{sec:WideField} discusses the need for wide-field surveys to fully span the breadth of astrophysics.  We discuss possible strategies for the Catalog and other exotica efforts in Section~\ref{sec:Strategies}.  Section~\ref{sec:Summary} is a summary of the paper.  A series of appendices presents the entries in each sample: Prototypes in Appendix~\ref{sec:AppendixPrototype}, with discussion of classification; Superlatives in Appendix~\ref{sec:AppendixSuperlative}; Anomalies in Appendix~\ref{sec:AppendixAnomaly}; and Controls in Appendix~\ref{sec:AppendixControl}.  Appendix~\ref
{sec:AppendixTotal} presents the full unified Catalog, with notes on data sources used.

\section{Concepts}
\label{sec:Concepts}

\subsection{Breadth, depth, and \SurveyCountLower{}}
\begin{figure*}
\centerline{\includegraphics[width=18cm]{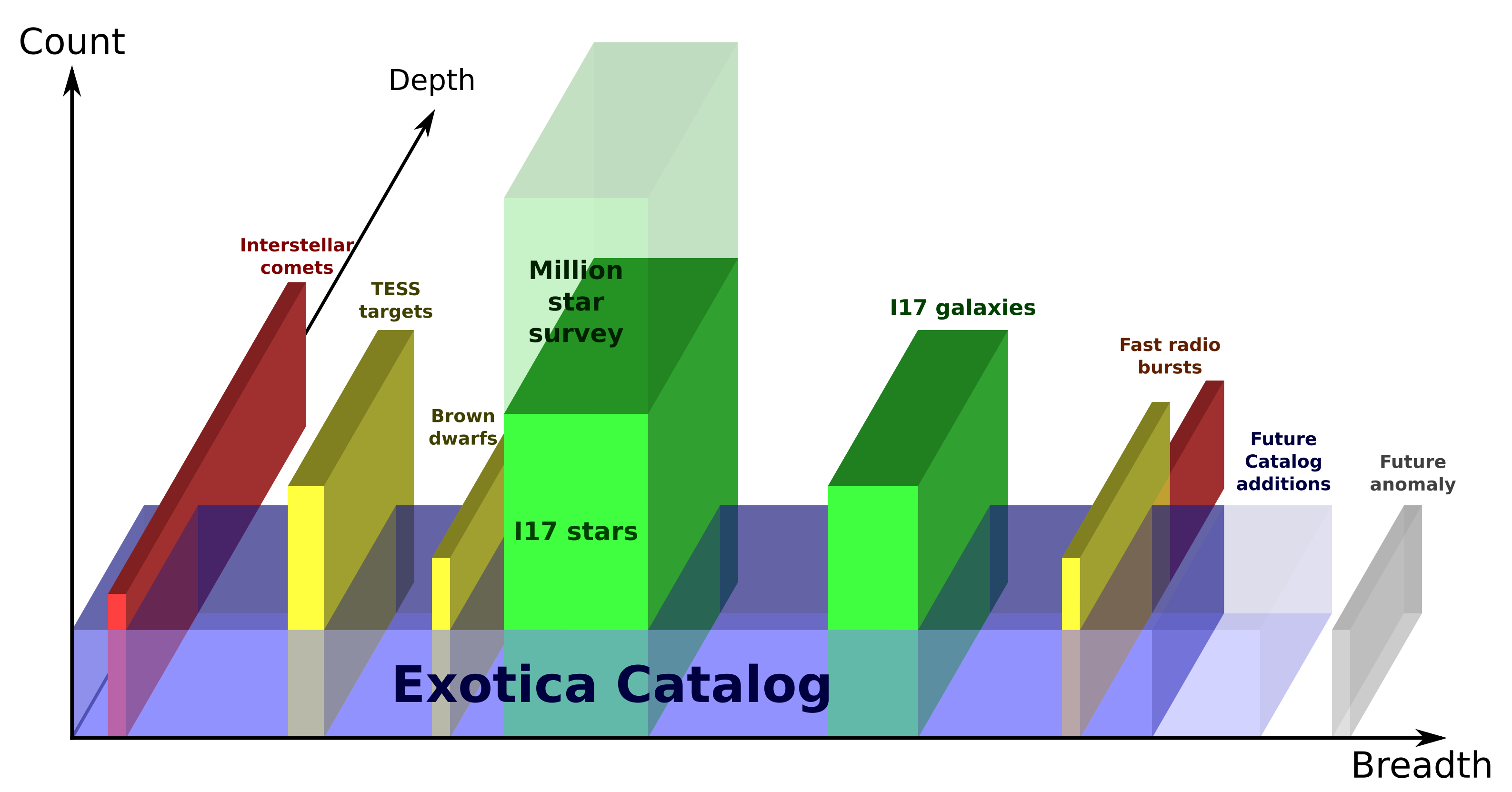}}
\figcaption{A cartoon of the three directions of target selection and the relative advantages of Breakthrough Listen's primary programs observing stars and galaxies (green), a survey of the Breakthrough Listen \ExoticaCatalog{} (blue), and some example campaigns.  Previous SETI surveys have generally aimed for \edit1{maximal} depth, achieving strong limits for a small number of similar targets, or \SurveyCountLower{}, achieving modest limits for a large number of similar targets.  Other exotica efforts can \edit1{include} high-depth (red) or high-\SurveyCountLower{} (gold) campaigns, but observations of the \ExoticaCatalog{} will be broad, achieving modest limits on a small number each of a wide variety of targets.  Future discoveries may be added to a later version of the catalog (pale blue), or prompt new campaigns that we cannot yet plan for (grey). \label{fig:SurveyDimensions}}
\end{figure*}

Each astronomical survey on a given instrument makes trade-offs.  To illustrate the differences of our programs, we distinguish between three measures of the extent of a targeted survey of individual axes. The program must balance the variety of observed object types, the number of each observed type of object, and how long to spend observing each individual object. We call these three dimensions breadth, \SurveyCountLower{}, and depth, respectively.  These three quantities can be loosely thought of as three different dimensions of parameter space, and a survey searches a bounded volume within that space, as depicted in Figure~\ref{fig:SurveyDimensions}.  A program can emphasize extent along one dimension over another, but because of limited observational resources, it cannot cover the entire realm of possibilities.  

The reader should be cautioned, however, that Figure~\ref{fig:SurveyDimensions} is not literal.  Each ``dimension'' can itself be multivalent (for example, increasing depth by increasing integration time versus cadence versus frequency coverage) and thus could actually be represented as a subspace with many dimensions \citep[as in][]{Harwit81,Djorgovski13,Wright18-Volume}.  Our emphasis here is evaluating target selection of a survey rather than its effectiveness for a given target \citep[see also][]{Sheikh19-Merit}.  Note also that breadth and \SurveyCountLower{} apply more to targeted surveys rather than wide-field surveys, which may be better parameterized with sky area \citep[c.f.,][]{Djorgovski13}.

Breadth, \SurveyCountLower{}, and depth each emphasize different levels of confidence in our ideas about for where we can make desired discoveries.  If we are very confident that a phenomenon, such as ETIs, are very common around a particular kind of object, like G dwarfs, we should push for high depth.  Depth is essential when we are sure the signals will be faint, since shallow observations cannot then be successful.  If instead we believe that a phenomenon is very rare, but are still sure of the environments that generate it, then we should aim to examine a large number of objects, in hopes that some of its signatures will turn out to be bright.  Finally, if we have no idea where we are likely to find a phenomenon, it makes sense to have a broad survey.  After all, we do not want to keep missing something otherwise obvious because we never happen to look. Shifting our emphasis from depth to \SurveyCountLower{} to breadth allows for increasing levels of serendipity.

The guiding assumption of many previous SETI efforts has been that ETIs are likely to live around sunlike stars and are not vastly more powerful than our own.  This kind of technological society is the only one known to exist \citep[e.g.,][]{Sagan93}, and a conservative approach minimizes the amount of speculation piled upon the hypotheses that ETIs are common and that they broadcast brightly enough to be detected.  Some surveys have gone for the deep approach by examining a few nearby stars with high sensitivity \citep[as in][]{Rampadarath12}.  Drake's equation implies that it's unlikely for a given star to be inhabited now, unless the ETIs persist for billions of years or \edit1{have} interstellar travel.  For this reason, a common SETI approach is to examine a large number of sunlike stars, as with the HabCat of Project Phoenix \citep{Turnbull03-HabCat}.  

Few SETI surveys have sought to examine a broad range of possible habitats.  An important exception to this are the all-sky surveys, as done with Big Ear  \citep{Dixon85} or META \citep{Horowitz93}.  In a way, all-sky surveys allow for the ultimate breadth and \SurveyCountLower{} because they observe everything in the sky.  Even if a new phenomenon is completely unknown, an all-sky survey has a chance to pick it up.  In order to accomplish this, however, they tend to have a very \edit1{low} depth.  The limited number of SETI surveys of external galaxies may be considered broad to the extent the target galaxies presumably include all kinds of stellar and planetary phenomena, although the diversity of galaxies itself is usually limited \citep{Shostak96,Gray17}. As far as targeted surveys go, \citet{Harp18} is one of the few recent efforts that emphasize breadth; their targets included quasars, masers, pulsars, supernova remnants, and an Earth-Sun Lagrange point.

Surveys are constrained by the cost and ease of access to facilities. Breakthrough Listen has unprecedented access to powerful instruments for SETI purposes, allowing our program to stretch out in all three directions.  Our main efforts so far have concentrated on the nearby stars and galaxies listed in I17.  This is a relatively broad catalog in SETI terms, including stars of spectral type from B to M and class from dwarfs to giants, as well as galaxies with a wide range of luminosities and morphologies.  Our reach will be expanded immensely by our upcoming million star survey with MeerKAT, a commensal effort that will achieve the largest \SurveyCountLower{} of a targeted SETI search.  Nonetheless, its breadth is limited because the types sampled are not too rare, unconventional, or extreme: there are no X-ray binaries or blazars in the I17 sample, for example.  

To supplement the large but finite extent of I17 (green boxes in Figure 1), we have engaged in several additional programs that can extend along any of the three dimensions.  In some cases (yellow boxes), we have effectively appended an object class to I17 by observing a number of examples, as with brown dwarfs \citep{Price20}.  In others, we have focused intently on a single extraordinary object to a high depth (red boxes), as in our studies of the repeating FRB 121102 \citep{Gajjar18}.  To these efforts, we now add the \ExoticaCatalog{} (blue box), an effort to cover the full range of known astrophysical phenomena.  The broadness of this catalog is unprecedented in targeted SETI, with only all-sky surveys being even \edit1{broader}.  Given how little we know about ETI prevalence, forms, technology, and motivations, we believe that efforts along all three dimensions are necessary.

\subsection{Core motivations for an exotica SETI program}
\label{sec:CatalogMotivations}

We have several motivations in mind for observing exotic targets, and these have informed the kind of catalogs we have created:

\begin{itemize}
    \item \emph{Motivation I}: constraining the possibility of different kinds of intelligence living in non-Earthly habitats.  Speculations about exotic habitats in the literature include: life living on habitable icy worlds around red giants \citep{Lorenz97,Lopez05,Ramirez16}, inside large carbonaceous asteroids \citep{Abramov11}, in Kuiper Belts \citep{Dyson03}, inside rogue planets \citep{Stevenson99,Abbot11}, or in the atmospheres of gas giants and brown dwarfs \citep{Sagan76,Sagan94,Yates17}.  Exotic life may be based on alternate biochemistries (\edit1{\citealt{Bains04}}; \citealt{Baross07}).  
    
    Intelligence does not need to be native to unusual habitats, as some locations may draw ETIs from their homeworlds for reasons of energy collection, curiosity, or isolation.  Some phenomena might be modulated or harnessed to act as beacons \citep[e.g.,][]{Cordes93,Learned08,Chennamangalam15}.  The postbiological universe paradigm also suggests that a spacefaring intelligence could be very different from its biological origins, with very different needs \citep{Scheffer94,Dick03}.  The practicality of some kinds of megastructures may depend on their environment or the phenomenon they are harnessing \citep{Semiz15,Osmanov16}.  Thus, there could be inhabited environments that seem inhospitable to us, like the central engine of an AGN or the outskirts of a galaxy \citep[some examples include][]{Dyson63,Cirkovic06-Cold,Vidal11,Inoue11,Lingam20}.  This goal motivates us to examine a wide variety of phenomena, both typical and extreme examples.
    
    \item \emph{Motivation II}: constraining the possibility that some astrophysical phenomena or objects are themselves artificial, a possibility suggested at least as early as the 1960s by \citet{Kardashev64}.  Blue straggler stars and fast radio bursts are examples of classes posited as engineered in the literature \citep{Beech90,Lingam17}.  Not just entire source classes, but individual mysterious objects or small subclasses might be artificial as well.  Examples of these anomalies include Boyajian's Star and Przybylsky's Star \citep{Wright16-Transit} and Hoag's Object \citep{Voros14}.  Although non-artificial explanations are far likelier and frequently plentiful, there is a small chance that we are throwing away evidence of ETIs that is staring us in the face because it does not fit our preconceptions \citep[e.g.,][]{Cirkovic18-Astro}.  This goal motivates us to examine rare, unusual subtypes of astronomical phenomena, as well as anomalous sources that defy explanation.
    
    \item \emph{Motivation III}: constraining the possibility that some natural phenomena mimic ETIs.\footnote{\editbfTwo{Given the murkiness of the concept of ``intelligence'', it is conceivable there are unknown phenomena that are not clearly artificial or natural, like the biological radio transmissions suggested by \citet{Raup92}. There could thus be a gray zone between this motivation and Motivation I, although history leads us to expect most ETI mimics will be clearly non-artificial.}}  Pulsars were briefly if unseriously considered possible contenders for alien signals because of their regular radio signals \citep[for a historical perspective on the SETI context, see][]{Penny13}.  The Astropulse survey seeks nanosecond long radio pulses from ETIs \citep{Siemion10}, but brief pulses are known to be generated by the Crab Pulsar \citep{Hankins03}, and it's possible that evaporating primordial black holes and relativistic fireballs produce similar signals \citep{Rees77,Thompson17}.  Thus, we want to deliberately seek out objects that are likely to generate unusual signals naturally.  This goal motivates us to examine extreme objects \edit1{and those} with nonthermal emission mechanisms.
    
    \item \emph{Motivation IV}: using the unique Breakthrough Listen instrumentation for general astrophysical interest.  Previous efforts along these lines include our observations of fast radio bursts \citep{Gajjar18,Price19-FRB}.  This goal motivates us to examine a wide range of sources, not just stars and galaxies that are hospitable to life.
    
    \item \emph{Motivation V}: constraining the possibility that some unexpected systematics generate false positives for ETIs.  These might include instrumental problems, problems with analysis, or especially radio frequency interference (RFI).  A claimed detection of an ETI, or even an unusual natural phenomenon, will lead to considerable skepticism.  By conducting observations where we expect nothing at all, we learn about the behavior of the instrument system.  This goal motivates us to examine empty spots on the sky, or unphysical ``sources'' like the zenith.
\end{itemize}

\subsection{Campaigns and catalogs}
\label{sec:CampaignsVsCatalogs}
Previously we have focused on targets classes that are typical of SETI, for which the nearest members were well-known.  In contrast, exotica include the more dynamic side of astrophysics.  The list of known astrophysical phenomena, and proposed links between them and ETIs, is always growing.  Some of the phenomena that fall under the auspice of exotica include violent and energetic objects that emit bright transients, like pulsars and active galactic nuclei.  Others are very faint or small, so faint that new nearby examples are constantly being discovered, like the coolest brown dwarfs and ultrafaint galaxies.  Either way, a program that observes exotica for SETI reasons needs to be more flexible than one that observes nearby stars and galaxies. 

Breakthrough Listen has two basic approaches to observe exotic objects.  The first is a series of short \emph{campaigns}, each dedicated to a particular object or object class.  If someone proposes a phenomenon is actually artificial or claims detection of ETIs, we follow up on it by observing it.  By focusing on just a few objects each, these campaigns allow us to peer deeply to lower flux levels, constraining transmitters with lower equivalent isotropic radiated power (EIRP).\footnote{EIRP is the luminosity of an isotropically radiating object at the same distance and with the same flux as a source.}  Unlike the catalogs of stars and galaxies, these programs are developed as new opportunities and discoveries arise, a more dynamic approach than having a fixed catalog.  Some examples are discussed in Section~\ref{sec:Campaigns}.

The second is a \emph{catalog} of ``exotic'' objects, including those that are extreme or interesting from an astrophysical perspective, and those that are just unusual to typical SETI searches.  The catalog is a wide mix of objects, but with few members of each type: it is more broad than deep.  On the other hand, the \ExoticaCatalog{} is intended to be a more permanent fixture of Breakthrough Listen.  Nonetheless, we anticipate revisions and additions as further new phenomena are discovered and classified.

\section{\texorpdfstring{The \ExoticaCatalogUpper{}: Surveying the breadth of astrophysical phenomena}{The \ExoticaCatalog{}: Surveying the breadth of astrophysical phenomena}}
\label{sec:CatalogDivisions}
\begin{figure}
\centerline{\includegraphics[width=9cm]{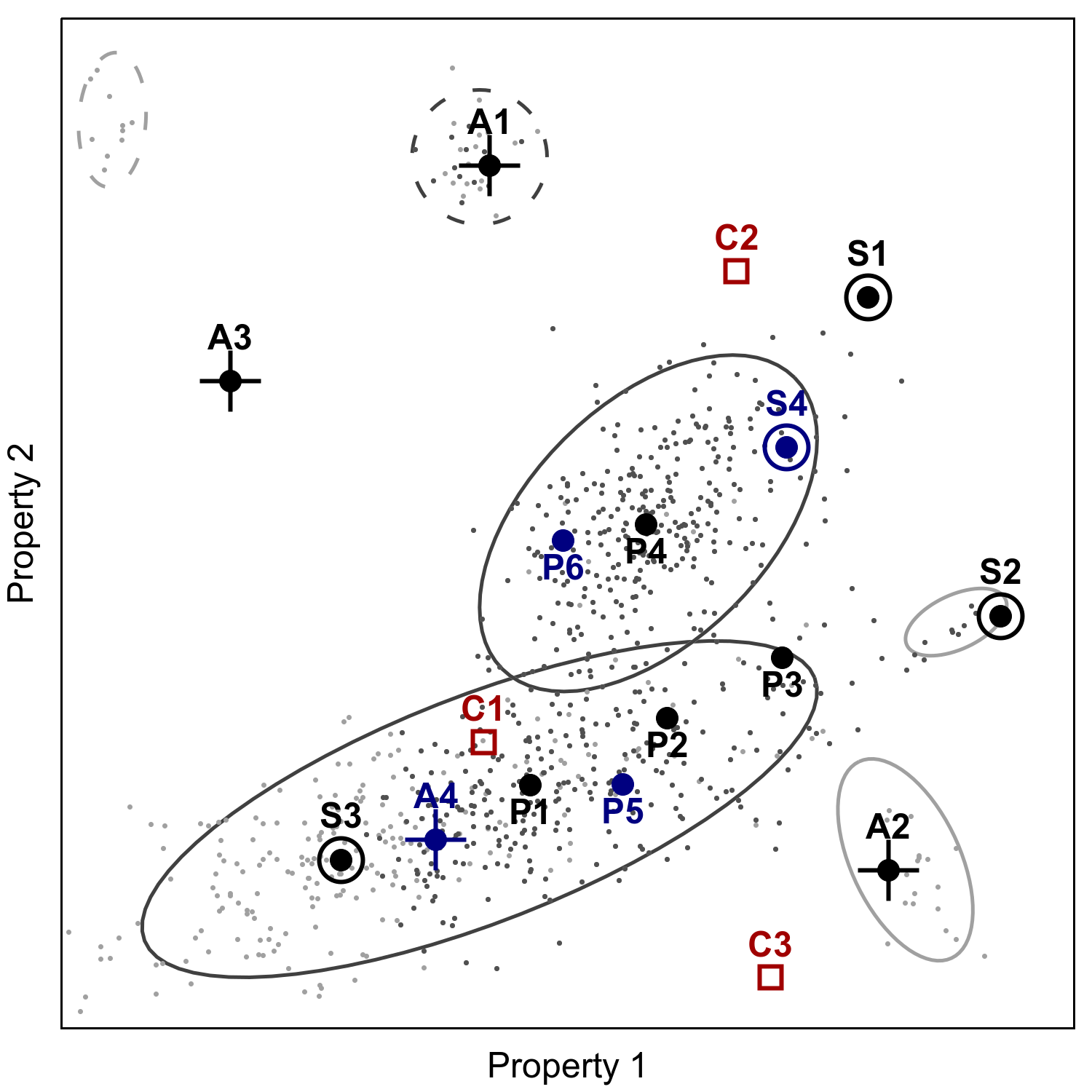}}
\figcaption{\edit1{Hypothetical illustration of parameter space for a population of objects, showing the relations between different subpopulations (ellipses) and the Prototype (P), Superlative (S), and Anomaly (A) samples.  Prototypes (plain dots) probe the ``cores'' or bulk of the subpopulations; Superlatives (ringed dots) probe their ``rims'' (as observed); Anomalies (crossed dots) include the seeming outliers.  Observational selection effects give us a biased sample of the objects, where unobserved objects (light grey points) may be concentrated in some regions and observed objects (dark grey points) in other.  While some subpopulations may be recognized (dark, solid outlines), others may be unrecognized because too few examples are known (light outlines) or they are too extreme to be linked with the other objects (dashed outlines).  In addition, some objects in the Exotica catalog may be selected by other criteria than those plotted (blue).  A small Control (C) sample (open squares) includes ``objects'' that turned out to be unreal or mundane upon examination; they may have originally been thought prototypes, superlatives, or anomalies. }\label{fig:PSADiagram}}
\end{figure}

\edit1{Although Motivations I--IV have different relationships with ETI, they together suggest a broad program of searching the variety of objects in the Universe.  Objects of a given type have properties that generally cluster in parameter space (Figure~\ref{fig:PSADiagram}).\footnote{\citet{Djorgovski13} calls these spaces Measurement Parameter Spaces and Physical Parameter Spaces, and differentiates them from the Observable Parameter Spaces that include the ``Cosmic Haystack'' discussed in SETI.}  These clusters represent different subpopulations.  We assume that a new phenomenon (like ETI) could have three qualitative relationships to the subpopulations: it might be found in typical members of one of these subpopulations; they might be biased towards extreme values of a property and generally be found in the rims of the clusters; or they may appear isolated outside of any recognized subpopulation.  This three-way division is reflected in the Exotica catalog with the Prototype, Superlative, and Anomaly samples, respectively (represented by the sources labeled with P, S, and A in the figure).}

\edit1{In all cases, our understanding of the diversity of targets will be filtered through selection biases, observational capabilities, and our theoretical categories.  A subpopulation itself may span a wide range of characteristics, and thus no single target may represent its entire subclass.  In those cases, we may artificially divide the span into several ``bins'' and choose representative examples from each bin (as in P1, P2, and P3 in Figure~\ref{fig:PSADiagram}), as we do with the stellar main sequence, despite forming a continuum.  This also allows us to impose some constraints when the new phenomenon occurs for a restricted parameter space region within the subpopulation.  Objects may have extreme properties because they genuinely are superlative among the actual population, in turn because they are simply at the tail of a distribution (S1) or they actually are part of an unrecognized new subclass (S2); or just because we cannot detect the many more extreme objects (S3).  Even ``false'' or apparent superlatives are useful in expanding the range of explored parameter space, as if we added extra ``bins'' to a sequence, since interesting phenomena may be biased in the population even while being common.  Finally, anomalous outlier objects may be examples of an observed group whose sources are not yet identified (A1), the first example of an otherwise unobserved subpopulation (A2), or be genuinely unique (A3).  As theory and observations advance, these outliers may be explained and added to the inventory of subpopulations.  All these different possibilities are interesting for at least one of the motivations for the Exotica catalog, although which motivation depends on why an object is a Prototype, Superlative, or Anomaly.  Adding to the complexity is the number of possible dimensions of parameter space, so that some objects may be selected by criteria not evident in any one visualization (P5, P6, S4, and A4).}

\edit1{Thus, t}o address the core motivations (Section~\ref{sec:CatalogMotivations}), the \ExoticaCatalog{} has four parts: 
\begin{itemize}
\item A Prototype sample including one of each type of astrophysical phenomenon (Motivations I and IV).  We emphasize the inclusion of many types of energetic and extreme objects like neutron stars (Motivations II and III), but many quiescent examples are included too. 
\item A Superlative sample that includes objects of known types that are on the tail ends of the \edit1{observed} distribution of some properties, to better span the range of objects in the Universe (Motivations I, II, and IV).
\item An Anomaly sample that includes inexplicable sources noted in the literature (Motivations II, III, and IV).  A subsample includes previously published ETI candidates (Motivations I, II, and III).
\item \edit1{Finally, a} Control sample that includes ``sources'' that are unphysical, like the zenith, or objects that have been revealed to be mundane or nonexistent.  \edit1{These nonexistent sources may have originally been proposed as typical, extreme, or outlier objects (Figure~\ref{fig:PSADiagram}).} These targets will help us get a better handle on systematics (Motivation V).
\end{itemize}.

In addition, we plan on doing occasional pointings at random positions on the sky.  That will ensure we do not miss any completely unknown phenomenon because we are looking at known objects, allowing a non-biased sample of the sky (Motivation I).  Postbiological ETIs living in interstellar space are a possible example of a hostless phenomenon.

\edit1{Despite the Prototype/Superlative/Anomaly division, the \ExoticaCatalog{} results in only a very coarse sampling of relevant parameter spaces.  It is possible that conditions for the evolution of ETIs (or other interesting phenomena) are very stringent, and they will only be found in a very small subset of the space (c.f. speculation that ETIs evolve only around ``Solar twins''; \citealt{Fracassini88}).  Even if the ``window'' is in the core of the parameter space clouds, the \ExoticaCatalog{} could easily miss them.  This is a fundamental issue with its low-\SurveyCountLower{} approach. A high-\SurveyCountLower{} survey is needed to constrain these possibilities by covering parameter space with a fine grid and selecting objects from each bin.  This grid approach was applied to the stellar color-magnitude diagram in I17, although the grid was limited in extent, resulting in more extreme stellar objects being excluded.}\footnote{\edit1{As a volume limited sample, I17 necessarily would fail to include some rarer types of stars like O dwarfs or red supergiants, although some \editbfTwo{common} stellar objects like white dwarfs and brown dwarfs fell outside the grid.}}

\section{The Prototype sample: A treasury survey}
\label{sec:Prototypes}
The Prototype Sample is the largest portion of the full \ExoticaCatalog{}.  The objects and source classes are listed in Table~\ref{table:PrototypeSample} in Appendix~\ref{sec:AppendixPrototype}.

\subsection{Classification of objects}
\label{sec:Classification}
The Prototypes form the bulk of the \ExoticaCatalog{}, and represent our effort to have ``one of everything''.  The range of things we would like includes both long-lived objects and localized phenomena.  To build this catalog, we need some classification system to enumerate the different kinds of things.  We start with a high-level classification system with 13 main categories listed in Table~\ref{table:Phyla}.  These categories have extremely different evolutionary mechanisms and power sources, and they might each have a chapter in an undergraduate textbook or have an entire advanced textbook to themselves.  Nonetheless, they are narrower than the three ``kingdoms'' proposed by \citet{Dick13}.  In analogy with the ``kingdoms'', we call these categories \emph{phyla}, emphasizing both the shared fundamental traits and the vast diversity within them, and allowing for higher level ``root'' categories.\footnote{Of course, the use of ``phylum'' does not imply a biological relationship in terms of inheritance from a common ancestor.  Instead the term refers to the older usage of morphological similarities.  Nor do the phyla necessarily fit in a tree-like hierarchy within \citet{Dick13}'s kingdoms as zoological phyla generally do: arguably stellar groups and the ISM straddle his Stellar and Galactic kingdoms.}

ETIs would probably need distinct engineering methods to harness members of different phyla.  They span the \citet{Kardashev64} (K64) scale, which groups technological ETIs broadly by the amount of power they consume.  On the K64 scale, Type I ETIs harness the amount of power available on a planet, Type II ETIs harness the power available from a sun, and Type III ETIs harness the power available from a galaxy, with proposed extensions to allow for levels beyond the range I--III and fractional levels.  The power usage increases by a factor of order ten billion between each K64 level.  Table~\ref{table:Phyla} gives a very rough estimation of the K64 rating available in each phylum.

There are edge cases like brown dwarfs that might fit in several categories, which we have placed according to convention or our judgement.  To the natural categories we have added \emph{technology}.  Human artifacts in space are part of the optical and radio sky, and are thus part of the astrophysical landscape now \edit1{(e.g., \citealt{Maley87,Combrinck94,Tingay13-Sats,Hainaut20,Corbett20}; note the inclusion of artificial satellites in \citealt{Gott05})}.  Our primary aim in SETI is to find \edit1{(or rule out)} other examples of this category.  We have also added a category for reference points not corresponding to any physical points.  Transients are also discussed separately (Section~\ref{sec:Transients}).

\begin{deluxetable*}{p{0.05\linewidth}lp{0.5\linewidth}l}
\tabletypesize{\scriptsize}
\tablecolumns{4}
\tablecaption{Phyla of astronomical phenomena: a high-level classification system used in the \ExoticaCatalog{}\label{table:Phyla}}
\tablehead{\colhead{K64 Rating} & \colhead{Phylum} & \colhead{Characteristics} & \colhead{Example} }
\startdata
\onehalf    & Minor bodies	             & Solid, small, typically irregular shape, modification mostly from cratering after initial $^{26}$Al differentiation & 1I/'Oumuamua\\
I	        & Solid planetoids          & Hydrostatic equilibrium; solid\editbfOne{, possibly with liquid oceans}; round; geology plays key role in interior evolution; \editbfOne{thin atmosphere with insignificant mass compared to body} & Titan\\
I	        & Giant planets	           & Hydrostatic equilibrium; fluids dominate mass and evolution; \editbfOne{larger than Earth}; typically high internal heat luminosity; formation by gas accretion \editbfOne{onto} solid core & Jupiter\\
II	        & Stars	                   & Hydrostatic equilibrium; plasma; powered by nuclear fusion or sometimes gravitational \editbfOne{contraction}; no solid cores; includes brown dwarfs and protostars by convention & Sun\\
II	        & Collapsed stars          & Supported by degeneracy pressure if at all; luminosity \editbfOne{primarily} from release of stored thermal, rotational, or magnetic energy & Sirius B\\
\editbfOne{II}          & Interacting binary stars & Evolution of component stars affected by mass transfer; substantial luminosity from accretion, surface nuclear burning, or shocked outflow; often compact object is mass recipient & SS 433\\
II\onehalf  & Stellar groups     	   & Gravitationally bound collections of stars, which do not otherwise interact for the most part; little to no dark matter & 47 Tuc\\
II\onehalf  & Nebulae and ISM          & Diffuse gases and plasmas; includes flows of matter to/from stars and diffuse galaxy; generally not self-bound & Orion GMC\\
III	        & Galaxies	               & Gravitationally bound, dominated by dark matter; generally contain vast numbers of stars, gas, and possibly a central black hole; gravitationally bound & \editbfOne{M81}\\
III	        & Active galactic nuclei   & Powered by accretion onto a supermassive black hole; includes gas flows on and off SMBH, frequently with jets and particle-filled bubbles & Cygnus A\\
III	        & Galaxy associations      & Dark matter dominates internal gravitation; gravitationally bound collections of galaxies and intracluster medium & Virgo Cluster\\
III\onehalf & Large-scale structures   & Unbound or loosely bound structures on $\gg$Mpc scales; includes high-order arrangements of galaxies and diffuse gas (IGM); non-virialized & Shapley Supercluster\\
Any	        & Technology	             & Structures built intentionally, frequently for processing of matter, energy, and information; so far only known to exist on/near Earth & \emph{Voyager 1}\\
\hline
\nodata     & Reference	               & Sky locations only important relative to observer & Solar antipoint
\enddata
\tablecomments{The ``phyla'' are used to group objects in the Prototype and Superlative samples by shared physical traits.  From a SETI perspective, they indicate the need for very different techniques necessary for astroengineering, as reflected in the amount of used power measured by the \citet{Kardashev64} scale rating of on left.}
\end{deluxetable*}

The phyla, while covering the breadth of astrophysics, are too coarse to form an exhaustive Prototype list by themselves.  Obviously, hugely different possibilities for habitability and exploitation exist for an asymptotic giant branch (AGB) star and a red dwarf, for example, or a white dwarf and a black hole.  Thus, we have broken down these categories into much more fine-grained types for the Prototype catalog.  Some formal classification schemes already exist in the literature, most notably \citet{Harwit81} and \citet{Dick13}.  There are also three classification systems designed as metadata for organizing the literature that nonetheless have played a huge role in astronomical research: the old AAS journal keyword list\footnote{\url{https://journals.aas.org/keywords-2013/}}, Simbad's object classification system \citep[described in][]{Wenger00}, and most recently the Unified Astronomy Thesaurus \citep{Frey18}.\footnote{Hosted at \url{http://astrothesaurus.org/blog/}.  We consulted version 3.1.0 during the writing of this paper.}

We found none exactly fit our needs, usually being too coarse-grained in most respects (e.g., not emphasizing the great differences between interacting and detached binary stars), or too detailed for a subset of phenomena (e.g., the many narrow types of dwarf novae).  Nevertheless, these systems did serve to ensure coverage of the gamut of astronomical objects.  To supplement these classification systems, we consulted review articles and textbooks, which frequently describe classification systems used for particular types of astrophysical phenomena, and how they are distinguished.  Discovery papers announcing new classes of phenomena also are useful for assembling a list of classes.  \edit1{Trimble's Astrophysics in 1991--2006 series (starting with \citealt{Trimble92} and ending with \citealt{Trimble07}) provided general overviews of the entire field for those years and highlighted newly discovered phenomena.}

Astrophysical phenomena can be classified according to a wide range of characteristics.  An object's composition, evolution, environment, and kinematics could all affect their attractiveness as a habitat for ETIs, or indicate new phenomena at work.  \edit1{One} example \edit1{is the} asteroids, which may be classified according to orbit or composition.  We therefore included several overlapping classification systems.  Some Prototypes do double duty, acting as representative examples of several of these overlapping classes, to help keep the number of objects manageable.  

Some classes of objects are excluded, usually because they are too impractical to observe.  These included diffuse ``objects'' like the interstellar medium as a whole and very large structures like the Fermi Bubbles \citep{Su10}.  In addition, we required classes to have at least one example that was fairly well established to serve as the Prototype.  For example, carbon planets, open cluster remnants, and dark matter minihalos are omitted because there have no confirmed examples.

Although our goal is to have ``one of everything'' for the Prototype catalog, our focus is not entirely even.  We cannot be entirely consistent about how fine-grained each category is, as it relies on subjective judgements about how to distinguish types.  We have included finer subclasses when they may be relevant to SETI.  Among these are the non-interacting double degenerate binary stars, which have been proposed as possible ``gravitational engines'' \citep{Dyson63}.  We also included some subclasses because their members have unusual properties that might indicate or motivate ETI presence.  Hypervelocity stars, for example, might be good places for ETIs interested in extragalactic travel to settle.  A couple of large classes that lie on a continuum of properties, namely main sequence stars and disk galaxies, have been broken up to ensure good coverage over the possible range of environments they can host.

To complete a broad census of the cosmos, we also consider the possibility that habitability changes with cosmic time \citep[c.f.,][]{Loeb16-HabEvol}.  Luminous AGNs, explosive transients, and galaxies with immense star-formation rates and chaotic morphology were more prevalent at high redshift \citep[e.g.,][]{Elmegreen05,Madau14}, with unknown effects on the evolution of ETIs and their technological capabilities \citep[for additional speculation on some of these effects, see][]{Annis99-GRBs,Cirkovic08,Gowanlock16,Lingam19}.  We partly mitigate the sensitivity losses over the vast luminosity distances by selecting gravitationally lensed galaxies.  Our EIRP sensitivity will thus be boosted by about an order of magnitude \edit1{for the selected objects}.  There is also the potential for gravitational microlensing from foreground objects, which has been known to magnify individual $z \ga 1$ stars by over a thousand \citep{Kelly18}, allowing us to achieve better EIRP limits for small high-$z$ populations. 

\subsection{Prototype selection}
Each class's Prototype is selected to be a fairly typical example, even if the object type itself is unusual.  \edit1{The reason each object is chosen as the Prototype is given in the extensive notes to the Catalog available online.} When available, we generally pick objects that are well-studied and explicitly called a ``prototype'', ``archetype'', ``benchmark'', or something similar, like the ``prototypical'' starburst M82 \citep[e.g.,][]{Seaquist91,Leroy15}. This will provide us the greatest context if we discover something.  However, because  Breakthrough Listen has unique capabilities, these observations will not be redundant with previous studies.

Although not directly referred as such, stars have prototypes in the form of spectral standard stars, which we have used when practical.  Classes of variable stars (including interacting binaries, like cataclysmic variables) are named after well-known examples, and we generally adopt the eponymous object as the Prototype.  In two cases, we substituted another object if it is much closer and well-studied: TW Hydrae for T Tauri stars \citep{Zuckerman04}, and AG Car for S Dor-type Luminous Blue Variables \citep{Groh09}.

A principle we adopted, especially when explicit prototypes are lacking, is to use objects with high citation counts, which we take as a proxy for how well studied an object is.  We adopted this criterion by either selecting nearby objects filtered by object type in Simbad, or by viewing the objects in a prominent catalog for a type in Simbad \citep{Wenger00}.  Then we sorted by number of citations, and examined a few objects with the highest citation count.  We are mindful that well-cited objects may be particularly well studied because they host a rarer phenomenon or are anomalous rather than representative.\footnote{For example, $\omega$ Cen is one of the most cited ``globular clusters'' but likely is a dwarf galaxy nucleus; Cen A may be the most cited lenticular galaxy, but is not representative.}

A final consideration is that we pick objects that are easy to observe and set meaningful limits on.  Nearby objects are generally preferred.  Not only are they usually well-studied and well-known, but in a survey with limited integration time, we set tighter limits on novel signals from nearby objects.  We also try to pick Prototypes that have already been observed by Breakthrough Listen as part of its nearby stars and nearby galaxy programs, or other campaigns.  Since these have been selected by distance, already observed objects also tend to be nearby and frequently have high citation rates anyway.

\subsection{The role of the Prototype sample compared to previous standards}
\edit1{We expect that a list of ``prototypes'' as in the \ExoticaCatalog{} will be useful to the general astronomical community.  Although no single object can provide much information on the characteristics of a population (which requires a high-\SurveyCountLower{} study), single objects can be studied in greater detail with high-depth studies.  The literature on a single prototypical object can therefore provide insight into the core traits of a phenomenon.  Having prototypes is also useful for theorists, who generally want some sense about whether some new predicted effect is detectable or not.  Often the best chance for discriminating theories is found by deep observations of nearby, bright, well-studied objects.}\footnote{\edit1{An example is the long-predicted expectations of gamma-rays from starburst galaxies generated by cosmic rays in their interstellar media; the first detections were from the prototypical starbursts M82 and NGC 253 \citep{Acero09,Acciari09-M82,Abdo10-SBs}.}}

\edit1{The primary literature already contains standard systems and explicit prototypes for types of objects, and the Prototype sample may seem redundant for these.  Indeed, if one is interested in a narrower survey on only, say, asteroids, stars, or local galaxies of various morphologies, these are preferable.}

\edit1{Nonetheless, we believe the Prototype sample serves several useful purposes.  First, even when we use these primary standards, the sample has collected them in one place for easier reference.  This is important for a broad survey, of course -- standards for stars and galaxies generally do not refer to each other -- but may also be useful for a narrower survey, as there can be overlapping systems of classification based on different criteria (e.g., asteroid spectral types versus orbital classes).  Second, as classification systems evolve, even a ``standard'' object may be listed under contradictory types and it may not be clear which to use (for example, differing galaxy morphology types in \citealt{deVaucouleurs91} and \citealt{Buta15}).  We have done our best to choose targets whose classifications are stable.  Third, and most important, many of the object types listed in the sample do not have formalized classification systems with explicit standards.  References to the consensus prototypes may be scattered far and wide among hundreds of papers, and it is useful to compile these objects in one place.  Some types, particularly those discovered by large surveys or at high redshift, may be studied only in bulk with no individual object standing out.  This can lead to an ``embarrassment of riches'' where the literature provides useful population studies but no guidance on choosing an example of the phenomenon.\footnote{\edit1{For example, relatively few works examine a single $z \sim 3$ Lyman Alpha Emitter in depth as a representative of the class, despite their importance to galaxy evolution, with an exception being works on lensed galaxies.}}  For added utility, we are hosting online notes on the Catalog.\footnote{\edit1{Available at the Breakthrough Listen \ExoticaCatalog{} webpage: \url{http://seti.berkeley.edu/exotica/}}\editbfTwo{. The notes for version \CatalogVersion{} are hosted at Zenodo: \dataset[doi:10.5281/zenodo.4726253]{https://doi.org/10.5281/zenodo.4726253}}}  These describe our reasons for selecting a particular object as a Prototype, potential caveats for the selection, provide citations to works where they are given as prototypes or standards, and in some cases give alternates that may be better suited for observation.}

\subsection{The challenge of transients}
\label{sec:Transients}
Among the panoply of celestial phenomena, transients pose special challenges for any attempt to observe ``one of everything''.  Of course, everything in astronomy is transient on some scale: the planets, stars, galaxies, and black holes will all disperse over the next $10^{200}\ \yr$ \citep{Adams97}; life, intelligence, and technology too all are expected to perish in a $\Lambda$CDM universe \citep{Krauss00}.  ``Transients'' here has an observer-relative definition, referring to objects that evolve in some way, typically brightness, on a timescale comparable to or shorter than the observing program.  The importance of transients to our understanding of the cosmos has been recognized in the past few decades, and the past few years have seen a huge growth in detection capabilities and characterization, from radio (e.g., with the SETI-oriented Allen Telescope Array: \citealt{Croft11,Siemion12}; see also \citet{Murphy17,Amiri18} for other recent examples) to optical \citep[e.g.,][]{Law15,Chambers16}, X-rays \citep[e.g.,][]{Matsuoka09,Krimm13}, gamma-rays \citep[e.g.,][]{Ackermann16,Abdalla19}, neutrinos \citep[e.g.,][]{AdrianMartinez16,Aartsen17-Alerts}, and gravitational waves \citep[e.g.,][]{Abbott19-Alerts,Abbott19-GWCat}.

Classes of transient events already number in the dozens, with a variety listed in Table~\ref{table:Transients}.  They span all of the electromagnetic spectrum and all messengers, all timescales from nanoseconds to decades and longer, and occur among most of the phyla, including some more properly considered Anomalies (Section~\ref{sec:Anomalies}).  Although some are mundane visual effects, like eclipses, others contribute profoundly to cosmic evolution, like the $r$-process element birthsites in supernovae and neutron star mergers.  We excluded in Table~\ref{table:Transients} those phenomena that induce only small variability (e.g., planetary transits), or are continuously ongoing rather than episodic (e.g., variable stars).

If there were no constraints on observations, we would like to observe at least one of each of the transient classes in Table~\ref{table:Transients}.  This follows from a broad-minded perspective on the possible forms of ETIs.  Both known and hypothetical transients have been studied as possible technosignatures.  More speculatively, conceivably some ETIs live on a completely different timescale than humans: perhaps whole societies of rapidly evolving artificial intelligences or neutron star life rise and fall in moments \citep[c.f.,][]{Forward80}.  Practical considerations make observing every known transient class in the table impossible.  The most important reason for the difficulty is that we have to be pointing the telescope at the right point on the sky when the transient occurs.  In order to detect a member of a rare transient class, we would either have to commit our facilities to staring and waiting for an event, or use facilities that observe much of the sky.  Future wide-field instruments could be helpful (Section~\ref{sec:WideField}).  Although there is no such facility in radio above 1 GHz or optical associated with Breakthrough Listen, we have pilot programs with the low-frequency facilities LOFAR (LOw-Frequency ARray; \citealt{vanHaarlem13}) and MWA (Murchinson Widefield Array; \citealt{Tingay13-MWA}) with wide-field capabilities.

Some transients repeat.  These are listed in the first two sections of the Table.  The hosts of repeating transients in Table~\ref{table:Transients} represent object types in their own right listed in the \ExoticaCatalog{}.  A few transients occur at predictable times since their timing is the result of orbital motion: these include the periodic comets and OJ 287's flares.  We may schedule observations during examples of predicted transients, as resources permit.  Most are unpredictable, though, so simply knowing where they happen is of little help in ensuring a transient is observed.  We would need to rely on alerts, and observations of the event itself will be dependent on our access to the facilities.

Other transients occur only once, at least on human timescales, as listed in the second section of Table~\ref{table:Transients}.  With these, we not only are unable to schedule observations of a transient example, we generally cannot study the pre-transient progenitor.\footnote{For example, if some peculiar supernovae actually herald the self-destruction of a Type II ETI, it would be too late to detect them once the supernova went off.}   Any hope of observing them requires an alert and a flexible observing schedule.  While we may learn of an ongoing transient event through Astronomer's Telegrams \citep{Rutledge98}\footnote{\url{http://www.astronomerstelegram.org/}} and Gamma-ray Coordinates Network (GCN) circulars\footnote{\url{https://gcn.gsfc.nasa.gov/gcn3\_archive.html}}, we cannot guarantee that we will be able to slot in observations in time.  Furthermore, some transients are so short lived -- minutes or less -- that they would be over by the time we received any alert.  If the event occurs in the field-of-view of a radio telescope with a ring buffer, a trigger or prompt alert can allow us to beamform on it using temporarily stored voltages in the buffer \citep[e.g.,][]{Wilkinson04}.  Otherwise, we can only hope to observe the aftereffects of these short-lived transients.  It is also possible short-lived transients are technosignatures of a larger ETI society that is still observable after the event itself is over.  Hence, we include anomalous transients as objects in their own right in the Anomaly sample (Section~\ref{sec:Anomalies}). 

Finally, a few transients are common enough that they invariably will occur during our observations, as listed in the final part of the table.  Since these are generally not coming from the sources we are interested in observing, they in fact have the opposite problem of most transients: they are a form of interference that we cannot avoid even though we want to.

\section{The Superlative sample}
\label{sec:Superlatives}

The Superlative sample expands the reach of the \ExoticaCatalog{} survey across a wider range of parameter space \edit1{(Section~\ref{sec:CatalogDivisions})}.  These are objects that are among the most extreme in at least one major physical property, the record-breakers.  Perhaps ETIs, or unusual natural phenomena, are biased to very atypical examples of space objects, like the hottest planets, the lowest mass stars, or the richest galaxy clusters.  Extreme physical properties can also be a sign of different evolutionary histories or even new subtypes of phenomena.  For example, pulsars with very short rotation periods are the result of a mass transfer phase during their evolution \citep{Alpar82}.  Table~\ref{table:SuperlativeSample} in Appendix~\ref{sec:AppendixSuperlative} lists the members of the Superlative sample, and the ways they are superlative.

\subsection{Classification and Superlatives}
Classification is an important factor in determining whether an object is really a superlative at the tail end of its class's properties, the prototype of a new class, or even a unique anomaly.  In astronomy, some kinds of objects fall on a continuum in terms of properties, but are conventionally delineated by an arbitrarily chosen range.  An example is the spectral type of a star: there is a continuous range of temperature, but the boundaries of the spectral types themselves do not directly map onto different evolutionary trajectories or habitability.  Objects at the boundary of these classes have no special significance.  
We thus focus on the superlatives of easily distinguishable classes.  Thus, we give superlatives for the phyla (Section~\ref{sec:Classification}), which is our coarsest level of classification.  We include superlatives for some intermediate level classes, particularly within the Solar System and by distinguishing white dwarfs, neutron stars, and black holes.  Yet even the use of phyla does not completely solve the problem of overlapping categories.  The smallest sub-brown dwarfs have the same mass as the biggest giant planets (as is the case for the coldest \edit1{member of the} star \edit1{phylum} listed); galaxies overlap with stellar associations, with globular clusters forming a continuum with ultra-compact dwarf galaxies.

An object may have properties that are so outstanding that we consider it an Anomaly.  The basic distinction we make is that a Superlative should still be clearly a member of its class, governed by the same physical processes.  Superlatives are drawn for the tail of a class's distribution, while Anomalies \editbfOne{can} appear to lie far outside it, inexplicably so.

\subsection{Properties considered: Superlative in what way?}

An object can stand out in one of many possible quantities.  In principle, there are hundreds: the abundance of every element, absolute magnitude in every possible filter band or frequency, quantities describing internal structure, and velocities in different frames are all possibilities.  

Additionally, we could take advantage of known empirical relations between quantities, and include outliers from these relations as ``superlative''.  These superlatives with respect to relations can actually indicate new object classes: objects lying far off the stellar main sequence are the giants and white dwarfs, both fundamentally different from dwarf stars, for example. 

There are a great many such relations, however.  \edit1{The observable and physical properties of objects generally fall into ``clouds'' surrounding manifolds describing these relations in high-dimensional parameter space (Section~\ref{sec:CatalogDivisions}; \citealt{Djorgovski13}).  If we wish to include the entire ``rim'' of each cloud, t}he ``superlatives'' would \edit1{include \emph{all}} of the objects forming the boundary of that cloud, lying the furthest away from the manifold.  Indeed, searching for such outliers is one proposed method to search for new phenomena \citep{Djorgovski01}.  The number of possible combinations of variables is vast, each corresponding to one small piece of the boundary, and we conceivably could include them all.  It is impractical to select and observe all such objects.

We avoid proliferation by focusing on a small number of quantities that describe the basic properties of objects, which usually apply to most to all the phyla.  We consider the basic characteristics to be \emph{size}, quantified by radius\edit1{,} mass\edit1{, and number of members}; \emph{composition}, quantified by density and metallicity; and \emph{energetics}, quantified by bolometric luminosity and temperature.  When appropriate, mainly for objects with well-characterized orbits like planets, we include \emph{kinematics} or \emph{position}, quantified by space velocity and orbital semi-major axis and/or period.  Finally, we include \emph{era}, in the form of ages for planets and stars, and redshift for larger or brighter objects like galaxies.  Some additional properties are included for certain object classes when they fundamentally regulate an object's behavior and are easy to find in the literature.  For planets, we also consider some basic characteristics of the host star (mass, luminosity, surface temperature, and metallicity).  Magnetic fields and rotation periods are included for neutron stars.  \edit1{An organizational parameter (number of levels in a system's hierarchy) is also included for multiple stars.}

\subsection{Finding Superlatives in the literature}
We used several methods to try to find Superlative objects.  Starting a search for a Superlative involves reading through literature (generally as part of the Prototype and Anomaly search) and being alert for mentions of record-breaking objects.  Sometimes the search started outside the peer-reviewed literature.  Wikipedia maintains several lists of Superlative objects\footnote{These lists and others are themselves listed at \url{https://en.wikipedia.org/wiki/Lists\_of\_astronomical\_objects}.  Examples include the ``List of exceptional asteroids'' (\url{https://en.wikipedia.org/wiki/List\_of\_exceptional\_asteroids}), the ``List of most luminous stars'' (\url{https://en.wikipedia.org/wiki/List\_of\_most\_luminous\_stars}), and the ``List of most distant astronomical objects'' (\url{https://en.wikipedia.org/wiki/List\_of\_the\_most\_distant\_astronomical\_objects}).}; although we do not consider its entries as final, the objects it lists are generally among the most extreme, providing a place to start, and it includes links to relevant papers.  Sorting major catalogs on VizieR\footnote{\url{http://vizier.u-strasbg.fr/}}  \citep{Ochsenbein00} or exoplanet.eu\footnote{\url{http://exoplanet.eu/catalog/}} \citep{Schneider11} by key quantities also gave us candidates. When we find a paper describing a candidate Superlative, we look at citing papers in the Astrophysical Data System (ADS; \citealt{Kurtz00}), as anything that supersedes the object is likely to cite the previous record holder.  We also look for papers describing previous record holders, and the papers that cite those, especially if the Superlative's measurements are uncertain.

Ideally, we wish to use objects that are explicitly described as being superlative in some regard in the literature.  These can be announced by papers in journals like \emph{Nature} or \emph{Science}.\footnote{For example: the hottest planet \citep{Gaudi17}, the star with the shortest Galactic orbital period \citep{Meyer12}, and the most distant quasar \citep{Banados18}.  In other journals, the lowest albedo exoplanet \citep{Kipping11}, the slowest spinning radio pulsar \citep{Tan18}, and the densest galaxy \citep{Sandoval15}.}  In other cases, the record may not be heralded prominently as the subject of a paper, but we found it mentioned in a paper while searching for Prototypes or Anomalies: the most massive supercluster (Shapley), for example (\citealt{Kocevski06}, while researching the Great Attractor).  Review papers or compendiums sometimes highlight objects that are outliers and can be useful in this regard.\footnote{As when \citet{Lattimer19} mentions the possibility that black widow neutron stars are the most massive; MACS J0717.5+34 as having the brightest radio halo in \citet{vanWeeren19}; 65 UMa as being a rare septuplet system in \citet{Tokovinin18}.}  \edit1{Trimble's Astrophysics in 1991--2006 series contained sections explicitly highlighting extreme objects (even referring to them as ``Superlatives'' in \citealt{Trimble92}), although in practice we found many of them had either been included already or were superceded since publication.}\footnote{\edit1{We included HD 97950 as the densest Galactic open cluster, after finding it in \citet{Trimble97}.}}

Many cases are less certain.  Papers sometimes proclaim the extraordinary nature of an object without explicitly stating it is superlative.  One case is the Superlative R136 a1 is noted to have an extraordinary mass ($\sim 300\ \Msun$), the highest in recent literature, without being explicitly said to be the most massive known star \citep{Crowther10}.

Sometimes it seems that no one keeps track of certain extremes, and measurements may not be all that reliable.  While there are many cases of extremely low metallicity stars or galaxies being touted \citep[e.g., for Superlatives in our catalog,][]{Caffau11,Keller14,Simon15,Izotov18}, few papers describe very high metallicity stars or galaxies \citep[a few exceptions are][]{Trevisan11,Do18}, and none proclaim a single Superlative.  As we do want to include a superlative high metallicity star, we looked through several catalogs listing stellar metallicities: \citet{CayrelDeStrobel01}, Hypatia \citep{Hinkel14}, \citet{Bensby14}, PASTEL \citep{Soubiran16}, CATSUP \citep{Hinkel17}, PTPS \citep{DekaSzymankiewicz18}, and \citet{AguileraGomez18}.  These often gave contradictory measurements; a star that is the highest metallicity in one is relatively normal in another.  In this case, we looked for a star that was frequently among the highest metallicities, and reliably supersolar metallicity, settling on 14 Her (\citealt{Gonzalez99} comments on its high metallicity explicitly, supporting its selection).  As our goal is to sample the range of astrophysical phenomena, we thus prefer reliable outliers over including the most extreme objects with unreliable measurements.  Very large catalogs often include these extreme superlatives, but do not comment on them.  For example, PASTEL contains dozens of stars listed with [Fe/H] $> +1.0$, although these results are not replicated in other catalogs.  We disregard these cases as likely being due to model inadequacies or measurement errors.

Some superlatives are not included at all.  This is particularly the case when one class blends into another and the superlative depends simply on where the boundary is set, with many objects lying upon the border consistent with measurement errors.  The most massive planet and the least massive brown dwarf are excluded superlatives for this reason \citep[see the extensive discussion of the issue of classification in][]{Dick13}.  \edit1{Other reasons include the case when the potential Superlatives are heavily disputed, or when the objects have not been localized enough to follow up on -- both circumstances applying to the most massive stellar black holes, because of difficulties in modeling X-ray binaries and the lack of counterparts for gravitational wave events.}

Of course, we cannot guarantee that the Superlatives listed here are literally the most record-breaking objects in the Universe.  New record breakers are being discovered all the time.  We expect to update the list as these are discovered.

\subsection{On ``true'' and ``apparent'' Superlatives}
\edit1{An important caveat is that the diversity of known objects is limited by our observational capabilities (Section~\ref{sec:CatalogDivisions}).  Whole regions of parameter space are inaccessible, and thus some Superlative objects will change with time.  This is particularly true when considering the smallest, least luminous, or furthest objects.  Indeed, these currently unobservable objects may be both interesting to SETI and actually dominate the population: consider the case of old neutron stars, which are expected to be so numerous that one is within $\sim 10\ \pc$ \citep[e.g.,][]{Ofek09}, would have several properties interesting for astroengineering without the intense radiation environment, and any example of which would qualify as Superlative.  If we are looking for objects that stand out because they are engineered (as in Motivation II), or testing the hypothesis that ETIs are attracted to the most extreme objects only, then apparent Superlatives based on observational selection effects are of little use.  Nonetheless, as mentioned in Section~\ref{sec:CatalogDivisions}, merely apparent Superlatives are useful in that they expand the range of parameter space sampled.  Put another way, if observational selection effects are hiding many sources, then the average source that we know about -- and thus its representative in the Prototype sample -- are actually atypical and the listed Superlative lets us probe a more typical example. Apparent Superlatives allow us to search for new phenomena that are biased towards certain environments (Motivation I) but not requiring the most extreme conditions possible (ETIs who build structures around all neutron stars below a certain spindown luminosity, for example).}  

\edit1{Still, because they favor different motivations and test different hypotheses, we indicate our estimation of whether a Superlative is likely to be truly extreme among the cosmic population or not with the ``True?'' column in Table~\ref{table:SuperlativeSample}.  The ``true'' superlatives include those where we would expect to be able to detect an object that was much more extreme (e.g., the most massive galaxy cluster), or when there are good theoretical reasons to expect there are no objects that are much more extreme (e.g., the faintest hydrogen-burning star).  In some cases, we expect the listed Superlative to be the most extreme within the Solar System, Galaxy, or Local Group but are uncertain if more extreme examples exist too far away to probe; these are indicated by special symbols.  In many cases, though, we are unsure whether the Superlative is ``true'' or not, and these are marked with a question mark.}

\section{The Anomaly sample}
\label{sec:Anomalies}

Anomalies are phenomena with observable properties that do not easily fit into current theories, not even roughly.  Upon discovery, they frequently spur theorists to develop many new hypotheses and explore new mechanisms that might operate in the Universe (as happened with gamma-ray bursts and continues to happen with fast radio bursts; \citealt{Nemiroff94,Platts19}). Wild early speculation about alien engineering also tends to be associated with Anomalies, although to-date natural explanations have been eventually found.  Nonetheless, it has often been encouraged to examine Anomalies for evidence of ETIs and any distinct signs of alien intelligence will be an Anomaly upon discovery \citep{Djorgovski00,Davies10}.  Examining these anomalies does not only have to reflect a belief that they are alien engineering.  Anomalies may also induce natural phenomena that mimic ETIs and are interesting in that regard (Motivation III).  By examining these anomalies, we get a better sense of what to expect from the non-ETI Universe, and a better sense of what isn't normal.  Thus the presence of an object in the Anomaly sample is \emph{not} a statement of belief about artificiality; many of the objects are clearly natural if unusual, like Iapetus.

The targets in our Anomaly sample are listed in Tables~\ref{table:AnomalyNonSETISample} and~\ref{table:AnomalySETISample} in Appendix~\ref{sec:AppendixAnomaly}.

\subsection{Types of Anomalies}
\label{sec:AnomalyClasses}
We use a supplementary classification scheme for Anomalies that group them by the nature of their anomalousness.\footnote{For previous anomaly classification schemes, see \citet{Cordes06} and \citet{Norris17}.}  Our hope is that it will aid both further searches for anomalies in diverse datasets and inspire new ideas about technosignatures (for example, as Class I or V anomalies).

This system has six classes.
\begin{itemize}
\item Class 0 objects are those that are not anomalous from an astrophysical point of view, although they can be exotic in terms of SETI.  We define Class 0 anomalies as a likely member of a known, explained population even if classification is ambiguous, with no evidence for unknown phenomena at work.  Almost all of the Prototypes and Superlatives fall in Class 0.  An object can still have ambiguous classification without actually being mysterious; Pluto in the early 2000s was Class 0.  Alternatively, an object that is \edit1{unidentified} because \edit1{it has} multiple good explanations remains Class 0: IRAS 19312+1950 may be either a young star or an AGB, but seems readily explainable \citep{Nakashima11,Cordiner16}.

\item Class I anomalies are likely members of a known, explained population, with normal intrinsic properties, but located in an anomalous environment or context.  In other words, the object itself is not mysterious, only where it is.  This also includes objects with inexplicable kinematics.  Hot Jupiters were Class I anomalies until it was understood that gas giants could migrate \citep{Mayor95,Guillot96}.

\item Class II anomalies are likely members of a known, explained class, but whose properties quantitatively fall far outside the usual distribution.  Class II anomalies are qualitatively similar to other objects of the same type.  Millisecond pulsars, for a short time after discovery, were Class II because of their extreme spin \citep{Backer82}.  Candidate megastructures identified by infrared emission in excess of the expected value (as identified by the far--infrared radio correlation of galaxies, for example; \citealt{Garrett15,Zackrisson15}) or abnormally faint optical emission \citep{Annis99-K3,Zackrisson18} are examples in a SETI context.

\item Class III anomalies are likely members of a known, explained class, but displaying a qualitatively new and unexplained phenomenon.  These phenomena \edit1{generally are not even observed} around \edit1{most (or any other)} members of the class.  Saturn's rings are a classical example, as their nature as debris belts took centuries to understand despite Saturn clearly being a planet \citep{Dick13}.  Candidate SETI signals found in targeted surveys of nearby stars through the traditional methods of ultranarrowband radio emission or nanosecond optical pulses would be Class III anomalies.

\item Class IV anomalies are \edit1{members} of an unexplained or unknown phenomenon class.  They may be identified only in one waveband, without counterparts in any others.  Current explanations generally have plausibility issues.  Many of the famous discoveries of the past century were originally Class IV anomalies: among them, Galactic synchrotron emission \citep{Jansky33}, quasars \citep{Schmidt63}, pulsars \citep{Hewish68}, gamma-ray bursts \citep{Klebesadel73,Nemiroff94}, and now fast radio bursts \citep{Lorimer07,Platts19}.  They are, however, relatively rare.  Candidate SETI signals found in wide-field surveys are frequently Class IV, including the Wow! signal \citep{Dixon85,Gray02}.

\item Class V anomalies are objects or phenomena that appear to defy known laws of physics, whether the object itself is identifiable or not.  These are extremely rare.  In very special cases they may herald an upheaval in our understanding, as the Galilean satellites \edit1{and solar neutrino problem} did, but usually they are simply in error (e.g, the notorious case of candidate superluminal neutrinos in an early version of \citealt{Adam12}, which was a non-localized Class V anomaly).

\end{itemize}

These classes, all else being equal, roughly increase in order of mysteriousness.  Judgements about the classification of an anomaly will vary with new data and different emphasis.  For example, in the 18th century, was Saturn a planet with the anomalous feature of rings (Class III), or were the rings themselves an anomalous object (Class IV)? Other considerations are whether an Anomaly has some theoretical explanation, if a problematic one, and the confidence that the anomaly is real.  We list those with partial explanations as Class 0/$x$ in Table~\ref{table:AnomalyNonSETISample} (where $x$ is in the range I--V)\edit1{.  We} avoid \edit1{purported anomalies} whose existence \edit1{that we judge have} been rejected \edit1{by the astronomical community}.  

\subsection{Collecting Anomalies}
The search for Anomalies in the literature is even more associative and subjective than finding Prototypes and Superlatives.  While there is plenty of discussion of mysterious classes of objects like FRBs, to our knowledge, there hasn't been an attempt to create a list of anomalous objects.  An Anomaly might consist of a single object or event that doesn't attract a lot of outside attention, especially if it is poorly characterized.

A few Anomalies are well known in the SETI literature, including KIC 8462852 \citep{Boyajian16,Wright16-Transit,Abeysekara16,Wright18-Update,Lipman19} and 1I/'Oumuamua \citep{Bialy18,Enriquez18,Tingay18-Oumuamua,Bannister19}.  The abundance pattern of Przbylski's star, particularly the possible presence of short-lived radioisotopes \citep{Cowley04,Bidelman05,Gopka08}, has been informally discussed online as a possible technosignature \citep[c.f.,][]{Whitmire80}\edit1{, which is how we learned of it}.\footnote{See \url{https://sites.psu.edu/astrowright/2017/03/15/przybylskis-star-i-whats-that/} and subsequent articles in the series.} Others were known to us through general knowledge (e.g., the CMB cold spot, \citealt{Cruz05}).  Still other anomalies were found using the help of Astrophysical Data System, particularly while looking at citations and references to evaluate an object  \citep[ADS; ][]{Kurtz00}.  Papers that identify Anomalies frequently cite previous (explained or unexplained) Anomalies.\footnote{\citet{Vinko15} on ``Dougie'' citing \citet{Cenko13} on PTF11agg; \citet{Saito19} about VVV-WIT-07 citing \citet{Boyajian16} on KIC 8462852\edit1{; \citet{Narloch19} citing \citet{Butler98} on non-variable stars in the Cepheid variable strip, in turn leading us to OGLE LMC-CEP-4506.} }   Papers about Anomalies can cite well-known review articles or compendiums specifically so they can situate or differentiate their mysterious object from mundane phenomena.\footnote{\citet{Moskovitz08}\edit1{, about} (10537) 1991 RY$_{16}$\edit1{,} citing \citet{Bus02} on asteroid spectral classification.}  \edit1{As with Prototypes and Superlatives, we scanned Trimble's Astrophysics in 1991--2006 review series.\footnote{\edit1{Resulting in the inclusion of the anomalously polarized radio source J 06587-5558 (noted in \citealt{Trimble03}) and the anomalously non-variable star 45 Dra (an example of a phenomenon mentioned in \citealt{Trimble99}).}} Wikipedia also has a ``list of stars that dim oddly''\footnote{\edit1{\url{https://en.wikipedia.org/wiki/List_of_stars_that_dim_oddly}}} that includes several of the Anomalies, although we generally found them by using other sources and not all stars listed there are anomalous.}

The process was mainly serendipitous, however.  We did search ADS for terms like ``anomalous'', ``mysterious'', ``enigmatic'', and ``unusual'', although they tended to mostly turn up papers that were about relatively normal objects.\footnote{For a success, \citet{Demers01} refers to the apparent ``hole'' in UMi dSph as an ``enigma''.}  Those papers were mainly about minor puzzles that we do not consider inexplicable enough to rise to Anomaly status, although this is a subjective judgement.  We also had to ignore object classes with ``anomalous'' in the name, such as Anomalous X-ray Pulsars \edit1{(now identified as magnetars)}.  Similar searches on the Astronomer's Telegram archives netted several examples.\footnote{ASASSN-V J060000.76-310027.83 
 via \citet{Way19-13346}; ASASSN-V J190917.06+182837.36 via \citet{Way19-13106}; DDE 168 via \citet{Denisenko19-12638}.}
  Papers about recently identified mysterious objects appear in arXiv\footnote{\citet{Santerne19} on HIP 41378 f.} or journals like \emph{Nature} and \emph{Science}, where they garner attention.\footnote{\citet{DeLuca06} on RCW 106; historically, \citet{Schmidt63} on 3C 273, for example.}
  
Transients are frequently unexplained upon discovery, and they are well-represented in the sample.  Even though the events themselves are over, anomalous transients may hypothetically be technosignatures of K64 Type II--III ETIs.  We may then hypothesize there to be other technosignatures -- and other Anomalies -- coming from these societies, which would share the same field as the transients themselves did.  As a very speculative example, perhaps the large early-type galaxy M86 is one such location, as it hosts two Anomalous X-ray transients.\footnote{XRT 000519 and M86 tULX-1.}

\edit1{Some anomalies in the literature are not localized objects or are otherwise impractical to observe. These are listed in Table~\ref{table:ExcludedAnomaly} for readers interested in a more complete understanding of currently inexplicable phenomena.}

For this present catalog, we had to search for Anomalies through a literature scan.  There is great interest in searching for anomalous signals in instrumental data with machine learning\edit1{, however} \citep[e.g.,][]{Baron17,Solarz17,Giles19}.  In the coming years, these automated efforts will likely flag new objects with unusual properties that can be added to future versions of the \ExoticaCatalog{} \citep{Zhang19}.

\subsection{Previous SETI candidates}

A special subset of Anomalies, one especially relevant for Breakthrough Listen, are those identified by other SETI programs.  Signals identifiable as technosignatures essentially have to be anomalies, otherwise they could always be given a more mundane explanation and it would not be effective as a SETI search.  Indeed, the Earth, or the Solar System, would appear to be a Class III anomaly to radio astronomers with sensitive enough telescopes, due to its unnatural narrowband radio emission.  

Candidate signals identified by other SETI searches are included in the subcatalog listed in Table~\ref{table:AnomalySETISample}.  These include suggestions of narrowband radio emission \citep{Dixon85,Blair92,Horowitz93,Colomb95,Bowyer16,Pinchuk19}, nanosecond-duration optical pulses \citep{Howard04}, optical spectral features consistent with picosecond-cadence pulses \citep{Borra16}, infrared excesses compatible with waste heat from megastructures \citep{Carrigan09,Griffith15,Garrett15,Lacki16-K3}, and one example of an apparently vanished star \citep{Villarroel16}.  Most of the non-SETI Anomalies are not single-instance transients, and therefore their anomalous natures can be confirmed by independent groups.  The SETI candidates have a more uncertain existence, though, and may \edit1{have been} mundane (e.g., RFI). 

Despite not widely being considered strong evidence \edit1{of actual ETIs}, we include them because they match previous specific predictions about technosignatures.  In addition, Breakthrough Listen has followed up on claimed SETI candidates in the past, to-date verifying none \citep{Enriquez18,Isaacson19}.  Finally, the wide variety of technosignatures covered helps ameliorate the difficulty of learning about anomalies with unfamiliar techniques or subfields.  When a great many candidate ETIs are given, we include only a few examples, chosen according to the brightness of the stellar host \citep{Borra16}, the evaluated quality of the candidate \citep{Howard04,Carrigan09,Griffith15,Zackrisson15}, or the signal-to-noise ratio of the event \citep{Horowitz93,Colomb95,Bowyer16}.

\section{The Control sample}
\label{sec:Controls}
The final group of targets are those we observe with the expectation of \emph{not} observing anything new.  Instead, they provide a baseline set of observations for comparison with any positive detection of a new phenomenon, a control sample for the SETI experiment.  If some subtle systematic or source of interference is generating false positives, it should eventually show up during observations of appropriately chosen control sources.  We could thus rule out the false positive by comparison with the results of the Control sample.

\subsection{The Control sample}
The first part of the Control sample is along the lines of the other samples, a list of astrophysical targets (Table~\ref{table:ControlSample} in Appendix~\ref{sec:AppendixControl}).  These are targets that were once thought to represent a new or anomalous phenomenon but that turned out to have very mundane explanations.  We can simulate our response to transients or the discovery of an anomaly by looking at purported transients that turned out to be banal.  

As an example, \citet{Bailes91} reported the first discovered pulsar planet around PSR B1829-10.  If Breakthrough Listen had been operating in 1991, we surely would have scheduled a rapid campaign to observe PSR B1829-10, similar to our campaigns observing 1I/'Oumuamua and FRB121102, because it was among the first exoplanets reported in the literature.  A year later, however, the same authors traced the pulsar timing signal to an error in the treatment of Earth's orbit \citep{Lyne92}.  We would have acquired deep radio observations of a planet that turned out not to exist.  Any detection would likely have been the result of an error in our analysis, or perhaps have come from the pulsar itself.

Because of limited data and the difficulties of interpretation, there have been many mistaken claims in astronomy.  We chose the objects and phenomena in Table~\ref{table:ControlSample} on a few principles.  First, there should be consensus that the claimed discovery was in fact in error, preferably by the \edit1{original} authors themselves.  Second, most of the claimed discoveries should have appeared to be confident enough to have provided a strong impetus for observations.  \edit1{The majority} appeared in the peer-reviewed literature, although KIC 5520878 \edit1{and GW100916 were} explained in the same paper that reported \edit1{them} \citep{Hippke15}.  \edit1{The exceptions are GRB 090709A, for which a claimed 8 second  quasiperiodicity was reported in several GCNs by multiple groups \citep{Markwardt09,Golenetskii09,Gotz09,Ohno09} before being dismissed \citep[see][]{deLuca10,Cenko10}; HIP 114176, a false star in the \emph{Hipparcos} catalog; and Swift Trigger 954840, which could likewise have prompted observations but was not in fact an astronomical event.}\footnote{\edit1{Swift Trigger 954840 was chosen from many false triggers simply because it was the most recent erroneous event when we consulted the event list.}} Third, we avoided picking bright or exceptional sources, where we might expect new phenomena, in favor of non-existent or anonymous targets.  An example of a source we excluded is the X-ray binary Cygnus X-3, which was widely considered to be a brilliant TeV--EeV gamma-ray and cosmic ray source in the 1980s before the early detections lost credibility (\citealt{Chardin89}; see also \citealt{Archambault13} and references therein for recent gamma-ray observations).  Similarly, we excluded Barnard's Star, a nearby star that had two candidate exoplanets in the 1960s--1970s \citep{vanDeKamp63,Choi13}, because it actually does have a planet, albeit one much smaller, listed as a Prototype \citep{Ribas18}.  Fourth, sky coordinates needed to be reported, which prevented us from including perytons, seeming atmospheric radio transients later explained as RFI from a microwave oven \citep{Petroff15}.

Our inclusion of a target in the Control sample is not meant to disparage the claimants.  Science has a long history of observational errors, some of which have prompted deeper studies that in turn revealed real phenomena.  The mistaken discovery of planets around PSR B1829-10 encouraged the announcement of the actual first pulsar planets around PSR B1257+12 \citep{Wolszczan92,Wolszczan12}.  The retraction then encouraged careful scrutiny of PSR B1257+12's planets, building a more solid case for their existence.  The apparent discovery also prompted theoretical work into the origins of pulsar planets.  Another early exoplanet announcement was HD 114762 B, an apparent warm super-Jupiter detected by the radial velocity method \citep{Latham89}.  In a case of terrible luck, this real object has in recent years been shown to be a red dwarf orbiting HD 114762 in a nearly face-on orbit \citep{Kiefer19}.\footnote{Interestingly, HD 114762 was \edit1{selected essentially} as a Control source, because it was thought that no actual planet could be massive and close enough to be detected around it, according to \citet{Cenadelli15}.  The possibility that some Control sources may themselves turn out to be interesting is discussed in the next section.}  Yet it turns out there really are warm and hot gas giants, including those much bigger than Jupiter, and they really could be detected through the radial velocity method, in defiance of the expectations of the time \citep{Latham12,Cenadelli15}.  Discoveries like these can open conceptual spaces, unexplored possibilities that were simply not thought of before.  In other cases, a purported anomaly can be studied well enough that it becomes the prototype of an object class.  This \edit1{seems to be happening} with NGC 1277, a prototypical red nugget galaxy studied because of the now-disputed extreme mass of its central black hole \citep{Trujillo14-N1277}.

\subsection{Could we just look at ``empty'' regions of the sky?}
Even with the Control sample, there is a danger that we could observe something real.  HD 117043 probably does not have potassium flares \citep{Wing67}, but for all we know it does have a planet with ETIs; it's even one of the I17 sample stars.  Likewise, PSR B1829-10 is a real radio pulsar, even if it does not host planets \citep{Lyne92}.  Most of these are relatively distant, so only a very bright ETI signal would be discovered, but it remains a possibility.  \edit1{Of the Control sample, only about half correspond to actual astronmical objects (the others being GW 100916, Hertzsprung's Object, HIP 114176, OT 060420,} the Perseus Flasher\edit1{, Swift Trigger 954840, and VLA J172059.9+385226.6)}.

A more stringent set of Control sources would be to look at empty places on the \edit1{celestial sphere} with no objects at all.  This is often not actually possible, though.  A famous example is the Hubble Deep Field, chosen to be away from any known bright sources, which has thousands of galaxies \citep{Williams96}.  The beam size of the GBT at 1 GHz is about $9\farcm$, which would cover thousands of galaxies and trillions of extragalactic planets, and only one would need to send a bright broadcast in our direction.

At high frequency, radio beam sizes are smaller.  It could be possible to slip a beam between known all stars and galaxies near 100 GHz with GBT, or 10 GHz with one kilometer baselines using MeerKAT.  Even this would not guarantee the non-existence of ETIs within the main beam.  Red dwarfs, brown dwarfs, and white dwarfs would generally not be detectable from many kiloparsecs away and might exist in the beam, although hard limits on their abundance on sightlines towards the Magellanic Clouds have been set with microlensing surveys \citep[e.g.,][]{Tisserand07}.  But for all we know, some ETIs actually exist in the depths of interstellar space, away from all stars \citep{Cirkovic06-Cold}.  Perhaps they broadcast from starships \citep[e.g.,][]{Messerschmitt15}, or ride along ubiquitous interstellar objects like 'Oumuamua.   Thus, there is no place on the sky that can be guaranteed to be free of ETIs.  The main utility of looking at ``empty'' sky regions as a control is to limit the prevalence of relatively faint signals from nearby, known objects.

\subsection{Unphysical targets}
The ultimate Control source is one that cannot possibly be an actual source at all.  Then if we do observe what looks like an ETI signal, we can be quite sure it is some systematic issue, whether instrumental or caused by interference.  An unreal source can be simulated by selecting an unphysical trajectory for where the telescope is pointed, and then treating the output data as if it came from a single location on the sky.  Sources outside of the Solar System near the celestial equator are bound to rise in the east and set on the west like all other non-circumpolar sidereal sources; to do otherwise would be to violate causality.  Even within the Solar System, the motion of objects is constrained.  

As an example, consider the zenith as a ``source''. It is a very special point from the viewpoint of the observer, convenient and easily attention-grabbing. But practically speaking, the zenith actually sweeps out a band on the sky.  No distant source beyond one light day could remain at the zenith because of causality.  Unless the observatory is on the equator, no object in Earth orbit would remain at the zenith.  Any such beacon would have to be both nearby and would require expensive, power-intensive maneuvers.  Furthermore, the zenith's sightline depends on the \emph{specific} location of the observatory, an improbable coincidence unless the makers mapped the Earth thoroughly, recognized radio telescopes, and somehow determined which ones were doing SETI. If needed, other possible unreal sources abound by choosing other implausible tracking trajectories: we could track faux ``targets'' that rise in the west and set in the east, that rise in the north and set in the south or vice-versa, that move along circles of constant sky altitude as if we were at a pole, or jitter randomly on the sky. 

The GBT and Parkes are able to stare at fixed altitude-azimuth positions to conduct drift scan surveys of the sky.  It is also possible to do raster scan surveys where the telescope is tracked along ``impossible'' paths to map the sky quickly.  The key difference is that we would analyze the data not as a scan moving across the sky, but as if the telescope's celestial coordinates were fixed: we would look for signals that persist for the entire ``observation'' instead of rising and falling as the beam sweeps over them. We must also be aware that signals might spill in from the sidelobes of the beam.

We will not usually have control over where the dishes of MeerKAT are pointed.  Nonetheless, we do have the power to electronically beamform within the primary field of view.  We could move one commensal beam to a new random location in the field every millisecond, making jumps that are about a degree long in an instant.  No celestial source could follow such a trajectory.

The use of ``fake sources'' is closely related to our extant techniques to constrain RFI in our radio observations.  Bright RFI from elsewhere in the sky can impinge on the detector through the dish's sidelobes.  During observations, we employ an ABABAB (or ABACAD) strategy: we spend five minutes pointed at a target and then move to look at a different target for five minutes.  We alternate between on- and off-target five-minute pointings for a total of three each.  Signals that are detected through the sidelobes will appear in both the ON- and OFF-pointings, whereas a genuine signal that is not exceedingly bright will disappear in the OFF-pointings \citep{Enriquez17,Price20}.  The advantage of using scans as ``sources'' compared to ON/OFF strategies is that we will be able to constrain systematic issues that persist for only a few seconds.  A somewhat related strategy is possible on Parkes, where the multibeam receiver points at seven sky locations simultaneously.  Interference coming in from the sidelobes will appear in several beams at once.\footnote{Sometimes, a bright celestial source can appear in multiple beams: the \citet{Lorimer07} burst was detected in three beams through their sidelobes.  Nonetheless, it was not detected in all of the beams simultaneously, allowing it to be localized to a position clearly in the sky.}

Analyzing the faux sources will require some updating of our metadata and software.  In addition, the unphysicality of the simulated ``sources'' depend on their movement on the sky, which only is significant if the signals are prolonged.  Thus, they are no better than the Control sample or random sky locations in looking for false positives in pulse searches.  Other facilities interested in using the \ExoticaCatalog{} may not even have the capability to do raster scans.  For these reasons, we maintain the Control sample as an easy to implement check for systematics.

\subsection{Technological sources: looking for trouble}
A final set of objects that could act as a check are the technology Prototypes (listed at the end of Table~\ref{table:PrototypeSample}).  Technology, particularly RFI, is the biggest source of false positives in SETI.  Our observations of the technology Prototypes should build a small library of RFI for comparison with potential ETI candidates.  Indeed, we already have serendipitous observations at the GBT that could contribute to such a library, including from GPS and Iridium satellites.  Unlike the other Controls, when we look where we expect to see no signals, with these Prototypes, we are looking where we do expect to see a signal, one that confounds our other observations, to better understand our systematics.

Not all technological RFI can be studied in this manner, though.  Aircraft and ground transmitters would not be covered by these observations.  Nonetheless, satellites remain one of the biggest contaminants, and are likely to become much worse with the launch of Internet satellite constellations.

\section{\texorpdfstring{Discussion of the full \ExoticaCatalogUpper{}}{Discussion of the full \ExoticaCatalog{}}}
\label{sec:Discussion}

\subsection{Target demographics}

\begin{figure*}
\centerline{\includegraphics[width=18cm]{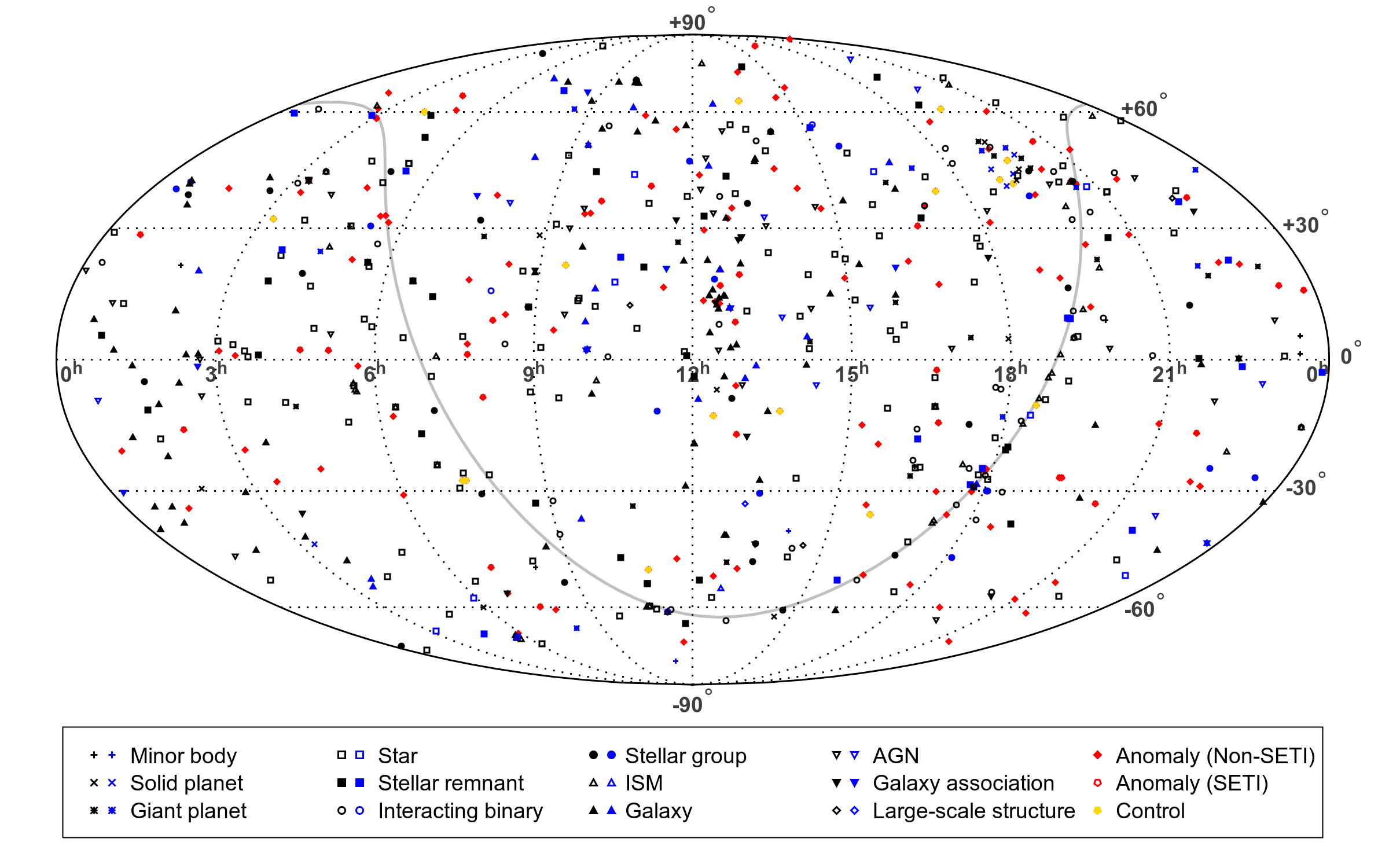}}
\figcaption{Map of objects in the \ExoticaCatalog{} in equatorial coordinates.  Each phylum is given its own symbol, with Prototypes in black and Superlatives in blue.  Anomalies are plotted in red and Control sources in gold.  The plane of the Milky Way is shown as the grey curve.\label{fig:Map}}
\end{figure*}

The full \ExoticaCatalog{} contains over \edit1{nine} hundred entries and more than \edit1{eight} hundred distinct targets.  It is about \edit1{40\%} of the size of the I17 nearby stars and galaxies catalog, and thus observing all objects would represent a significant effort on our part.  The targets are distributed across both hemispheres on the sky, with concentrations in Cygnus and the \emph{Kepler} field within it, the Galactic Center and the inner Galactic Plane, the Large Magellanic Cloud, and around the Virgo (and Coma) Cluster (Figure~\ref{fig:Map}).  The breakdown of the different catalogs into samples, into phyla, and objects we have already observed is given in Table~\ref{table:ExoticaSummary}.

\begin{deluxetable*}{lcccccc}
\tabletypesize{\scriptsize}
\tablecolumns{7}
\tablecaption{\ExoticaCatalog{} Summary\label{table:ExoticaSummary}}
\tablehead{\colhead{Property} & \colhead{Prototypes} & \colhead{Superlatives} & \multicolumn{2}{c}{Anomalies} & \colhead{Control} & \colhead{Total} \\ & & & \colhead{Non-SETI} & \colhead{SETI} & & }
\startdata
Entries                        & 598 & 187 & 125 & 36  & 17  & 963 \\ 
\hline
Distinct                       & 536 & 155 & 115 & 33  & 17  & 816 \\ 
...Minor body                  & 79  & 13  & 6   & 0   & 0   & 93  \\ 
...Solid planetoid             & 29  & 22  & 2   & 2   & 0   & 39  \\ 
...Giant planet                & 21  & 12  & 2   & 0   & 0   & 33  \\ 
...Star                        & 120 & 15  & 34  & 13  & 5   & 178 \\ 
...Collapsed star              & 35  & 24  & 8   & 0   & 2   & 64  \\ 
...Interacting binary star     & 47  & 3   & 1   & 0   & 0   & 50  \\ 
...Stellar group               & 29  & 18  & 9   & 1   & 1   & 54  \\ 
...ISM                         & 42  & 6   & 1   & 0   & 0   & 48  \\ 
...Galaxy                      & 81  & 23  & 3   & 10  & 0   & 116 \\ 
...AGN                         & 33  & 13  & 9   & 0   & 0   & 55  \\ 
...Galaxy association          & 14  & 9   & 0   & 0   & 1   & 23  \\ 
...LSS                         & 3   & 1   & 1   & 0   & 0   & 5   \\ 
...Technology                  & 15  & 0   & 0   & 0   & 0   & 15  \\ 
...Unknown                     & 0   & 0   & 42  & 9   & 0   & 50  \\ 
...Not real                    & 4   & 0   & 0   & 0   & 8   & 12  \\ 
\hline
Solar System                   & 103 & 26  & 5   & 0   & 0   & 118 \\ 
Sidereal                       & 433 & 129 & 110 & 33  & 17  & 698 \\ 
\hline
I17 stars                      & 49  & 2   & 2   & 1   & 2   & 55  \\ 
I17 galaxies                   & 15  & 0   & 1   & 0   & 0   & 16  \\ 
Other BL observed              & 2   & 0   & 3   & 0   & 0   & 4   \\ 
New targets                    & 470 & 153 & 109 & 32  & 15  & 741
\enddata
\tablecomments{Entries that are cross-listed in two different phyla are counted in the number of distinct targets in each phylum.}
\end{deluxetable*}

\begin{figure}
\centerline{\includegraphics[width=9cm]{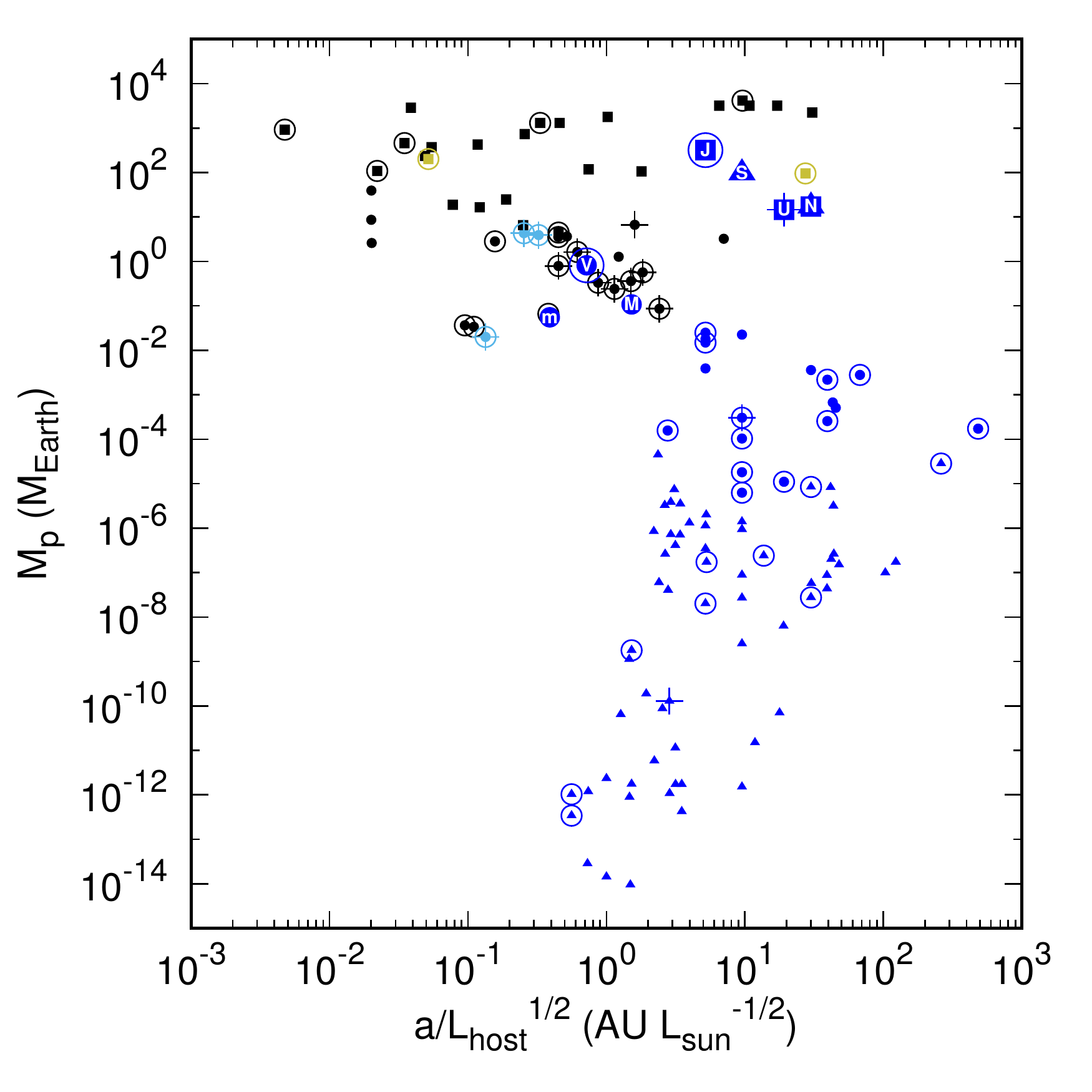}}
\figcaption{Diagram of planetary mass and stellar insolation for planets in the \ExoticaCatalog{}.  Solar System bodies are marked in deep blue, those around stellar remnants in sky blue, and those around protostars in gold.  Minor bodies (including planets hosting ring systems) are denoted by triangles, solid planetoids by circles, and giant planets by squares.  Bodies in the Superlative sample are marked by a ring around them, and those in the Anomaly sample are marked by an overlaid cross.\label{fig:PlanetMSD}}
\end{figure}

Over \edit1{160} ($\sim \edit1{20}\%$) distinct targets are in what \citet{Dick13} calls the Kingdom of the Planets: the minor bodies, the solid planetoids, and the giant planets.  Two of the most important properties of these bodies -- mass and stellar insolation -- are plotted in Figure~\ref{fig:PlanetMSD}.  The giant planets span over a factor of 100,000 in insolation ($\sim 20$ in temperature for a constant albedo), and the solid planets span nearly as large of a range.  The giant planets cover a full range of masses from super-Jovian to Neptunian and insolations from $\ga 1,000$ Earth to $0.001$ Earth \edit1{(and GJ 3483 B at $\sim 10^{-10}$ Earth insolation)}.  The solid planetoids we have chosen cover the range of temperate to hot temperatures and sub-Earths to super-Earths, with a tail of low-mass cold planetoids from the dwarf planets and large moons in the Solar System.  Minor bodies, almost all from the Solar System, cover temperate to cold environments and a factor of over ten billion in mass.

The \ExoticaCatalog{} will allow us to conduct a thorough initial reconnaissance of the Solar System.  All of the major planets\footnote{In Saturn's case, through its rings, which surround the planet.}, all IAU-recognized dwarf planets, and most of the larger moons\footnote{The Moon, the Galilean satellites, all of the large Saturn moons except Dione and Rhea, Miranda around Uranus, Triton around Neptune, and Charon around Pluto.} are included.  But an even greater coverage of the Solar System comes from the inclusion of the minor bodies, which form the third-largest phylum in the \ExoticaCatalog{} ($\sim \edit1{11}\%$).  There are a great many minor body Prototypes, which comes from the use of fine types.  This is partly because minor bodies form a large population that can be well-studied in the Solar System, but we include them also to ensure we examine objects from every region of the Solar System.  The importance for SETI is the possibility of ETI probes in the Solar System, which may have gone unnoticed (\citealt{Papagiannis78,Freitas85,Gertz16,Benford19}; see also \citealt{Wright18-PITS,Schmidt19} for more extreme possibilities).  Their radio presence will begin to be constrained with \ExoticaCatalog{} observations.   

\begin{figure*}
\centerline{\includegraphics[width=9cm]{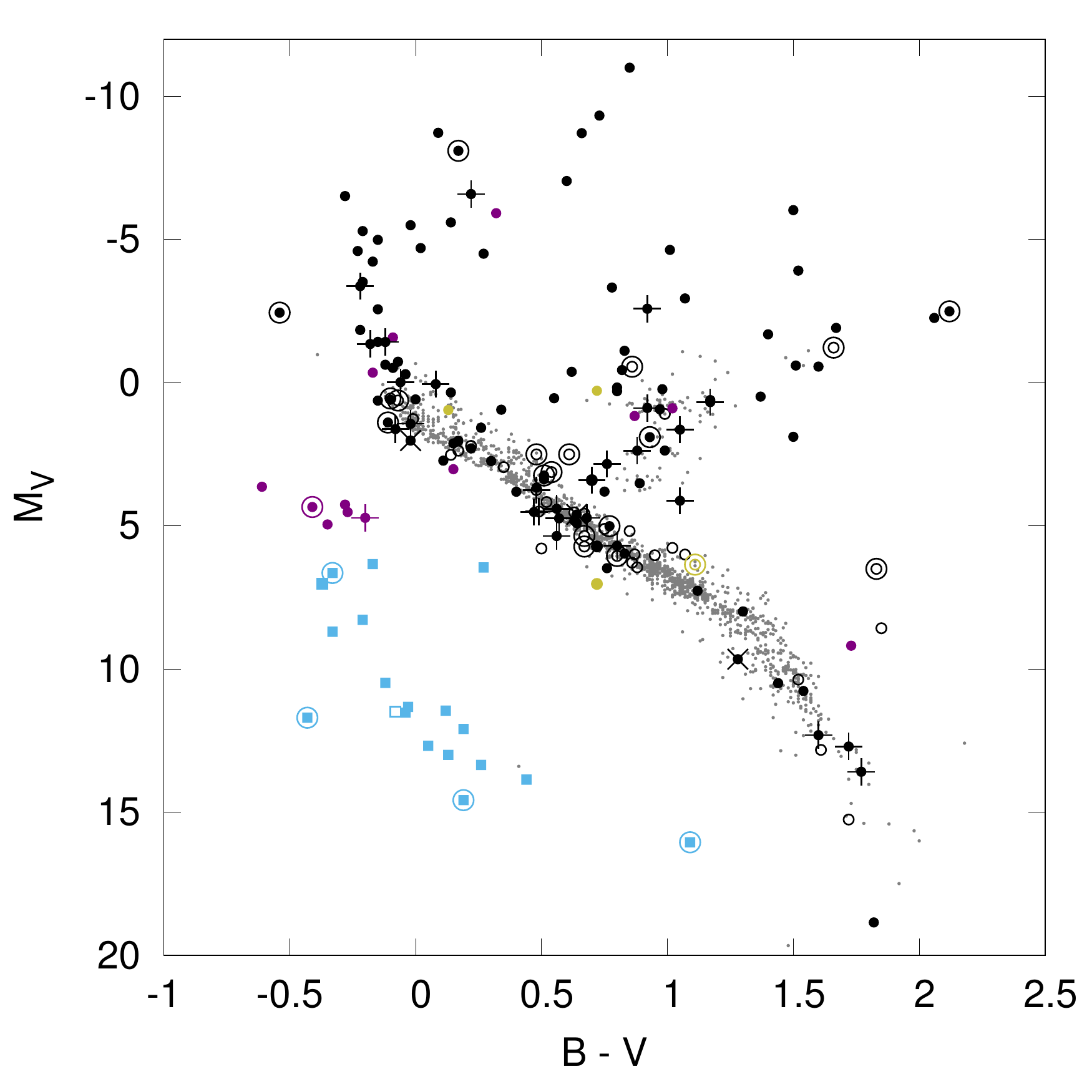}\includegraphics[width=9cm]{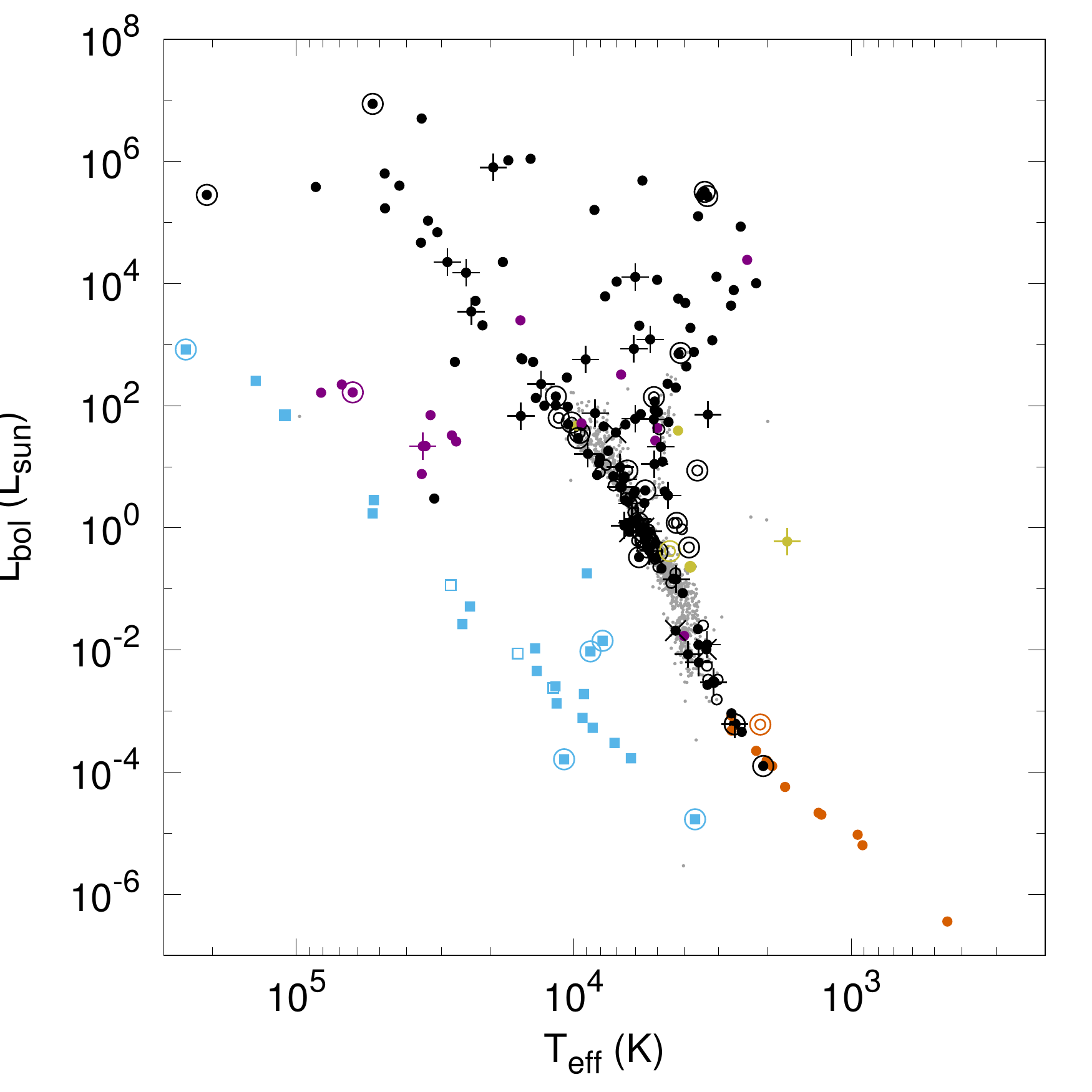}}
\figcaption{Color-magnitude (left) and HR (right) diagram of stars (circles) and white dwarfs (sky blue squares) in the \ExoticaCatalog{}.  Protostars and pre-MS stars are colored gold, brown dwarfs are dark orange, and post-interaction stars are dark violet.  The hosts of exoplanets are also included as open circles (stars) or open squares (white dwarfs).  Stars in I17 are shown as grey dots.  Superlatives and Anomalies sources are marked as in Figure~\ref{fig:PlanetMSD}.\label{fig:StellarHR}\label{fig:StellarCMD}}
\end{figure*}

The phylum most represented in the final catalog is stars, with \edit1{almost 180} distinct targets ($\sim \edit1{22}\%$).  This is partly driven by the fine classifications we use in the Prototypes catalog, particularly with the main sequence divided into narrow spectral subtypes.  The fact that we have already observed most of the main sequence Prototypes allow us this luxury, since we will not need to repeat most of those observations \citep{Enriquez17,Price20}.  Yet the stars also include a number of types not at all represented in I17.  We see this in Figure~\ref{fig:StellarCMD}, the color-magnitude diagram (CMD) and Hertzprung-Russel (HR) diagram of stars in the \ExoticaCatalog{}.  The \ExoticaCatalog{} expands the coverage of Breakthrough Listen to include supergiants and hypergiants at the top of the diagram, a region of parameter space entirely missed by the I17 sample.  O dwarfs are finally covered as well.  Sprinkled around the diagram are some unusual stellar types that result from binary evolution, which swell the ranks of stars in the Prototypes sample.  Among them are the hot subdwarfs (sdB and sdO), which sit in between the white dwarfs and hot main sequence on the left of Figure~\ref{fig:StellarCMD}.  The \ExoticaCatalog{} also includes some star types that are unusual in ways other than where they sit on the HR diagram, like \edit1{variable and} hypervelocity stars.  Not shown in the color-magnitude diagram are brown dwarfs, mostly invisible in optical light, which supplement the ones observed in \citet{Price20}.

Collapsed stars and interacting binary stars are smaller phyla, each with about \edit1{fifty to sixty} objects ($\sim \edit1{8}$ \edit1{and} $6\%$ \edit1{respectively}).  Collapsed stars have \edit1{a moderate representation} in the Prototypes sample -- they broadly fall into white dwarfs \edit1{(with both mass and spectral categories)}, neutron stars, and the hard-to-observe black holes -- but the compact objects power a rich variety of interacting binary stars.  There are also about \edit1{two-thirds as} many Superlative collapsed stars as Prototypical ones, especially since the well-studied pulsars allow for precise studies of their properties.  The \ExoticaCatalog{} includes five symbiotic systems \edit1{(excluding symbiotic X-ray binaries)}, twelve cataclysmic variables and related white-dwarf powered binaries, and \edit1{over} twenty X-ray binaries.  These \edit1{objects} have been practically ignored in SETI, but they are among the most powerful and dynamic objects in \edit1{the Universe}.  Including them allows us to start probing the idea that they draw energy-hungry ETIs \edit1{\citep[as in][]{Lacki20-Lens}}.  These extreme objects are also good bets to look for other new astrophysical phenomena.  A few more collapsed stars in detached binaries reside among the stellar associations.

Stellar groups and the ISM are also smaller phyla ($\sim \edit1{7}$ \edit1{and} $6\%$).  Even within the Galaxy, the simple open/globular cluster divide is supplemented by division of the globulars into distinct populations and the inclusion of Superlative clusters.  Observations of other galaxies have revealed star clusters different from those in the Galaxy, and have identified a number of Superlative clusters with densities and masses far more extreme than anything nearby.  To these are added detached stellar multiples and unbound stellar associations.  The ISM includes a panoply of clouds in different phases and on different scales.  Because of its diffuse nature, though, some of its most important features cannot be divided into discrete targets or are not practical to observe, particularly large structures of hot gas like the Local Bubble \citep{Snowden90}.  A variety of energetic phenomena drive a similar variety of expanding bubbles.  We note the inclusion of cosmic ray ISM features, particularly the Cygnus Cocoon \citep{Ackermann11-Cygnus}.

Galaxies form the second largest phylum ($\sim \edit1{14}\%$). Even though we use a relatively coarse division along the \citet{Hubble26} ``tuning fork'', they are supplemented by a rich variety of subtle morphological features \citep{Buta15} and peculiar galaxies \citep[c.f.,][]{Arp66}.  In fact, about one-fifth of the galaxy prototypes simply cover disk galaxy morphologies, such as lenses and rings.  Just as some abnormal stellar types are the result of binary interactions, we have peculiar galaxy types from their own interactions.  High redshift galaxies only increase the panoply, as do the technosignature candidates identified by previous SETI studies.

\begin{figure*}
\centerline{\includegraphics[width=9cm]{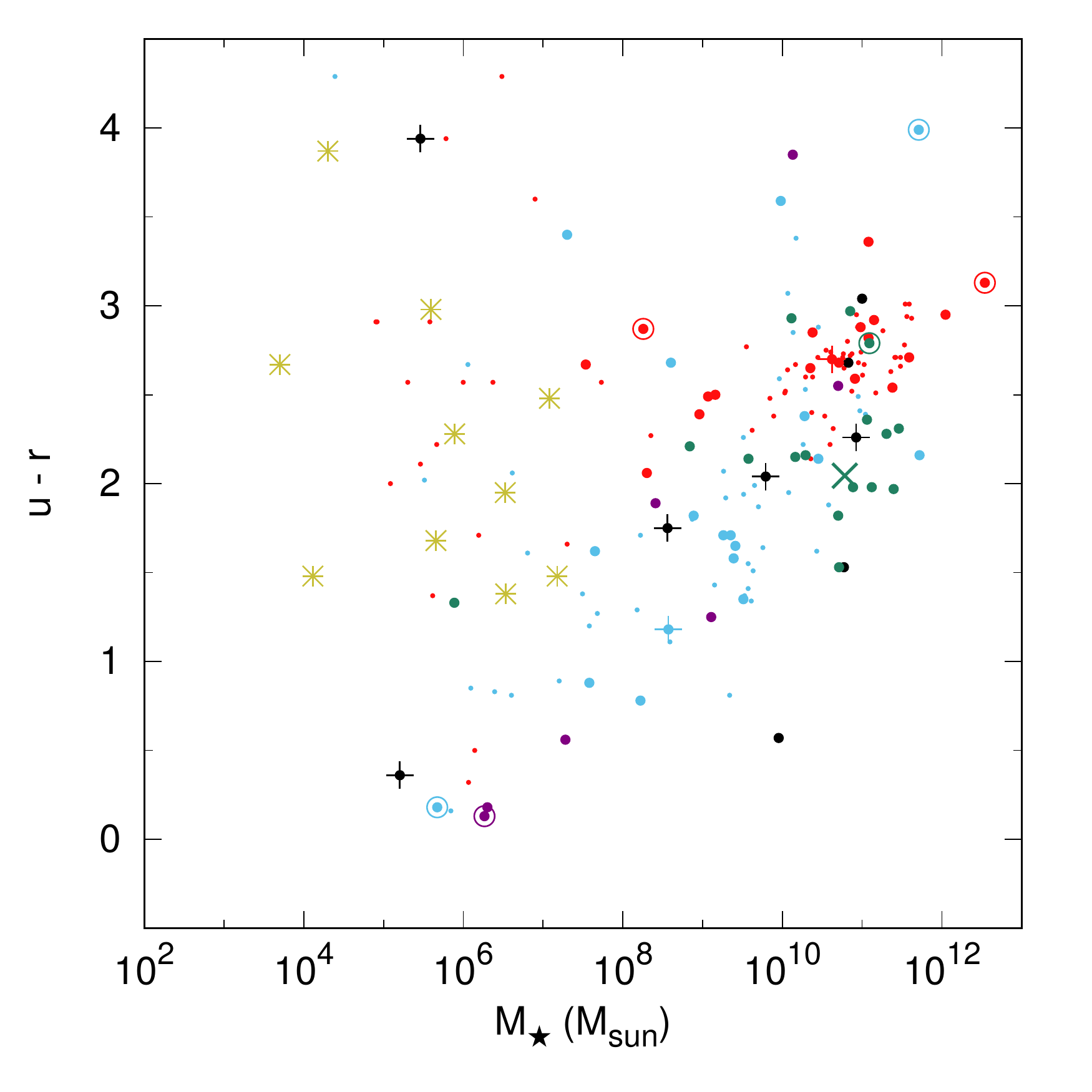}\includegraphics[width=9cm]{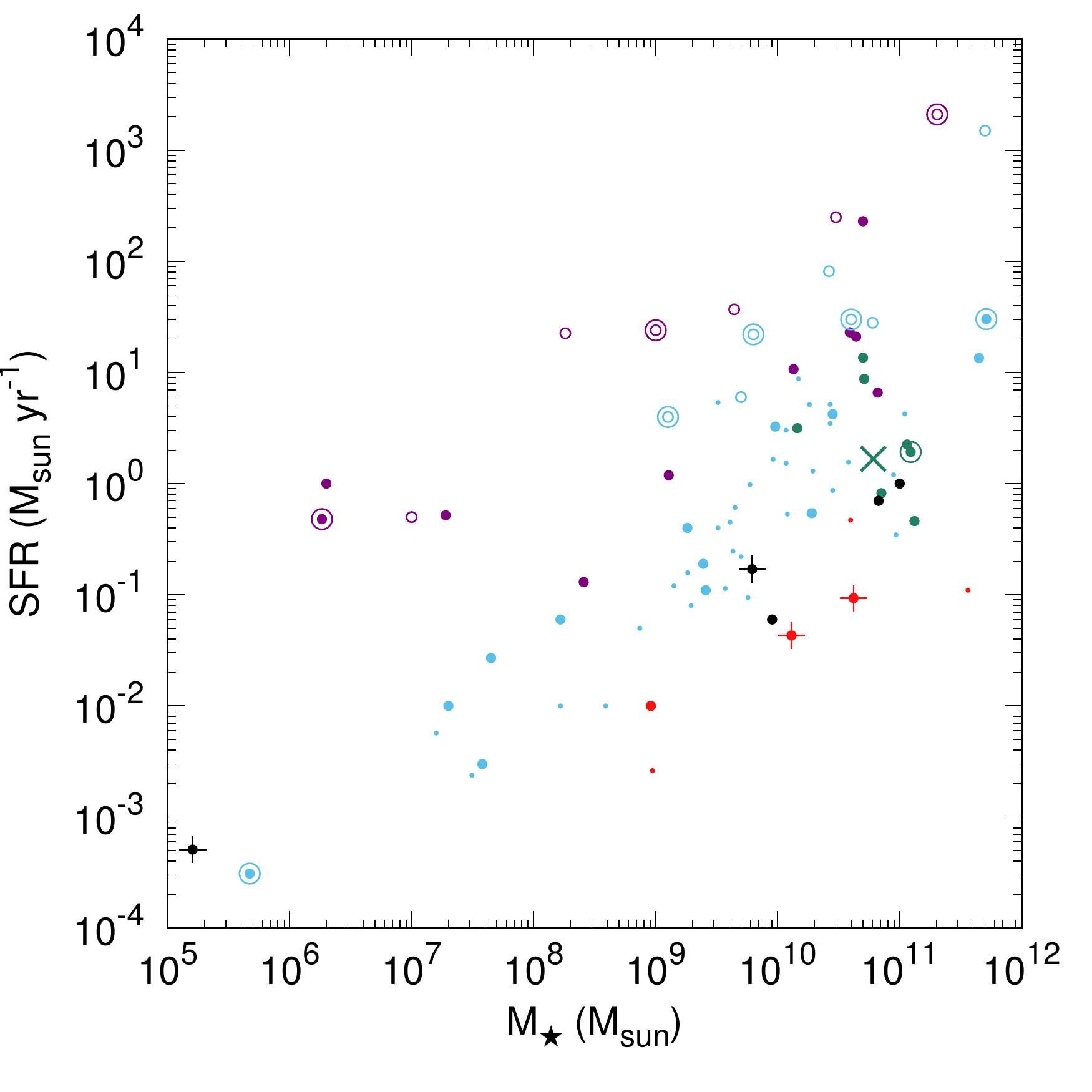}}
\figcaption{Mass-color (left) and Mass-SFR (right) diagram of galaxies in the \ExoticaCatalog{} for which we could find or derive data.  Galaxies classified as quiescent are shown in red, Green Valley galaxies in dark teal, main sequence galaxies in sky blue, and starbursts in dark violet.  Samples are marked as in Figure~\ref{fig:PlanetMSD}.  Galaxies in other categories are classified according to morphological type, with the remainders in black.  We also show I17 galaxies as small dots, and the Milky Way as the big teal X.  For the mass-color diagram, we include star clusters as gold asterisks, but exclude galaxies with $z \ge 0.1$ due to $k$-correction effects.  High-redshift galaxies are included in the mass-SFR diagram: those with $z < 0.5$ are marked with filled dots, and those with $z \ge 0.5$ are marked with open circles.\label{fig:GalaxyCMD}}
\end{figure*}

Figure~\ref{fig:GalaxyCMD} summarizes their gross properties, mass and color.  Included galaxies span a factor of $\sim 10^{10}$ in stellar mass.  In terms of gross properties, I17 samples much of the same parameter space, and in fact includes many more faint red (dwarf spheroidal) galaxies.  Note however that we were unable to find data to plot certain extreme dwarf galaxies like Segue 1, which are far smaller than the ones in I17.  We also are excited by the outliers in the CMD, particularly IC 1101 (red circled dot at far right), the Leoncino dwarf (black cross-dot at lower left), and NGC 4650A (isolated black dot at lower right).  More profound differences are visible in the mass-SFR diagram, where highly star-forming galaxies stand out with respect to the I17 smaple, particularly those at high redshift (open circles) and dwarf starbursts (purple dots at center-left), as does Coma P (circled blue dot in lower-left) and the Leoncino dwarf.  These are entirely new regions of parameter space to SETI.

Active galactic nuclei ($\sim 7\%$) and galaxy associations ($\sim 3\%$) are among the smaller phyla.  We might have gone with finer AGN classifications, with \citet{Padovani17} noting several dozen types posited over the years, but as per \citet{Padovani17}, we did not wish to cloud the Prototypes list with a menagerie of mostly overlapping types.  On the other hand, galaxy associations have a richer (if \edit1{still} small) representation than might have been expected.  A galaxy cluster, being virialized, can support some intriguing phenomena in its intracluster medium.  Superlative galaxy associations almost match the number of Prototypes, and include several rich galaxy clusters, which are unexplored in SETI.

The smallest phylum is large scale structure ($<1\%$).  Its variety is limited because it has only begun to form.  Furthermore, superclusters, walls, and voids have poorly defined boundaries and cover vast areas on the sky, and are excluded similarly to much of the ISM and IGM.  We do include the Great Attractor, formerly a striking anomaly, in the form of the Laniakea basin of attraction \citep{Tully14}.  Also included is the superlative Shapley Supercluster, the densest and richest region in the local Universe.  This supercluster is a massive collection of rich galaxy clusters, and has bound a region over 10 Mpc in radius; its gravitational pull actually entirely subsumes the flow from the Great Attractor \citep{Kocevski06,Chon15,Hoffman17}.  It is an exceptional place for aspiring Kardashev Type IV ETIs, and it will evolve into one of the most outstanding ``island universes'' as the accelerating cosmic evolution cuts off long-distance travel over the next trillion years \citep{Heyl05}.

Finally, about seventy ($\sim \edit1{9}\%$) targets stand apart from the conventional astrophysical phenomena.  \edit1{About t}wo thirds of them ($\sim 6\%$ of the catalog) are unidentified, consisting of both SETI-related and non-SETI related Anomalies.  Another fifth, in the unique phylum of technology, consists of our own technosignatures, which we may someday extend with non-Earthly examples.  Nine spurious and miscellaneous targets complete the Catalog.

Further discussion of specific issues in classification are in the Appendices~\ref{sec:AppendixPrototype} (Prototype),~\ref{sec:AppendixSuperlative} (Superlative),~\ref{sec:AppendixAnomaly} (Anomaly), and~\ref{sec:AppendixControl} (Control).  More extensive notes on the entries in the catalog and their selection will be available online.\footnote{At \url{http://seti.berkeley.edu/exotica}. \editbfTwo{The full appendices for version \CatalogVersion{} are also available at Zenodo: \dataset[doi:10.5281/zenodo.4726253]{https://doi.org/10.5281/zenodo.4726253}}}

We note that the entries in the catalog are not all independent entities.  Some are collective objects that contain other entries in the catalog, or are ``siblings'' in the same cluster.  In addition, some ``new'' objects in the \ExoticaCatalog{} are related to objects in the I17 catalog.  Examples include the several objects located in the Virgo Cluster, which itself is a Prototype.  In these cases, one observation may actually cover several objects, and they can be treated as all one object.  Table~\ref{table:Relationships} lists these relationships.

\subsection{Catalog presentation}
The objects in the catalog are listed in Table~\ref{table:FullSolSys} (targets within the Solar System) and Table~\ref{table:FullSidereal} (sidereal targets outside the Solar System).  Each distinct source has been assigned a unique identifier \edit1{that is simply an integer in the range 001 -- \NDistinct{}, with increasing values for objects that first appear in the Prototype, Superlative, non-SETI Anomaly, SETI Anomaly, and Control samples.}  These tables also include basic data about the sources, with references discussed in Appendix~\ref{sec:DataNotes}.

In addition, we will host an online database on the \ExoticaCatalog{} sources.  The database will include detailed notes motivating our selection of a particular object, links to the object on Simbad or other databases, and references for the object and its type.

\subsection{\editbfOne{The \ExoticaCatalog{}~and inhomogeneous observations}}

\edit1{In many ways, the \ExoticaCatalog{} is driven by theory, which both provides the motivation and guides the classification systems used to select Prototypes and Superlatives and evaluate Anomalies, and is intended for instruments, as a kind of ``treasury survey'' they may do.  As a high-breadth catalog, the \ExoticaCatalog{} represents an extremely diverse range of targets.  In general, the possible observations that can be done on themselves vary from object to object.  One clear case of this is the great difference between our knowledge of Solar System and sidereal objects.  Our space probes can journey to Solar System objects and even land on some, providing ``ground truth'' that is as yet not possible with objects outside the Solar System.  Whereas an exoplanet is generally understood in terms of its bulk properties (aside from those with temperature maps), Solar System bodies can be mapped with individual geographic regions and even individual landing sites studied.  Direct comparison can therefore be problematic.  Even in SETI, there are \emph{qualitatively} distinct technosignatures that can be sought in the Solar System, namely abandoned artifacts on planetary surfaces \citep{Davies12,Davies13}.}

\edit1{The observations will be relatively homogeneous for the instruments that Breakthrough Listen is currently using, however.  The biggest distinction among objects is whether they are resolved or not.  This distinction already exists in the I17 catalog, where some galaxies are resolved while other galaxies and all stars are not.  As with resolved galaxies in the I17 catalog, we expect we will observe a single pointing of the resolved objects and then add more pointings resource permitting.  All included Solar System objects except for the Sun and Moon are unresolved even with the GBT for $\nu \la 10\ \GHz$, and many to much high frequencies.  In all cases the data products will be the same for the radio instruments, a collection of spectral products and possibly some voltages, as was the case for our observations of 1I/'Oumuamua.  Of course, the sensitivity we will achieve for the objects in the Catalog will diverge by orders of magnitude, but this would be true even if we were observing low- and high-$z$ galaxies, for example.  We expect the same is true of other individual instruments that might observe the \ExoticaCatalog{}, including radio telescopes, to some extent X-ray telescopes (although planets will generally be resolved; c.f., \citealt{Bhardwaj07}), and gamma-ray and neutrino detectors \citep[e.g.,][]{Abdo11-Sun,Aartsen19-Sun}.}

\edit1{These considerations do result in the exclusion of one object that we would have included as a Prototype and an Anomaly (Table~\ref{table:ExcludedAnomaly}): the Earth itself.  Our ground-based radio instruments are only able to sample small regions of the Earth's radio environment.  We have no way of measuring the Earth's integrated radio emission, much less its optical technosignatures, except through the indirect method of moonbounce experiments \citep{McKinley12,DeMarines19}.}

\section{All-sky and random surveys}
\label{sec:WideField}

The least biased strategy of when and where to look is to look everywhere all the time, with an all-sky survey.  Even if ETIs inhabit objects we haven't even detected yet, we would still be able to see their activities.  Similarly, we would be able to look for natural objects that are completely unanticipated in our current paradigms \citep{Wilkinson16}.  Since the sky has already been mapped across the spectrum, such unanticipated sources either are intermittent, very faint, or are unrecognized in current data sets.  

The need for all-sky all-the-time surveys has been recognized in SETI and other fields of astronomy.  Ideally, we would be able to keep the data permanently from such sources.  This would both allow us to do searches for precursors when an anomalous event is triggered, or to find anomalies years later in re-analyses.  At present, no radio SETI facility exists with all these capabilities, despite the strong possibility that any ``beacons'' have a very low duty cycle to conserve energy.  In the future, dipole arrays may be constructed to fulfill this purpose \citep{Garrett17}.  In the optical and the near-infrared, the Pano-SETI experiment aims to monitor several steradians instantaneously for short pulses \citep{Cosens18}.  Outside of SETI, there are multiple instruments with wide-field high-cadence capabilities in the optical \citep{Law15}, X-rays \citep{Krimm13}, gamma-rays \citep{Atwood09,Abeysekara17-Transients}, neutrinos \citep{Aartsen17-Alerts}, and gravitational waves \citep{Abbott19-Alerts}.

Wide-field facilities allow us to set meaningful constraints on unanticipated events at the present.  Very wide-field capabilities are already granted by LOFAR and MWA \citep[e.g.,][]{Tingay16,Tingay18-GalAnticen}, and have been used for serendipitous constraints \citep{Tingay18-Oumuamua}.  MeerKAT, an array of 64 dishes of aperture 13.5 meters, will allow us to synthesize dozens of commensal beams within the relatively large primary field of view.  It will be possible to devote one beam to painting the entire primary dish field over and over, a miniature wide-field survey that is open to finding truly unknown sources at unknown locations.  The Breakthrough Listen backend will also perform incoherent summing of the dish voltages during observations.  Despite having less sensitivity ($S_{\nu}$ larger by a factor of $8$), the entire primary dish beam is covered, an instantaneous field thousands of times larger than a single synthesized beam.  The incoherent summing will be our first steps towards all-sky surveys, which have the maximal breadth allowed.  Because it needs to observe the extended Cherenkov light of air showers produced by gamma-rays, VERITAS always has a relatively wide field-of-view \citep{Weekes02}.  However, neither telescope's pointings are uniformly distributed across the sky.

A more limited unbiased search that is possible even on narrow field of view telescopes is to point at random targets on the sky.  Some completely random positions on the celestial sphere can be added as a supplement to the \ExoticaCatalog{}; they may be observed like any other target or with long integration times as a ``pencil beam'' survey to achieve sensitivity to faint flux levels or very intermittent sources.  This will also be possible with MeerKAT, which will perform long targeted observations of some sky regions during its nominal science operations.  We could devote synthesized beams to a random point in the dish field to do a pencil beam survey with every observation.  

The final possibility -- that anomalies already are present in extant datasets -- calls for keeping data until such time as innovative analyses can find them.  There has been plenty of precedent for this scenario: the long-term variability of Boyajian's Star was recovered in \emph{Kepler} data \citep{Montet16}; the Lorimer burst, the first reported FRB, happened in 2001 but was not reported until \citet{Lorimer07}; the first ANITA upwards neutrino shower was not reported for three years after its detection \citep{Gorham18}.  Breakthrough Listen is devoted to making its data publically available in accessible formats \citep{Lebofsky19}.  Moreover, on the analysis side, we are pursuing research in using machine learning to classify signals and look for anomalies \citep[e.g.,][]{Zhang19}.  These have the potential of evading our biases. 

\section{Exotica efforts and Breakthrough Listen}
\label{sec:Strategies}
As noted in Section~\ref{sec:CampaignsVsCatalogs}, Breakthrough Listen's exotica efforts will include both observations of the targets in the Catalog and a flexible series of campaigns.  In this section, we briefly discuss our engagement with exotica with both approaches.

\subsection{Strategy for observing the \ExoticaCatalog{}}
\label{sec:CatalogStrategy}

The full \ExoticaCatalog{} includes over \edit1{eight} hundred distinct sources, \edit1{over} 40\% of the number of targets in the I17 list.  Priority is given to the I17 targets, however, with the \ExoticaCatalog{} nominally getting $\la 10\%$ of telescope time.  This is especially a problem since many of the sources are extended.  An additional constraint is disk space, which prohibits keeping the raw voltages from radio observations of the entire Catalog.  We present some strategies for dealing with these issues, which may be useful to other programs aiming to observe the \ExoticaCatalog{}.

The core of our strategy is to develop a prioritization system for the targets.  With GBT, Parkes, and APF, we are free to select targets without the constraints of commensal observing.  Furthermore, the GBT has a number of frequency bands to observe in, and we can choose which ones to emphasize.  Thus, we do not need to carry out the full suite of observations we perform on the I17 stars and galaxies, and conversely we can spend more time on high-priority targets.  Table~\ref{table:PrioritizationRanks} breaks down the radio data products for one possible prioritization plan, as might be implemented on the GBT.  If all objects are observed with this plan, the \ExoticaCatalog{} would take $\sim 800\ \hr$ to observe and would generate $\sim \edit1{2.0}\ \PB$ of data ($\sim \edit1{1.6}\ \PB$ of filterbank files).  For comparison, the $\sim 1,700$ stars in the I17 catalog observed in all bands would take $\sim 8,000\ \hr$ and would generate $14\ \PB$ of data.  The amount of time spent at each rank is $150 \endash \edit1{350}\ \hr$.  Restricting our \ExoticaCatalog{} efforts to the low- and mid-frequency bands observed in \citet{Price20} would roughly halve the observing time, mostly spent on low priority targets.

\begin{deluxetable*}{lccp{0.15\linewidth}p{0.15\linewidth}p{0.15\linewidth}p{0.15\linewidth}}
\tabletypesize{\scriptsize}
\tablecolumns{7}
\tablecaption{Example prioritization system\label{table:PrioritizationRanks}}
\tablehead{\colhead{Rank} & \colhead{\SurveyCountTitle{}} & \colhead{Total data} & \colhead{Filterbank products} & \colhead{Raw voltages} & \colhead{Observed bands} & \colhead{Extended target strategy}}
\startdata
Maximal  & 12         & $\sim 0.6\ \PB$  & $3 \times 5\ \min$ ABACAD             & $5\ \min$ in a low-, medium-, and high-$\nu$ bands for at least one pointing     & All    & $\sim 5$ pointings, including core and axes\\
High     & 30         & $\sim 0.4\ \PB$  & $3 \times 5\ \min$ ABACAD             & $1\ \min$ in medium-frequency band for one pointing                              & All (one pointing), three (other pointings) & $\sim 3$ pointings\\ 
Medium   & 60         & $\sim 0.3\ \PB$  & $3 \times 5\ \min$ ABACAD             & None                                                                             & Three  & $\sim 3$ pointings\\
Low      & \editbfOne{$\sim 700$} & \editbfOne{$\sim 0.6\ \PB$}  & As available                          & None                                                                             & Single & $1$ pointing
\enddata
\tablecomments{The ABACAD (and related ABABAB) strategies) involve pointing the telescope on- and then off- target to constrain RFI; see \citet{Price20} for further information. The data rates assume that half the objects at each rank are extended, for an order-of-magnitude estimate.}
\end{deluxetable*}

As yet, prioritization is subjective.  We intend to favor extreme phenomena: those with high luminosities and nonthermal emission across the spectrum. These are more likely to display new astrophysical phenomena.  Higher luminosities also make them better suited for modulation as beacons by ETIs. Thus, we will include proportionally more pulsars, interacting binaries, and AGNs. We specifically will put maximum priority on the primary calibrators at radio frequencies and high energies: Cygnus A and the Crab Nebula, respectively \citep{Perley17,Kirsch05}.  The Crab Nebula specifically is known to flare \citep{WilsonHodge11,Tavani11,Abdo11-CrabFlare}. Another thing we have in mind is to achieve the widest breadth in astrophysical environments and object type.  We will assign high-to-maximal priority to at least one object of every phylum priority list.  We also intend to prioritize Superlatives with extreme luminosities and several Anomalies.

The APF is capable of collecting high resolution spectra of exotica targets.  Target acquisition is challenging for targets fainter than $V \sim 12$ and those with large motions across the sky.  As such, we expect to only be able to observe a limited subset of the catalog.  Such a search for narrow wavelength signals would be different than that in \citet{Lipman19} since the targets are producing almost no astrophysical optical signal.

Our time on MeerKAT and VERITAS is commensal; thus, we will be unable to choose where the telescopes are pointed or prioritize different targets.  Some of the Southern targets are likely to be primary-user targets of MeerKAT during its nominal observations.  In these cases, we will be able to re-use the primary synthesized beam data for our observations.  Others will likely enter the field of view during large sky area surveys.  Additionally, Breakthrough Listen's backend will allow us to form dozens of commensal beams within the dish field of view.  While most of these beams will be dedicated to the one million star effort of Breakthrough Listen, several will typically be available for exotica observing efforts.  From time to time, we may substitute a suitable object in the MeerKAT field for a given Prototype by using supplementary catalogs.  VERITAS has a wide field of view, with diameter of several degrees \citep{Weekes02}.  Many of the objects in the \ExoticaCatalog{} have been primary targets of VERITAS.  Others are clustered near the targets, like M81 and NGC 3077 near M82 \citep{Acciari09-M82}, or the galaxies in the Virgo and Perseus Clusters \citep{Acciari08-M87,Acciari09-N1275}.  Thus, for many of the Northern \ExoticaCatalog{} targets, it will simply be a matter of collecting VERITAS's commensal observations and organizing them.

The large angular extent of many of the objects in the catalog pose another challenge, especially in high-frequency radio observations where the beams of the telescopes are very small.  It would be impractical to map the full extent of the sources with the GBT and Parkes.  For the low priority sources, we will observe only the core or flux peak of the target, as a proxy for the potential for activity.  There will be additional pointings for higher priority sources, resource permitting, generally including two at the ends of the semimajor axis.  In some cases where there are multiple well-defined interesting spots, we will target these (e.g., the hotspots of radio galaxies).  Parkes' multibeam receiver will also be useful, since it covers seven times the sky area, although its frequency range is limited.  The electronic beamforming capabilities of MeerKAT will afford us more flexibility; we will be able to dedicate a synthesized beam to ``painting'' the entire primary field of view, allowing us to construct low sensitivity maps.  The APF's nominal field of view of 1 square arcsecond will limit its use for extended exotica objects.  VERITAS should be able to observe all but the largest targets in their entirety at any given moment.  

Another part of our strategy will be to opportunistically take advantage of available time.  As we complete observations of the I17 sample with our main radio facilities, the GBT and Parkes, more time will be freed up for exotica. Another tactic is to make use of the ON/OFF strategies we use to constrain RFI: we can use an exotica source as an OFF-source during observations of a program star or vice-versa, whereas previously we have used stars.  The OFF observations themselves can be used for technosignature searches, using the ON-observations to check for RFI \citep{Price20}. Additionally, the remaining GBT observations are mostly at high frequency, and are sensitive to the weather.  When weather conditions preclude high frequency observations, we can observe exotica sources at low frequency.  A similar strategy may be employed at other facilities, perhaps taking advantage of time when the Moon is up for at least the brighter sources.

Finally, some of the targets in the \ExoticaCatalog{} have already been observed as part of previous studies. These will not need to be re-observed. This too could be adapted to other facilities, adapting the catalog to include sources that have already been covered.

\subsection{Exotica campaigns and Breakthrough Listen}
\label{sec:Campaigns}
Several focused programs to observe unusual classes of objects have been completed, and more are ongoing. These single- or few-target campaigns are complementary to both our primary high-\SurveyCountLower{} effort in observing stars and galaxies over a hundred nearby galaxies and also the high-breadth catalog.  These have included:
\begin{itemize}
\item \emph{Interstellar asteroids: A test case for SETI} -- The interstellar objects 1I/'Oumuamua \citep{Bannister19} and 2I/Borisov \citep{Guzik20} were observed by us as they passed through the Solar System.  Starfaring ETIs might very slowly diffuse through the galaxy by hitching rides on comets. More pragmatically, our observing campaigns for 1I/'Oumuamua and 2I/Borisov act as rehearsals if a convincing ETI probe is detected.  The 'Oumuamua observations included two hours each in L, S, C, and X-bands, allowing us to cover the entire rotation cycle of the object to an EIRP sensitivity of 0.08 W, comparable to some Bluetooth or WiFi transmitters \citep{Enriquez18}.  We also observe Solar System asteroids that have been postulated to be captured interstellar objects, the first being the observation of (514107) 2015 BZ$_{509}$ on Parkes for ten minutes of total integration \citep{Price19-BZ509}.  These observations are being supplemented by observations of other candidate captured interstellar objects, effectively adding a new class of objects to the Breakthrough Listen program.

\item \emph{FRB 121102 and other FRBs: Transient astrophysics} -- The Breakthrough Listen backend can also be applied to phenomena thought to be natural.  In \citet{Gajjar18}, we observed the repeating FRB 121102 with the backend on the GBT for an entire hour.  On average, one burst was detected every three minutes, achieving the high-
est frequency detections of the anomalous object; we also independently confirmed the high rotation measure discovered by \citet{Michilli18}.  The long time devoted to these observations allowed us to observe and characterize many bursts, which would not have been possible with a five minute observing time.  This highlights the advantage of the flexible nature of campaigns, allowing us to achieve high depth.  The backend of Parkes was able to record a real-time fast radio burst, FRB 180301, in commensal mode \citep{Price19-FRB}.  Significantly, we were able to capture raw voltages on this anomalous object, greatly increasing the analysis possibilities.  We expect to observe other FRBs in the future.

\item \emph{Ultracool dwarfs: Extending I17} -- We have devoted time to specifically observe five nearby ultracool dwarfs of spectral types late M to early Y with Parkes, as reported by \citet{Price20}.  The observations of each object was otherwise similar to those of the stars in the catalog.  The purpose of these observations was to extend the completeness of I17, effectively supplementing it with a new object type.  In the future, it might be possible to further increase the I17 catalog's breadth by running short campaigns on other object classes, such as nearby white dwarfs \citep[c.f.,][]{Gertz19}.

\item \emph{SETI campaigns on stars: Following up on SETI candidates} -- Campaigns have also been undertaken on the facilities to test claims of technosignatures.  \citet{Borra16} reported the existence of a spectral comb in the Sloan spectra of dozens of Sun-like stars, which they interpreted as resulting from coherent pulses repeating on picosecond timescales.  The two brightest of these stars were observed with the APF, but no signs of a spectral comb were found in the spectra \citep{Isaacson19}.  Other campaigns to follow up on individual objects that have been noted by SETI researchers include optical observations of Boyajian's Star with VERITAS \citep{Lipman19}, radio observations of the Random Transiter \citep{Brzycki19}, and radio follow up of an ETI candidate signal from Ross 128 \citep{Enriquez19}.  These observations were scheduled rapidly, within four days for Ross 128. With campaigns, new high-priority targets can be followed up rapidly, without waiting for the catalog to be updated.
\end{itemize}

These efforts demonstrate the uses of campaigns to supplement the \ExoticaCatalog{}: (1) rapid follow-up of new and time-sensitive discoveries; (2) deep observations of high-priority targets; (3) supplementing the primary program to expand its breadth; and (4) moderate \SurveyCountLower{} programs of unconventional targets.  These advantages should apply to other programs as well.

\section{Summary}
\label{sec:Summary}
Breakthrough Listen intends to spend $\sim 10\%$ of its observing time on exotic objects.  There are many reasons to search for technological intelligence in unconventional places.  Unearthlike or nonbiological entities will not be constrained to live in Earthly habitats hospitable to lifeforms like us.  It is also conceivable that some kinds of seemingly natural phenomena are the result of alien engineering.   Yet there are good motivations for observing unusual objects even if ETIs cannot possibly live there.   Extreme, energetic objects are more likely to produce unusual signals, particularly transients, that might be confused with artificial signals.  Breakthrough Listen has unique instrumentation, and observation of a broad range of objects would benefit the general astronomy community.  Finally, there could be unaccounted for systematic errors in our systems that give false positives.  Observing exotic objects and empty regions on the sky allow us to constrain these possibilities.

Whereas the nearby star and galaxy catalogs aim for depth and high \SurveyCountLower{}, efforts to look for exotica can be more broad, trying to peek at everything that might be interesting.  Breakthrough Listen's approach to exotica is twofold.   First, from time to time, we have dedicated programs that observe one to a few objects of one class.  Although we sample only a few types of object this way, these types are very diverse.   Efforts include observations of the interstellar minor body 'Oumuamua and the repeating fast radio burst FRB 121102.  By focusing these efforts on just a few objects, we can achieve deep sensitivity with the long integration times devoted to them.

The other prong is the \ExoticaCatalog{}, which is the main focus of this paper.  The \ExoticaCatalog{} aims to achieve the widest possible breadth, including the most diverse range of celestial phenomena possible.  The catalog itself contains four samples:
\begin{itemize}
\item The Prototype sample includes an archetypal object of each type of astrophysical phenomenon.
\item The Superlative sample list record-breaking objects with the most extreme values of basic properties like mass.
\item The Anomaly sample contains objects and phenomena that are inexplicable by current theories, including candidate positive events from other SETI efforts.
\item A small Control sample includes objects once thought to be anomalous or noteworthy but that turned out to be banal or non-existent, and will be used as a check on systematics and to rehearse observations.
\end{itemize}
There are \NDistinct{} objects in the total \ExoticaCatalog{}, which \edit1{will be} sorted into different levels of priority.  About a dozen sources will be observed with maximum priority and will include raw voltage data in multiple bands.  Most will be low priority and will be observed as allowed by time.

We hope that the \ExoticaCatalog{} will prove useful to other efforts, both within SETI and outside it, in characterizing the whole panoply of objects in the known Universe.

\acknowledgments
{Breakthrough Listen is managed by the Breakthrough Initiatives, sponsored by the Breakthrough Prize Foundation. 

This work would not have been possible without the extensive use of a number of online databases.  It has made thorough use of the SIMBAD database, the VizieR catalogue access tool, and the cross-match service XMatch, all operated at CDS, Strasbourg, France.  We have also benefited immensely from the NASA Astrophysics Data System Bibliographic Services and the arXiv preprint server. This research has made use of the NASA/IPAC Extragalactic Database, which is funded by the National Aeronautics and Space Administration and operated by the California Institute of Technology.

\editbfOne{We thank the referee for their comments, and we thank Jill Tarter and Ken Shen for suggestions and clarifications.}}

\clearpage
\appendix
\restartappendixnumbering

\section{The Prototype sample}
\label{sec:AppendixPrototype}

\subsection{Minor bodies}
\label{sec:MinorBodies}
We classify Solar System minor bodies according to both orbital family and composition, with a small number of additional subtypes.  Minor bodies of specific compositions might be selected by ETIs for mining \citep[c.f.,][]{Papagiannis78}.  From a SETI perspective, orbital families might be targeted by ETI probes to provide a unique vantage point \edit1{over} bodies like the Earth, or because they are dynamically stable for long periods of time and could accumulate a large number of artifacts \citep[e.g.,][]{Benford19}.  There is a large overlap in some cases \edit1{between spectral and orbital groups} \citep[as in][]{DeMeo14}, as with the E-belt and E-type asteroids, for which we use the same Prototype. 

For asteroids, our spectral-type system is largely taken from \citet{Tholen84} \citep[see also][]{Tedesco89}.  We selected those types considered the most significant by \citet{Tholen84}, adding those unique to one or a few members.  Some intermediate classes that blend into larger ``complexes'' in the more recent \citet{Bus02} taxonomy were omitted.  In choosing the Prototypes, we were guided by the classifications of \citet{Tholen84}, \citet{Tedesco89}, and \citet{Bus02}.

The comet orbital classifications were informed by \citet{Levison96}.

``\edit1{Distant minor planets}''\edit1{, adapting the ``distant objects'' term used by the Minor Planet Center,}\footnote{\edit1{\url{https://www.minorplanetcenter.net/iau/mpc.html}}} refer to outer Solar System bodies beyond the Jupiter Trojans that are not comets.  The spectral type system is that of \citet{Barucci05} and \citet{Fulchignoni08}, with the latter guiding our Prototype selection.  The division into orbital groups is based on the system in \citet{Gladman08}, which we consulted especially when selecting Scattered Disk and Detached objects.  We aimed to select Prototypes that are almost certainly minor bodies and not dwarf planets, as indicated by a ``probably not dwarf planet'' designation on Mike Brown's website.\footnote{\url{http://web.gps.caltech.edu/~mbrown/dps.html}}

The small classification system for satellites into regular, irregular, and ``collisional shards'' in \citet{Burns86} informs our classification.

\citet{Hughes18} informed our grouping of debris disks into cold, warm, and hot/exozodis.

\subsection{Solid planetoids}
\label{sec:SolidPlanetoids}
Planets were classed according to size and stellar insolation.  \emph{Kepler} result papers classifies planets by radius into: Earths ($<1.25\ \REarth$), Super-Earths ($1.25 \endash 2\ \REarth$), Neptunes ($2 \endash 6\ \REarth$), Jupiters ($6 \endash 15\ \REarth$), and non-planetary ($>15\ \REarth$) \citep{Borucki11,Batalha13}.  We adjusted this scheme by: (1) setting the boundary between solid super-Earths and giant Neptunes at $1.5\ \REarth$, except when density is known \citep{Rogers15,Fulton17}; (2) adding a sub-Earth category for radii $<0.75\ \REarth$ to cover planets like Mars where the habitability prospects are likely different \citep[e.g.,][]{Wordsworth16}; (3) using the \citet{Weiss14} relation to translate the radii categories into mass categories when no radius is available, using the mean $\sin i$ of $\pi/4$ as a guide.  The mass categories for solid planetoids are sub-Earths ($\la 0.4\ \MEarth$), Earth ($\sim 0.4 \endash 2\ \MEarth$), and super-Earths ($\sim 2 \endash 4.5\ \MEarth$, unless densities are known).   

We use the terms ``hot'', ``warm'', ``temperate'', and ``cold'' to group by insolation.  Cold planets are outside the conventional habitable zone (roughly taken to be $\la 0.25$ Earth), temperate planets are within the conventional habitable zone ($\sim 0.25 \endash 2$ Earth), warm planets are interior to the habitable zone but with insolations $\la 100$ Earth, and hot planets have insolations $\ga 100$ Earth.  The distinction between ``warm'' and ``hot'' carries over from the giant planets, where warm planets around Sunlike stars are defined by period or semimajor axis \citep{Dong14,Huang16,Petrovich16}. \edit1{The insolation range for ``temperate'' planets was chosen somewhat arbitrarily, partly to include Mars on the outside and leave no gaps with the ``warm'' planets as defined in the literature.  \citet{Kopparapu14} finds habitable zone boundaries that range from 0.8--1.4 Earth on the inside (generally near 1.0 AU) to 0.2--0.4 Earth on the outside.  All the ``temperate'' Prototypes have insolations of 0.39--1.1 Earth, a more conservative range, except for the temperate Jupiter HD 93083b with 1.8 Earth.}

Dwarf planets in the Solar System are classed according to the spectral type and orbit, similarly to minor bodies (Appendix~\ref{sec:MinorBodies}). \edit1{With the exception of Sedna, only the IAU-recognized dwarf planets (Ceres, Pluto, Eris, Makemake, Haumea) are listed as dwarf planets.}

We added a ``geological'' classification intended to very roughly sample the diversity of surface environments and histories in the Solar System, excluding the Earth itself.

\subsection{Giant planets}
Giant planets are classed according to insolation and size (see Appendix~\ref{sec:SolidPlanetoids}).  Where possible we use densities to distinguish between ice giants and gas giants.  We also added a ``Superjovian'' category to cover a better range of planetary masses.  In the literature, the minimum mass for ``superjovians'' has been defined as $1$, $3$, and $5\ \MJupiter{}$ \citep{Clanton14,Johnson09,Currie14}.  We arbitrarily are guided by a threshold of $\sim 3\ \MJupiter$, and allow $M \sin i$ values of $2 \endash 10\ \MJupiter$.

\edit1{An additional subcategory classifies giant planets according to (non-main sequence) stellar host, with a final entry for a resonant chain system.}

\subsection{Stars}
Our estimation is that the major distinction between different types of stars is based on evolutionary stage and stellar mass.

Low mass protostars and pre-main sequence stars are classed numerically in the literature as 0, I, II, and III according to obscuration, where II and III correspond to T Tauri stages \citep{Lada87}.  We retain the distinction between class 0 and 1 protostars and T Tauri stars.  High mass protostars are given their own categories.  

Sub-brown dwarfs are too small to have fused deuterium but are believed to have formed similarly to stars\edit1{, as opposed to giant planets} \citep{Caballero18}.

Brown dwarfs and main sequence (MS) stars are classed by Harvard spectral type.  Each spectral type is divided into early (0-3), mid (4-6), and late (7+) subdivisions.  Where possible we chose spectral standards as Prototypes \citep{Morgan73,Kirkpatrick91,Walborn02,Kirkpatrick05,Burgasser06}.  We also favored stars in I17, because we have already observed a wide range of B through mid-M dwarfs.  For brown dwarfs, we also tried to choose those with a mass that was clearly below the hydrogen-burning limit.

Covering post-MS evolution is more complicated.  Stars that are actually in distinct evolutionary stages can appear on the same place in the HR diagram, such as the red giant branch and the asymptotic giant branch.  To start, we divided post-MS stars into mass groups with qualitatively different evolution:
\begin{itemize}
    \item \emph{Very low mass stars} (initial mass $\la 0.2\ \Msun$) are not predicted to have a red giant phase, but no isolated post-MS examples are known \citep{Laughlin97}\edit1{.}
    \item \emph{Low mass stars} (initial mass $\sim 0.2 \endash 2.2\ \Msun$) pass through a red giant branch (RGB) phase terminating in a \edit1{helium flash in their degenerate cores}.  The RGB star rapidly ($\ll 10^6\ \yr$) transitions to a Core Helium Burning (CHeB) phase before ascending the Asymptotic Giant Branch (AGB).  The ultimate remnant is a CO white dwarf.
    \item \emph{Intermediate mass stars}\footnote{\edit1{\citet{Karakas14} calls these lower-intermediate mass stars.}} (initial mass $\sim 2.2 \endash 7\ \Msun$) ascend the RGB but do not have a helium flash, settle gradually into the CHeB phase, possibly executing a ``blue loop'', before ascending the AGB.  The ultimate remnant is a CO white dwarf \citep{Karakas14}.
    \item \emph{Transitional mass stars}\footnote{This term does not appear in the literature; we use it to indicate they have characteristics similar to smaller intermediate mass stars (no core collapse supernova) and massive stars (later stages of nuclear burning, possible neutron star remnants).  \citet{Karakas14} calls these stars \edit1{middle}- and \edit1{massive}-intermediate mass stars.} (initial mass $\sim 7 \endash 11\ \Msun$) proceed similarly to the intermediate mass stars until the end of the AGB phase, whereupon they begin carbon burning as a super-AGB phase.  Those with lower mass end as an ONeMg white dwarf, while larger ones may undergo electron capture supernovae and possibly leave behind neutron stars \citep{Karakas14,Woosley15,Jones16}.  The division between these two fates is not precisely known, so we do not make the distinction.
    \item \emph{Massive stars} (initial mass $\sim 11 \endash 40\ \Msun$) undergo later stages of core nuclear burning.  They switch between being red and blue supergiants during these later stages \citep{Gordon19-YHyGs}.  They may have pronounced mass loss that transforms them into Wolf-Rayet stars \citep{Clark12-BlueHyGs}.  Massive stars end with a core collapse leaving behind a neutron star or black hole \citep{Heger03}.
    \item \emph{Very massive stars} (initial mass $\ga 40\ \Msun$) leave the main sequence but are unable to become red supergiants\edit1{, probably because of their extreme mass loss \citep{Humphreys79,Woolsey02}}.  They instead become blue supergiants and blue hypergiants, \edit1{and after the mass loss,} Wolf-Rayet stars \citep{Clark12-BlueHyGs}.  The stellar remnant is a black hole or nothing at all \citep{Heger03}.
\end{itemize}
ETIs living around stars in different groups would face different challenges when adapting to post-MS evolution (for example, the post-helium flash contraction would require large-scale migration over just a few millennia to remain in the habitable zone).  For low- and intermediate-mass stars, we preferred to use \emph{Gaia} benchmark stars with well-determined masses \citep{Heiter15}. 

The evolutionary stages are supplemented with stars with atypical characteristics, including chemically peculiar stars, Be stars with decretion disks, pulsar-like stars, Population II stars, and a collection of pulsational variables.  Peculiar stars that are the result of stellar mergers are emphasized because of their diverse and unique evolutionary histories \citep[e.g.,][]{Jeffery08-Types,Heber16}.  

\subsection{Collapsed stars}
These are divided into white dwarfs, neutron stars, and black holes.  

White dwarfs are mainly grouped according to mass \citep{Liebert05} \edit1{or spectral type \citep{Sion83}} with supplemental subtypes based on evolutionary, composition, magnetic, \edit1{or} variability characteristics.

Neutron stars are grouped primarily by their rotation rate and magnetic field.  These parameters also control the emission we observe and are related to evolutionary state \citep[e.g.,][]{Alpar82,Olausen14}.

Black holes are grouped by detection method.  Only a few detached black holes are known \edit1{with firm positions, and many candidates are disputed \citep[e.g.,][]{ElBadry20,vandenHeuvel20,Bodensteiner20}}, constraining our choice of Prototypes.

\subsection{Interacting binary stars}
We group interacting binary stars powered by accretion by the nature of the mass donor and that of the recipient:
\begin{itemize}
    \item \emph{Semidetached and contact binaries} -- both components are stars \edit1{(i.e., not stellar remnants)}.
    \item \emph{Symbiotic stars} -- donor is a giant, recipient is a small early-type star or a white dwarf \citep{Belczynski00}. \edit1{Symbiotic systems including neutron stars are listed under X-ray binaries.}
    \item \emph{Cataclysmic variables} (CVs) -- donor is a late-type dwarf star, recipient is a white dwarf.  They are further divided by variability/eruption characteristics \citep{Osaki96,Schaefer10} and white dwarf magnetic field interaction with the accretion disk \citep{Patterson94}.  Closely related are the AM CVn binaries, where the donor is a helium star or white dwarf, and the close binary supersoft sources where the donor is a higher mass subgiant \citep{Kahabka97}.
    \item \emph{X-ray binaries} -- The recipient is a neutron star or black hole.  They are further divided based on the mass of the donor (low mass or high mass), and still further by the mode of mass transfer and other characteristics \citep[see especially][]{Reig11,Kaaret17}. \edit1{X-ray binaries with an extreme super-Eddington accretion rate or luminosity (including the ultraluminous X-ray sources) are given their own special subcategory.  Two more empirical types where the recipient's nature is indeterminate round out the category.}
\end{itemize}.  

In a few cases, stellar outflows rather than accretion dominates the system.  These include systems where shocks dominate the luminosity \edit1{(like colliding wind binaries)}, and the spider pulsars where a formerly-accreting neutron star ablates its companion \citep[e.g.,][]{Dubus13,Roberts13-Spiders}.

\subsection{Stellar groups}
We include non-interacting binary and multiple stars with other stellar groups like star clusters.  As in \citet{Eggleton08}, they are distinguished from clusters by their hierarchical organization.  Binary systems are well represented in the I17 catalog, and we do not try to capture all combinations of stellar types or separations.  We specifically include double degenerate systems, \edit1{however,} which are not included \edit1{in} I17, and have the potential to be sites of ETI activity \citep{Dyson63}.  \edit1{A few binaries classified by how they are detected from Earth or physical effects (heartbeat, eclipsing and self-lensing, chromospherically active) are included when they indicate distinct physical phenomena could be exploited for observation coordination by ETIs; visual, astrometric, photometric, and beaming binaries are excluded, however.}

Globular clusters are classified according to the orbit classification of \citet{Mackey05}; additional subtypes are based on internal structure and luminosity.  Other stellar clusters are divided into massive super star clusters, \edit1{``faint fuzzies'',} nuclear clusters (including former nuclei of dwarf galaxies), and open star clusters.

Some unbound stellar associations are also included, when well-studied examples were not too large on the sky to be practically observed.  

\subsection{Interstellar medium and nebulae}
The interstellar medium (ISM) as a whole has a complicated turbulent structure, although it is classically divided into hot, warm, and cool and cold phases \citep[e.g.,][]{Cox05}.  Some structures in the interstellar medium are too large to practically study with our facilities: the hot ISM, the loops, the warm ionized medium, and so on.  In terms of the general ISM, we focus on molecular clouds and HII regions, which are relatively compact.  These are classed according to column density into translucent and dark clouds, with the dark clouds further divided by scale.  Molecular clouds have self-similar structure, and although they are sometimes labeled as complexes, clouds, clumps, and cores in the literature, the divisions are arbitrary \citep[see][]{Wu10}.  HII regions formed within the molecular clouds are also included and likewise grouped by density/size \citep{Habing79}.   

Most of the entries in this phylum are structures produced by outflows from central engines and their interaction with the general ISM.  These are grouped according to the nature of the central engine.

Additionally, we include a bubble of cosmic rays to represent the nonthermal ISM, and two circumgalactic medium clouds that we judged practical to observe.

\subsection{Galaxies}
We regard the fundamental distinction between galaxies as based on their specific star-formation rate (sSFR), the ratio of star-formation rate and stellar mass.  There is a natural division in this parameter plane that is robust out to high redshift \edit1{($z$)} into quiescent galaxies, intermediate galaxies, ``main sequence'' star-forming galaxies, and starbursts \citep{Brinchmann04,Elbaz11,Speagle14,Renzini15}.  The first three categories translate into different features on a color-magnitude diagram: the red sequence, green valley, and blue cloud, respectively \citep[e.g.,][]{Strateva01,Wyder07}.  The abundance of phenomena that might affect galactic habitability or could be used for astroengineering, like core collapse supernovae, is tied to sSFR.  At low redshift, quiescent galaxies are associated with early-type morphologies (ellipticals, spheroidals, and lenticulars), while main sequence galaxies are associated with late-type morphologies (late-type spirals and irregulars), but the correlation is not exact and we specifically include outliers as subtypes.  At \edit1{$z \sim 0$}, there is also a correlation with environment, with quiescent galaxies more often located in clusters, although we again include outliers.

Among the quiescent galaxies, there appears to be a robust division of most large ellipticals into two types: boxy/cored and disky/coreless.  The division is based on surface brightness profiles, shape, and X-ray emission \citep{Kormendy09}.  \edit1{A related division (not included in this version of the catalog) is between fast rotators and slow rotators, where (massive) slow rotators have the more boxy isophotes, brighter X-ray emission, and central light deficits of the boxy/cored ellipticals and the disky/coreless properties correlated with the fast rotators \citep{Emsellem07,Emsellem11,Sarzi13}.  The prototypes for the boxy/cored and disky/coreless galaxies were chosen to be unambiguous slow and fast rotators according to \citet{Emsellem11}.} Small early-type galaxies tend to be divided into high-density compact galaxies and low-density dwarf galaxies.  There is a vigorous debate in the literature about which are more likely to be the analogs of large ellipticals, and which form a separate sequence \citep[e.g.,][]{Graham03,Kormendy12}.  Dwarf quiescent galaxies are arbitrarily divided into dwarf elliptical, spheroidal, and ultrafaint simply to cover a full range in mass. 

Green valley galaxies are a heterogeneous class, and are here classed mainly by the mode of their passage through the ``valley'' \citep{Salim14,Schawinski14}.

The blue main sequence galaxies are very diverse.  Note the characteristic sSFR decreases with time since the Big Bang \citep{Speagle14}: a galaxy that would be classified as main sequence at $z \sim 2$ would be considered a starburst now\edit1{, and thus their $z \sim 0$ analogs are placed among the starbursts}.  In the present-day Universe, these are classified coarsely by morphology (see below for discussion of fine morphology types), with a few subtypes each of late spirals and irregulars.  We chose Prototypes mainly based on having consistent morphological types between \citet{deVaucouleurs91}, \citet{Karachentsev13}, \citet{Ann15}, and \citet{Buta15}. 

High-redshift star-forming galaxies have been classified mainly by spectrophotometric characteristics (e.g., BzK galaxies satisfy certain criteria in $(B - z)$ and $(z - K)$ colors).  These subtypes can include both main sequence and true starburst galaxies.  Our choice of Prototypes is constricted by the need for them to be gravitationally lensed to boost our sensitivity.  In some cases we were unable to find a likely candidate of a common class of high-redshift galaxies (particularly, no lensed main sequence Lyman \edit1{Alpha E}mitters or main sequence Lyman Break Galaxies).  

Starbursts are those galaxies with sSFRs significantly above the typical sSFR of star-forming galaxies at their redshift \citep{Elbaz11}.  We group them into nuclear starbursts occurring in the centers of larger galaxies, and dwarf starbursts occurring in small galaxies.  We also specifically include some relatively nearby starbursts noted to have properties analogous to high-$z$ galaxies, in addition to some lensed starbursts at high redshift, to further constrain the possibility that habitability evolves with time.

Disturbed galaxies broadly fall into three types \edit1{of unrelated origins}: the ring galaxies (which themselves have diverse origins), interacting galaxies, and galaxies affected by ram pressure stripping in an intracluster medium.

We add a catchall class of ``morphological subtypes'', which includes examples of galaxies hosting many kinds of features, particularly those found in galactic disks.  There are many morphological classification schemes for disk galaxies.  The basic sequence from early to late types is universal \citep[e.g.,][]{Hubble26} and is included in the previous classes.  In many traditional systems, the disk galaxies are classified by the strengths of their bar patterns \citep{deVaucouleurs91,Graham19}.  \citet{vanDenBergh76} instead classifies them according to the prominence of their arms from spirals to ``anemic'' galaxies to lenticulars \citep{Kormendy12}.  Spiral arms themselves come in a great many varieties, from grand design to flocculent varieties \citep{Elmegreen82}.  Added to this are many other obvious morphological features in disk galaxies: rings, pseudo-rings, lenses, plumes, and more, all with a number of variants \citep{Buta15}. Each feature adds another dimension to parameter space.  The resultant morphological types are lengthy and can vary from paper to paper.  To avoid combinatorial explosion, we just have a list of possible features \citep[see][for detailed discussion of these features]{Buta15}.  Some galaxies here are Prototypes for several types to minimize the number of galaxies observed.  Prototypes are chosen by their classification in \citet{Comeron14} and \citet{Buta15}, especially if they are given as explicit examples of a morphology in Table 1 of \citet{Buta15}.

Finally, we include \edit1{three} types of galaxies defined \edit1{by their relationships with their cosmic environment}.

\subsection{AGNs}
There are a plethora of classification schemes for AGNs, as reviewed in \citet{Padovani17}.  We use a canonical division of AGNs into the major divisions of LINERs, Seyferts, radio galaxies, quasars, and blazars as the foundation \citep[e.g.,][]{Peterson97}.  Except for LINERs, these are further subdivided into the common categories inspired by optical and radio characteristics \citep[as in][respectively]{Osterbrock77,Fanaroff74,Kellermann89,Ghisellini11}.  These main object types represent different luminosities, radio-loudness, and viewing angle, with some admitted overlap between the classes \citep{Urry95,Padovani17}.

Additional classes were added to cover objects with unusual spectral or morphological features, or the presence of multiple supermassive black holes.

A few auxiliary objects related to AGNs have also been included: megamasers and AGN relics (\edit1{a} voorwerp and \edit1{a} fossil AGN).

\subsection{Galaxy associations}
Galaxy associations are mainly classified by richness, from isolated pairs of galaxies through groups and clusters.  Only compact and ``fossil'' groups are included due to practicality considerations, as neither \edit1{of these types} is vastly larger than a galaxy \citep{Hickson93}\edit1{, whereas nearby galaxy groups cover too much of the sky to practically observe}.  A very simple galaxy cluster classification scheme is used, based on relative symmetry and richness \citep[compare with the more elaborate systems in][]{Bahcall77}.  A high-redshift protocluster, a grouping which has not yet virialized at the time of observation, is in the sample.

To the structures themselves, we also included examples of features in the intracluster medium (ICM) of galaxy clusters, both thermal and nonthermal \citep[e.g.,][]{Markevitch07,vanWeeren19}. 

\subsection{Large-scale structures}
Most large-scale structures \edit1{(which include superclusters, voids, and ``Great Walls'' of galaxies)} are too diffuse and large to observe practically.  The included ``attractor'' and ``repeller" points are not physical objects, but instead indicate local sinks and sources in the peculiar velocity of galaxies.  They do roughly correspond to a dense group of clusters and a void, respectively \citep{Hoffman17}, and may draw the attention of ETIs as special places.

\subsection{Technology}
Which active satellites are available for observation will depend on new launches and re-entries.  Although we list some major classes of satellite, the selection of \edit1{most of the individual} sources will be opportunistic.

\subsection{Not real}
We include the Solar antipoint as a ``source'' because of its special significance in SETI.  The Earth transits the Sun as seen by observers at stars in this direction.  It has been suggested that ETIs that observe Earth transits would be especially motivated to broadcast in our direction because they know a habitable planet exists; furthermore, the transit itself can be used for synchronization \citep{Shostak04,Heller16,Sheikh20}.

\edit1{One each of the stable Earth-Moon and Earth-Sun Lagrange points is included.  These points have been proposed as locations where probes may reside \citep{Freitas80,Benford19}.}

\edit1{Like the Solar antipoint, the Galactic anticenter is included as a special point on the sky.  \citet{Benford10-Receive} proposes that ETIs would preferentially beam transmission to and away from the Galactic Center, since it defines a natural axis or corridor.  Thus, we might then expect to see transmissions from ETIs further out in the disk than us from the location of the Galactic anticenter.}

\subsection{\editbfOne{Tables}}

Table~\ref{table:PrototypeSample} lists the entire Prototype sample, organized by the type of objects they are supposed to represent.

We also list transient phenomena in Table~\ref{table:Transients}, organized according to their predictability. Specific examples listed may not display the transient phenomenon at any one given time.

\startlongtable
\begin{deluxetable}{lllccc}
\tabletypesize{\scriptsize}
\tablecolumns{6}
\tablecaption{Prototype \editbfOne{s}ample\label{table:PrototypeSample}}
\tablehead{\colhead{Type} & \colhead{Subtype} & \colhead{Prototype} & \colhead{ID} & \colhead{Solar?} & \colhead{Ref}}
\startdata
\cutinhead{Minor body}
Asteroid                     & A-type                           & 446 Aeternitas              & 001  & \checkmark & \nodata\\
                             & C-type                           & 52 Europa                   & 002  & \checkmark & B02    \\
                             & D-type                           & 624 Hektor                  & 003  & \checkmark & \nodata\\
                             & E-type                           & 434 Hungaria                & 004  & \checkmark & B02    \\
                             & M-type                           & 16 Psyche                   & 005  & \checkmark & \nodata\\
                             & O-type                           & 3628 Bo\v{z}n\v{e}mcov\'a   & 006  & \checkmark & B02    \\
                             & P-type                           & 420 Bertholda               & 007  & \checkmark & \nodata\\
                             & Q-type                           & 1862 Apollo                 & 008  & \checkmark & B02, T84\\
\enddata
\end{deluxetable}

\startlongtable
\begin{deluxetable*}{p{2cm}p{3cm}lcp{2cm}p{2cm}cc}
\tabletypesize{\scriptsize}
\tablecolumns{8}
\tablecaption{Properties of transient phenomena\label{table:Transients}}
\tablehead{\colhead{Phylum} & \colhead{Type} & \colhead{Example host} & \colhead{Host ID} & \colhead{Duration} & \colhead{Recurrence} & \colhead{Emission} & \colhead{Ref}}
\startdata
\cutinhead{Periodic and predictable transients}
Minor body               & Periodic comets                  & 1P/Halley            & 032      & Weeks--Months         & Years--Centuries      & O            & \\ 
                         & Stellar occultation              & \nodata              & \nodata  & Milliseconds--Minutes & Infrequent            & O            & \\ 
Solid planetoid          & Total solar eclipse              & Sun/Moon             & 150, 101 & Minutes               & Decades               & O            & \\ 
                         & Lunar eclipse                    & Moon                 & 101      & Hours                 & Months                & O            & \\ 
                         & Lunar/planetary occultation (ingress/egress) & Moon/Aldebaran       & 101, 168 & Milliseconds          & Infrequent            & O            & \\ 
Star                     & \editbfOne{Ultracool pulsar pulse}           & TVLM 513-46546       & 227      & Minutes               & Hours                 & R            & 1 \\
                         & \editbfOne{mCP pulsar pulse}                & CU Vir               & 228      & Minutes               & Day                   & R            & 2 \\
Collapsed star           & Pulsar pulse                     & Crab Pulsar          & 266      & Milliseconds          & Milliseconds--Seconds & RIOX$\gamma$ & 3 \\
\enddata
\end{deluxetable*}

\clearpage
\onecolumngrid

\section{The Superlative sample}
\label{sec:AppendixSuperlative}

\restartappendixnumbering

All objects in the Superlative sample, with their relevant properties, are listed in Table~\ref{table:SuperlativeSample}.

\startlongtable
\begin{deluxetable}{p{2.0cm}p{2.0cm}llcccc}
\tabletypesize{\scriptsize}
\tablecolumns{7}
\tablecaption{Superlative \editbfOne{sample}\label{table:SuperlativeSample}}
\tablehead{\colhead{Type} & \colhead{Property} & \colhead{Value} & \colhead{Superlative} & \colhead{ID}  & \colhead{Solar?} & \colhead{\editbfOne{True?}} & \colhead{Ref}}
\startdata
\cutinhead{Minor body}
All minor bodies        & $\albedoGV$  (darkest)         & $0.027^{+0.006}_{-0.007}$        & 1173 Anchises             & 537  & \checkmark & \checkmark & 1      \\
                        & $\albedoGV$  (brightest)       & $0.76^{+0.18}_{0.45} \endash 0.88^{+0.15}_{-0.06}$ & (55636) 2002 TX$_{300}$   & 538  & \checkmark & \checkmark & 2      \\
                        & $\MHostStar$ (smallest)        & $8^{+7}_{-3}\ \MJupiter$         & Cha 110913-773444         & 539  &            & ?          & 3      \\
Interplanetary minor bodies & $a$ (closest)                  & $0.5553\ \pm 0.0002\ \AU$        & 2019 LF$_6$               & 540  & \checkmark & \ding{53}/S? & 4      \\
                        & $Q$ (closest)                  & $0.65377 \pm 0.00012\ \AU$       & 2020 AV$_2$               & 016  & \checkmark & \ding{53}/S? & 5      \\
                        & $q$ (furthest)                 & \editbfOne{$80.40 \pm 0.09\ \AU$}            & 2012 VP$_{113}$           & 541  & \checkmark & \ding{53}  & 6      \\
                        &                                & \editbfOne{$65.2 \pm 0.2\ \AU$}              & \editbfOne{541132 Lele\={a}k\={u}honua} & 052  & \checkmark & \ding{53}  & \nodata\\
Minor satellites        & $\Radius$ (largest)            & \editbfOne{$210 \pm 7\ \km$}                 & Proteus                   & 542  & \checkmark & \checkmark & 7      \\
\enddata
\end{deluxetable}

\clearpage
\onecolumngrid

\section{The Anomaly sample}
\label{sec:AppendixAnomaly}

\restartappendixnumbering

All objects in the Anomaly sample are listed in Table~\ref{table:AnomalyNonSETISample} (if not a previous SETI candidate) and Table~\ref{table:AnomalySETISample} (if a previous SETI candidate). Table~\ref{table:ExcludedAnomaly} lists anomalies that were not included in the catalog for practical reasons.

\startlongtable
\begin{deluxetable}{p{3.0cm}p{3.0cm}lcccc}
\tabletypesize{\scriptsize}
\tablecolumns{7}
\tablecaption{Anomal\editbfOne{y} (Non-SETI) \editbfOne{sample}\label{table:AnomalyNonSETISample}}
\tablehead{\colhead{Type} & \colhead{Description} & \colhead{Anomaly} & \colhead{Class} & \colhead{ID}  & \colhead{Solar?} & \colhead{Ref}}
\startdata
\cutinhead{Class I}
Pulsar planets                 & Unknown formation mechanism, rare for PSR               & PSR B1257+12                     & I (III) & 094  &            & 1       \\
Paradoxical ELM WDs            & ELM white dwarf in too wide binary to be formed         & KIC 8145411                      & 0/I     & 332  &            & 2       \\
                               &                                                         & HE 0430-2457                     & 0/I     & 664  &            & 3       \\
Peripheral MSP binaries        & MSP-WD binary unexpectedly at edge of globular          & PSR J1911-5958A                  & I       & 665  &            & 4       \\
                               & PSR-star binary at edge of globular cluster             & PSR J1740-5340                   & I       & 666  &            & 5       \\
Nuclear cluster stars          & Stars in hostile Galactic Center environment, where tides could prevent star formation & S0-2                             & I       & 667  &            & 6       \\
                               &                                                         & IRS 16C                          & I       & 668  &            & 7       \\
Nuclear subcluster             & Apparent extremely dense star subcluster within inner parsec, possibly within tidal disruption limit & IRS 16E                          & I       & 604  &            & 8       \\
\enddata
\end{deluxetable}

\startlongtable
\begin{deluxetable*}{lp{3.0cm}llcc}
\tabletypesize{\scriptsize}
\tablecolumns{6}
\tablecaption{Anomal\editbfOne{y} (SETI) \editbfOne{sample}\label{table:AnomalySETISample}}
\tablehead{\colhead{Type} & \colhead{Description} & \colhead{Program} & \colhead{Candidate} & \colhead{Class} & \colhead{ID}}
\startdata
\cutinhead{Class II}
IR excess stars                & MIR-bright star                                         & C09             & IRAS 16406-1406                  & \editbfOne{II/III}  & 770 \\
                               &                                                         & C09             & IRAS 20331+4024                  & \editbfOne{II/III}  & 771 \\
                               &                                                         & C09             & IRAS 20369+5131                  & \editbfOne{II/III}  & 772 \\
                               & MIR-bright star cluster                                 & \GHAT           & IRAS 04287+6444                  & 0/II    & 773 \\
                               & Microwave-excess star                                   & L16             & UW CMi                           & 0/II    & 774 \\
IR excess galaxies             & MIR-bright galaxy                                       & \GHAT           & WISE J224436.12+372533.6         & II      & 775 \\
                               & MIR-radio correlation outlier                           & Ga15            & UGC 3097                         & II      & 776 \\
                               &                                                         &                 & NGC 814                          & II      & 777 \\
\enddata
\tablecomments{All sources in this sample are sidereal.}
\tablerefs{\editbfTwo{B92: \citet{Blair92}; Big Ear: \citet{Dixon85}, \citet{Gray12};} BT16: \citet{Borra16}; C09: \citet{Carrigan09}; \GHAT: \citet{Griffith15}; \editbfTwo{Ga15: \citet{Garrett15}; Harvard OSETI: \citet{Howard04};} L16: \citet{Lacki16-K3}; META: \citet{Horowitz93}; META II: \citet{Colomb95}; P19: \citet{Pinchuk19}; SERENDIP III: \citet{Bowyer16}; V16: \citet{Villarroel16}; Z15: \citet{Zackrisson15}; Z18: \citet{Zackrisson18}}
\end{deluxetable*}

\startlongtable
\begin{deluxetable*}{p{3.0cm}p{4.0cm}lcp{3.5cm}c}
\tabletypesize{\scriptsize}
\tablecolumns{6}
\tablecaption{\editbfOne{Anomalies excluded from \ExoticaCatalog{}}\label{table:ExcludedAnomaly}}
\tablehead{\colhead{Type} & \colhead{Description} & \colhead{Anomaly} & \colhead{Class} & \colhead{Reason excluded} & \colhead{Ref}}
\startdata
\cutinhead{Class II}
NS-BH mass gap object           & Gravitational wave event involving $2.6\ \Msun$ compact object, in between the expected mass of a neutron star and stellar mass black hole            & GW190814      & II       & Insufficiently localized & 1\\
Pair SN mass gap black hole     & Gravitational wave event involving $85_{-14}^{+21}\ \Msun$ black hole, in mass range where pair instability supernovae are expected to leave no remnants  & GW190521      & II       & Insufficiently localized & 2\\
\cutinhead{Class III}
Biotic planet                   & Planet with biosphere of poorly constrained origin & Earth & 0/III & Impractical to observe as whole & 3\\
Fermi Bubbles                   & Kiloparsec-scale bipolar bubbles visible in radio, X-rays, and gamma-rays extending out of Galactic Center                             & Fermi Bubbles & 0/III/IV & Too large on sky         & 4\\
\cutinhead{Class IV}
Milagro hot spots               & TeV cosmic ray hotspots located in tail of heliosphere                  & Milagro region A  & 0/IV & Too large on sky & 5\\
                                &                                                                         & Milagro region B  & 0/IV & Too large on sky & 5\\
Great Annihilator               & Region of excess positron annihilation emission around Galactic Center  & Great Annihilator & 0/IV & Too large on sky & 6\\
Galactic Center GeV excess      & Region of excess GeV gamma-ray emission around Galactic Center          & (Galactic Center) & 0/IV & Somewhat large on sky & 7\\
\enddata
\tablerefs{(1) \citet{Abbott20-190814}; (2) \citet{Abbott20-190521Astro,Graham20,DePaolis20}; (3) \citet{Sagan93}; (4) \citet{Dobler08,Su10,Kataoka13}; (5) \citet{Abdo08}; (6) \citet{Knodlseder05,Prantzos11}; (7) \citet{Hooper11,Ackermann17}; (8) \citet{Seiffert11,Singal18}; (9) \citet{Chang08,Adriani09}; (10) \citet{Aartsen14,IceCube18-Disc,Murase18}; (11) \citet{Kotera11}; (12) \citet{Brin83,Cirkovic09,Cirkovic18-Book}; (13) \citet{Rubin80,Famaey12}; (14) \citet{Clowe06,Spergel07}; (15) \citet{Riess98,Perlmutter99}; (16) \citet{Verde19}}
\end{deluxetable*}

\clearpage
\onecolumngrid
\section{The Control sample}
\label{sec:AppendixControl}
\restartappendixnumbering

All objects in the Control sample are listed in Table~\ref{table:ControlSample}.

\startlongtable
\begin{deluxetable*}{lp{5.0cm}ccp{5.0cm}cc}
\tabletypesize{\scriptsize}
\tablecolumns{7}
\tablecaption{Control \editbfOne{sample}\label{table:ControlSample}}
\tablehead{\colhead{Object} & \colhead{Original explanation} & \colhead{Class} & \colhead{Ref} & \colhead{Current explanation} & \colhead{Ref} & \colhead{ID}}
\startdata
CSL-1                          & Cosmic string as gravitational lens                     & 0/IV    & 1       & Galaxy pair                                             & 2       & 800 \\
GRB 090709A                    & \editbfOne{GRB with 8 sec quasi-periodic oscillations}              & III     & 3       & GRB with no periodicity                                 & 4       & 801 \\
GW100916 (A)                   & First BH merger observed in gravitational waves         & 0       & 5       & \editbfOne{Planned blind injection}                                 & 5       & 802 \\
GW100916 (B)                   &                                                         & 0       & 5       &                                                         & 5       & 803 \\
HD 117043                      & Potassium line-emitting stellar flares                  & III     & 6       & Matches in observatory                                  & 7       & 804 \\
\editbfOne{Hertzsprung's Object}           & Bright optical transient                                & IV      & 8       & Static electric discharge on photographic plate         & 9       & 805 \\
HIP 114176                     & \editbfOne{Star in the \emph{Hipparcos} catalog}                    & 0       & 10      & \editbfOne{Scattered light from bright star}                        & 10      & 806 \\
KIC 5520878                    & \editbfOne{RR Lyr with prime number signal in autocorrelation of period lengths} & III     & 11      & \editbfOne{Fluke from natural two period variability}               & 11      & 807 \\
\enddata
\tablecomments{All sources in this sample are sidereal.}
\tablerefs{(1): \citet{Sazhin03}; (2): \citet{Agol06}; (3): \citet{Markwardt09,Golenetskii09,Gotz09}; (4): \citet{deLuca10,Cenko10}; (5): \citet{Evans12-Swift}; (6): \citet{Barbier62}; (7): \citet{Wing67}; (8): \citet{Hertzsprung1927,Klemola83}; (9): \citet{Schaefer83}; (10): \citet{Perryman97}; (11): \citet{Hippke15}; (12): \citet{Molnar17}; (13): \citet{Socia18,Kovacs19}; (14): \citet{Gaidos16,Coughlin16}; (15): \citet{Katz86}; (16): \citet{Halliday87,Corso87,Maley87,Schaefer87,Borovicka89}; (17): \citet{Bailes91}; (18): \citet{Lyne92}; (19): \citet{Shamir06}; (20): \citet{Shamir06,Smette06,Nemiroff06}; (21): \citet{Scholz04}; (22): \citet{Farihi05-J1549}; (23): \citet{Lipunov20,Gropp20}; (24): \citet{Gropp20}; (25): \citet{Schmadel96}; (26): \citet{Ofek10}}
\end{deluxetable*}

\clearpage
\onecolumngrid
\section{\texorpdfstring{The full \ExoticaCatalogUpper{}}{The full \ExoticaCatalog{}}}
\label{sec:AppendixTotal}
\restartappendixnumbering

\subsection{Notes on data sources}
\label{sec:DataNotes}
\onecolumngrid
Much of the data used in Figures~\ref{fig:PlanetMSD}--\ref{fig:GalaxyCMD}~comes from papers on individual sources on the literature.

For Solar System bodies, we consulted the Jet Propulsion Laboratory's Solar System Dynamics pages\footnote{\url{https://ssd.jpl.nasa.gov/}}, particularly the Small Solar System Browser\footnote{\url{https://ssd.jpl.nasa.gov/sbdb.cgi}}.  When masses were unavailable, we estimated them by assuming that objects interior to Jupiter had density $3\ \gcm3$ and the rest had density $2\ \gcm3$.

We relied on Simbad data for the bulk of Table~\ref{table:FullSidereal}.  Stellar data was partly based on \emph{Gaia} distances, colors, and extinctions \citep{Gaia18-Summary}; extinctions from \citet{Savage85} and \citet{Gudennavar12}; PASTEL effective temperatures and surface gravities \citep{Soubiran16}; \emph{Hipparcos} photometry and distances \citep{Perryman97}; and individualized references.  For the I17 stars plotted in Figure~\ref{fig:StellarCMD}, it was impractical to find individualized sources; we supplemented with data from \citet{Holmberg07}, \citet{Takeda07}, CATSUP \citep{Hinkel17}, and \citet{Swihart17}.  Frequently, we had to calculate the luminosity and/or surface temperature from other quantities (mass, radius, bolometric flux, angular size, and distance).  \edit1{When no other data was available for these quantities, we estimated bolometric corrections and effective temperatures using the color conversions of \citet{Flower96}, as corrected by \citet{Torres10}.}

{\footnotesize \textbf{Additional references for Figure~\ref{fig:PlanetMSD} as listed in Table~\ref{table:FullSidereal}}: (13) \citet{Santerne18}; (15) \citet{Buldgen19}; (17) \citet{AngladaEscude16}; (20) \citet{Ribas18}; (22) \citet{Dittmann17}; (24) \citet{Diaz16,Tuomi13}; (26) \citet{Crida18}; (28) \citet{Gonzales19,Wang17}; (30) \citet{Armstrong20}; (33) \citet{Konacki03,Pavlov07}; (35) \citet{vonBraun12}; (38) \citet{Hartman11}; (39) \citet{Charbonneau09}; (41) \citet{Lovis05}; (44) \citet{Brahm16}; (46) \citet{Bouchy05,Boyajian15}; (48) \citet{delBurgo16}; (50) \citet{Cochran11}; (52) \citet{Lacour19,Marois08,Marois10}; (54) \citet{Santos01,Wittenmyer09}; (56) \citet{Hebrard10,Liu18}; (58) \citet{Bakos07}; (60) \citet{Hatzes06,OGorman17}; (63) \citet{Doyle11,Moorman19}; (365) \citet{JontofHutter15}; (369) \citet{Bernkopf12,Feng17,Pepe11}; (373) \citet{Zhou17}; (375) \citet{LibbyRoberts20,Masuda14}; (376) \citet{Gaudi17}; (378) \citet{Barclay12}; (380) \citet{Donati16}; (382) \citet{Currie18}; (384) \citet{Sato12}; (387) \citet{Hollands18,Luhman12,Rodriguez11}; (388) \citet{Isella16,Pinte18}.}

{\footnotesize \textbf{Other references for the stellar CMD in Figure~\ref{fig:StellarCMD} as listed in Table~\ref{table:FullSidereal}:} (2) \citet{Rappaport16}; (5) \citet{Macias18}; (6) \citet{Heiter15}; (7) \citet{Ballering17}; (36) \citet{Reid04}; (42) \citet{Stassun17}; (45) \citet{Brahm16}; (49) \citet{delBurgo16}; (55) \citet{Tinney11}; (59) \citet{Bakos07}; (68) \citet{Tannirkulam08}; (82) \citet{Dieterich14,Lepine12}; (85) \citet{Diaz19}; (86) \citet{Heiter15}; (87) \citet{Heiter15}; (88) \citet{Heiter15}; (89) \citet{Ribas10}; (90) \citet{Heiter15}; (92) \citet{Swihart17}; (96) \citet{Swihart17}; (98) \citet{Baines18}; (99) \citet{Gordon18-O}; (101) \citet{Markova18}; (102) \citet{Baines13}; (108) \citet{Heiter15}; (111) \citet{For10}; (112) \citet{Benedict11}; (114) \citet{For10}; (115) \citet{Ohnaka19}; (124) \citet{daSilva06}; (125) \citet{Heiter15}; (127) \citet{Bennett96}; (128) \citet{Heiter15}; (129) \citet{Torres15}; (131) \citet{Heiter15}; (133) \citet{Cruzalebes13}; (139) \citet{Harper08}; (140) \citet{vanGenderen19}; (141) \citet{Zhang12-VYCMa}; (144) \citet{Morris04,vanDerHucht01}; (146) \citet{North07}; (148) \citet{Clark12-BlueHyGs,Crowther06}; (149) \citet{Groh09}; (150) \citet{Damineli19,Shull19}; (153) \citet{AllendePrieto16}; (154) \citet{Kochukhov10}; (158) \citet{Heiter15}; (161) \citet{Heiter15}; (162) \citet{Heiter15}; (165) \citet{Kaye99}; (167) \citet{Gordon19-B}; (170) \citet{ODonoghue97}; (178) \citet{Guinan86}; (180) \citet{Howell13,Yudin02}; (182) \citet{Jeffery01}; (184) \citet{Geier17}; (185) \citet{Geier17}; (187) \citet{Plez05}; (188) \citet{Gordon18-O}; (189) \citet{Heber08}; (193) \citet{Holberg16}; (199) \citet{Preval13}; (203) \citet{Sahu17}; (204) \citet{Holberg16,Hollands18}; (208) \citet{Holberg16}; (232) \citet{Torres02}; (233) \citet{Torres02}; (363) \citet{Mamajek12}; (370) \citet{Bernkopf12}; (371) \citet{Zhou17}; (374) \citet{Zhou17}; (386) \citet{Stassun17}; (390) \citet{Crowther10,Doran13}; (394) \citet{Geier17}; (396) \citet{Gordon18-OHIR,Zhang12-NMLCyg}; (403) \citet{Koposov20}; (407) \citet{Dahn04}; (409) \citet{Werner15}; (429) \citet{Tokovinin18}; (476) \citet{Vos18}; (477) \citet{Vos18}; (488) \citet{Hawkins16}; (489) \citet{Trundle01}; (490) \citet{Geier17}; (492) \citet{Masseron20}; (494) \citet{Geller17,Mathieu03}; (500) \citet{Boyajian16}; (510) \citet{Andrews96}; (511) \citet{Brown08}; (513) \citet{BailerJones11}; (542) \citet{Zackrisson18}; (545) \citet{Sandage97}; (546) \citet{Rayner09}; (547) \citet{Cvetkovic11}; (548) \citet{RuizDern18}; (560) \citet{Farihi05-J1549}.  Supplementary data for the I17 stars comes from: \citet{Baines18,Heiter15,Swihart17}.}

{\footnotesize \textbf{Other references for the HR diagram in Figure~\ref{fig:StellarHR} as listed in Table~\ref{table:FullSidereal}:} (1) \citet{Bonnefoy13}; (2) \citet{Rappaport16}; (3) \citet{Sokal18}; (4) \citet{Bodman17}; (5) \citet{Macias18}; (6) \citet{Heiter15}; (7) \citet{Ballering17}; (8) \citet{Peterson06}; (9) \citet{Weinberger08}; (10) \citet{Su07}; (11) \citet{Jura03}; (12) \citet{Gansicke19}; (14) \citet{Santerne18}; (16) \citet{Buldgen19}; (18) \citet{AngladaEscude16}; (19) \citet{SanchisOjeda13}; (21) \citet{Ribas18}; (23) \citet{Dittmann17}; (25) \citet{Tuomi13}; (27) \citet{Bourrier18}; (29) \citet{Gonzales19}; (31) \citet{Armstrong20}; (32) \citet{Rappaport12}; (34) \citet{Borucki12}; (37) \citet{vonBraun12}; (40) \citet{Charbonneau09}; (43) \citet{Schwarz07,Stassun17}; (45) \citet{Brahm16}; (47) \citet{Poppenhaeger13}; (49) \citet{delBurgo16}; (51) \citet{Cochran11}; (53) \citet{Marois08}; (55) \citet{Tinney11}; (57) \citet{Hebrard10,Liu18}; (59) \citet{Bakos07}; (61) \citet{OGorman17}; (64) \citet{Doyle11}; (65) \citet{Ceccarelli00}; (68) \citet{Tannirkulam08}; (69) \citet{Leggett17}; (71) \citet{Leggett17}; (72) \citet{delBurgo09}; (73) \citet{Dieterich18,King10}; (74) \citet{Faherty14,Garcia17}; (76) \citet{Faherty14,Garcia17}; (78) \citet{Dupuy09}; (79) \citet{Cushing08}; (80) \citet{Basri96,Basri99}; (81) \citet{Dieterich14}; (83) \citet{Dieterich14}; (84) \citet{Dieterich14}; (85) \citet{Diaz19}; (86) \citet{Heiter15}; (87) \citet{Heiter15}; (88) \citet{Heiter15}; (89) \citet{Ribas10}; (90) \citet{Heiter15}; (91) \citet{Boyajian12-AFG}; (92) \citet{Swihart17}; (93) \citet{Zhao09}; (94) \citet{Jones15}; (95) \citet{Monnier12}; (96) \citet{Swihart17}; (97) \citet{McCarthy12}; (98) \citet{Baines18}; (99) \citet{Gordon18-O}; (100) \citet{Blomme11}; (101) \citet{Markova18}; (103) \citet{Baines18}; (104) \citet{Li19}; (105) \citet{Heiter15}; (106) \citet{David15}; (107) \citet{Rau18}; (108) \citet{Heiter15}; (109) \citet{Heiter15}; (110) \citet{Gray16}; (111) \citet{For10}; (112) \citet{Benedict11}; (114) \citet{For10}; (115) \citet{Ohnaka19}; (116) \citet{Libert10}; (117) \citet{Groenewegen12,Menten12}; (118) \citet{Justtanont13}; (120) \citet{Witt09}; (122) \citet{Hadjara18}; (123) \citet{Torres15}; (124) \citet{daSilva06}; (126) \citet{Halabi15,Heiter15}; (127) \citet{Bennett96}; (128) \citet{Heiter15}; (129) \citet{Torres15}; (130) \citet{Natale08}; (131) \citet{Heiter15}; (132) \citet{Neilson16}; (133) \citet{Cruzalebes13}; (134) \citet{Groenewegen18}; (136) \citet{Pablo17}; (137) \citet{Zorec09}; (138) \citet{Aufdenberg02}; (139) \citet{Harper08}; (140) \citet{vanGenderen19}; (142) \citet{Wittkowski12,Zhang12-VYCMa}; (145) \citet{Morris04}; (146) \citet{North07}; (147) \citet{Tramper15}; (148) \citet{Clark12-BlueHyGs,Crowther06}; (149) \citet{Groh09}; (151) \citet{Hillier01,Mehner19,Shull19}; (153) \citet{AllendePrieto16}; (154) \citet{Kochukhov10}; (155) \citet{Ciardi07}; (156) \citet{Burgasser08}; (157) \citet{AngladaEscude14}; (158) \citet{Heiter15}; (159) \citet{Mortier12}; (160) \citet{Heiter15}; (161) \citet{Heiter15}; (162) \citet{Heiter15}; (163) \citet{Christlieb02,Norris13}; (164) \citet{Woodruff04}; (165) \citet{Kaye99}; (166) \citet{DeRidder99}; (167) \citet{Gordon19-B}; (168) \citet{Pietrukowicz17}; (169) \citet{Blanchette08}; (170) \citet{ODonoghue97}; (171) \citet{Latour11}; (172) \citet{Kervella16}; (173) \citet{Hallinan06}; (174) \citet{Kochukhov14}; (175) \citet{Loebman15}; (177) \citet{Leiner16}; (179) \citet{Guinan86,Jetsu93,Korhonen99}; (181) \citet{GarciaHernandez11,Howell13}; (182) \citet{Jeffery01}; (183) \citet{Fossati10}; (184) \citet{Geier17}; (185) \citet{Geier17}; (186) \citet{Sener14}; (187) \citet{Plez05}; (188) \citet{Gordon18-O}; (189) \citet{Heber08}; (190) \citet{Brown05-HVS}; (191) \citet{Kaplan14-NLTT11748}; (192) \citet{Bedard17}; (193) \citet{Holberg16}; (194) \citet{Holberg16}; (195) \citet{Jahn07}; (196) \citet{Raddi18}; (197) \citet{Shen18}; (199) \citet{Preval13}; (200) \citet{Holberg16,Hollands18}; (201) \citet{BischoffKim19}; (202) \citet{Bohlin08}; (203) \citet{Sahu17}; (204) \citet{Holberg16,Hollands18}; (205) \citet{Dreizler96}; (206) \citet{Dufour08}; (207) \citet{Gansicke10}; (208) \citet{Holberg16}; (209) \citet{Romero12}; (210) \citet{Serenelli19}; (228) \citet{PortoDeMello08,Pourbaix16}; (229) \citet{Ratzka09,Schaefer20}; (230) \citet{Bergeron89}; (232) \citet{Torres02}; (233) \citet{Torres02}; (234) \citet{Hoard10}; (236) \citet{Masuda19}; (237) \citet{Xiang20}; (361) \citet{Luhman05}; (364) \citet{Mamajek12,Mentel18}; (366) \citet{JontofHutter15}; (367) \citet{Barclay13}; (368) \citet{Bernkopf12}; (370) \citet{Bernkopf12}; (372) \citet{Rappaport13}; (374) \citet{Zhou17}; (377) \citet{Gaudi17}; (379) \citet{Esteves15}; (381) \citet{Donati16}; (383) \citet{Currie18}; (385) \citet{Stock18}; (386) \citet{Stassun17}; (389) \citet{Natta04}; (391) \citet{Crowther10,Crowther16}; (393) \citet{vonBoetticher17}; (395) \citet{LaPalombara19}; (396) \citet{Gordon18-OHIR,Zhang12-NMLCyg}; (398) \citet{Wittkowski17}; (399) \citet{Keller14}; (400) \citet{vonBraun14}; (401) \citet{Caffau11}; (402) \citet{Peissker20-ApJ}; (403) \citet{Koposov20}; (405) \citet{Hermes13}; (407) \citet{Dahn04}; (409) \citet{Werner15}; (411) \citet{Brinkworth13}; (412) \citet{Kilic12}; (423) \citet{Best17}; (425) \citet{Tehrani19}; (427) \citet{Burdge19}; (476) \citet{Vos18}; (477) \citet{Vos18}; (480) \citet{Habibi17}; (482) \citet{Martins07}; (487) \citet{Mkrtichian08}; (488) \citet{Hawkins16}; (489) \citet{Trundle01}; (491) \citet{Green11}; (492) \citet{Masseron20}; (495) \citet{Mathieu03}; (496) \citet{Spezzi11}; (498) \citet{Andrews16}; (500) \citet{Boyajian16}; (501) \citet{Rappaport19}; (502) \citet{Gaia18-Summary}; (503) \citet{Lyubimkov10}; (504) \citet{Pilecki18}; (505) \citet{Jarvinen18}; (506) \citet{McCollum19-13111}; (507) \citet{McCollum19-12849}; (509) \citet{Fokin04,Stankov03}; (511) \citet{Brown08}; (512) \citet{Lyubimkov10}; (513) \citet{BailerJones11}; (515) \citet{Gaia18-Summary}; (517) \citet{Gaia18-Summary}; (519) \citet{Kowalski10}; (520) \citet{Krticka07,Landstreet07,Shultz19}; (542) \citet{Zackrisson18}; (547) \citet{Cvetkovic11}; (558) \citet{Gaidos16}.  Supplementary data for the I17 stars comes from: \citet{AguileraGomez18,AngladaEscude14,AngladaEscude16,Baines18,Bonnefoy13,Boyajian12-AFG,Dieterich14,Dupuy09,Hadjara18,Heiter15,Jones15,Kolbas15,McCarthy12,Monnier12,Rau18,Swihart17,vonBraun14,Zhao09}.}

Galaxy data was partly based on NED redshifts and photometry from 2MASS \citep{Skrutskie06}, \citet{deVaucouleurs91}, and Data Releases 9 and 12 of the Sloan Digital Sky Survey \citep{Ahn12,Alam15}.  For I17 galaxies, we relied mainly on the B-band magnitudes in I17 itself and the photometry in \citet{Mateo98}.  Because photometry in $u$ and $r$ bands was frequently unavailable, we relied heavily on the color transformations of \citet{Blanton07}, \citet{Jester05}, and Lupton's equations\footnote{As presented at \url{http://classic.sdss.org/dr6/algorithms/sdssUBVRITransform.html}.} to derive the approximate colors for use in Figure~\ref{fig:GalaxyCMD}.  Redshifts were converted to luminosity distances assuming $H_0 = 70\ \kms\ \Mpc^{-1}$, $\Omega_m = 0.3$, and $\Omega_{\Lambda} = 0.7$.  In some cases, the stellar mass was calculated from $K_s$ absolute magnitudes using the conversion of \citet{Cappellari13}.  I17 galaxy star-formation rates were largely calculated from GALEX ultraviolet and IRAS total infrared luminosities \citep{Bai15,Sanders03}, using the corrections of \citet{Hao11}, with additional data from \citet{Licquia15-sSFR}, \citet{Jarrett19}, and \citet{McConnachie12}.

{\footnotesize \textbf{Other references for the galaxy CMD in Figure~\ref{fig:GalaxyCMD} as listed in Table~\ref{table:FullSidereal}:} (238) \citet{Baumgardt18,Harris10}; (240) \citet{Baumgardt18,Harris10}; (242) \citet{Harris10}; (245) \citet{Harris10}; (248) \citet{Baumgardt18,Harris10}; (263) \citet{Shaya94}; (280) \citet{Jarrett19}; (281) \citet{Jarrett19}; (282) \citet{Jarrett19}; (283) \citet{Boselli15}; (284) \citet{Jarrett19}; (286) \citet{Norris11,Norris15}; (287) \citet{Jarrett19}; (289) \citet{McConnachie12}; (290) \citet{Trujillo14-N1277}; (293) \citet{Thilker10}; (294) \citet{Jarrett19}; (295) \citet{Jarrett19}; (296) \citet{McConnachie12,Prugniel98}; (297) \citet{Wei10}; (299) \citet{Peeples08}; (301) \citet{Jarrett19}; (303) \citet{Jarrett19}; (304) \citet{Jarrett19}; (307) \citet{Jarrett19}; (308) \citet{Dale07,Jarrett19}; (310) \citet{Jarrett19}; (312) \citet{Karachentsev04}; (322) \citet{Jarrett19}; (326) \citet{Annibali13}; (327) \citet{Loose86}; (329) \citet{Micheva17}; (338) \citet{Croft06}; (339) \citet{Boissier16}; (340) \citet{Pandya18}; (342) \citet{Trujillo17}; (344) \citet{Spavone10}; (347) \citet{Kenney14}; (348) \citet{Jarrett19}; (349) \citet{Jarrett19}; (350) \citet{Jarrett19}; (351) \citet{Jarrett19}; (352) \citet{Jarrett19}; (353) \citet{Matthews99}; (432) \citet{Buzzoni12,Platais11}; (433) \citet{Meylan01}; (436) \citet{Sandoval15}; (439) \citet{Baumgardt18,Harris10}; (442) \citet{Baumgardt18,Harris10}; (448) \citet{Sandoval15}; (458) \citet{Brunker19}; (460) \citet{Izotov18}; (461) \citet{Jarrett19}; (465) \citet{Ogle16,Ogle19}; (485) \citet{Caldwell14}; (522) \citet{Mateo98,McConnachie12}; (523) \citet{Jarrett19}; (524) \citet{Filho18}; (543) \citet{GomezLopez19}.  Supplementary CMD data for I17 galaxies comes from: \citet{deVaucouleurs68,Lauberts82,Licquia15-mag,Mateo98,Prugniel98}.}

{\footnotesize \textbf{Other references for the $\Mstar \endash {\rm SFR}$ plot in Figure~\ref{fig:GalaxyCMD} as listed in Table~\ref{table:FullSidereal}:} (264) \citet{Varenius16}; (287) \citet{Jarrett19}; (294) \citet{Jarrett19}; (295) \citet{Jarrett19}; (298) \citet{ErrozFerrer13}; (300) \citet{Gu06,Peeples08}; (301) \citet{Jarrett19}; (303) \citet{Jarrett19}; (304) \citet{Jarrett19}; (305) \citet{Ogle16,Ogle19}; (307) \citet{Jarrett19}; (309) \citet{Mateo98}; (311) \citet{Madore18,Mateo98}; (313) \citet{Karachentsev04,Meier01}; (314) \citet{Gladders13}; (316) \citet{DiTeodoro18,Yuan11}; (318) \citet{Girard18}; (320) \citet{MarquesChaves18}; (322) \citet{Jarrett19}; (324) \citet{CortijoFerrero17}; (325) \citet{Cluver08}; (326) \citet{Annibali13}; (328) \citet{Loose86,Summers01}; (329) \citet{Micheva17}; (330) \citet{Vanzella16}; (332) \citet{Berg18}; (334) \citet{DessaugesZavadsky15}; (336) \citet{Swinbank10,Zhang18-LensedSB}; (338) \citet{Croft06}; (339) \citet{Boissier16}; (342) \citet{Trujillo17}; (344) \citet{Spavone10}; (345) \citet{Finkelman11-Acc}; (346) \citet{Brandl09,Lahen18}; (348) \citet{Jarrett19}; (350) \citet{Jarrett19}; (351) \citet{Jarrett19}; (352) \citet{Jarrett19}; (450) \citet{Ma16-SPT0346}; (452) \citet{Toba20}; (454) \citet{Oesch16}; (456) \citet{Lam19}; (458) \citet{Brunker19}; (460) \citet{Izotov18}; (461) \citet{Jarrett19}; (463) \citet{Ogle16,Ogle19}; (465) \citet{Ogle16,Ogle19}; (467) \citet{Yuan17}; (468) \citet{Bayliss20}; (470) \citet{Cava18}; (523) \citet{Jarrett19}; (524) \citet{Filho18}; (541) \citet{Vaddi16}; (544) \citet{ErrozFerrer13}.}

\subsection{Tables of the catalog}
\label{sec:FullCatalogTables}

Table~\ref{table:FullSolSys} lists all Solar System sources in the Exotica Catalog. Table~\ref{table:FullSidereal} lists all sidereal sources in the Exotica Catalog. Table~\ref{table:Relationships} lists relationships between sidereal sources in the Catalog, and between Catalog sources and objects in I17.

\startlongtable
\begin{deluxetable*}{lp{2.5cm}cp{2cm}cccccccc}
\tabletypesize{\scriptsize}
\tablecolumns{12}
\tablecaption{The \ExoticaCatalog{}: Solar System targets\label{table:FullSolSys}}
\tablehead{\colhead{ID} & \colhead{Name} & \colhead{Samples} & \colhead{Phyla} & \colhead{Primary} & \colhead{$a$} & \colhead{$e$} & \colhead{$i$} & \colhead{$a_{\odot}$} & \colhead{$\MOID_{\oplus}$} & \colhead{$\Theta$} & \colhead{Ref} \\ & & & &  & & & & \colhead{($\AU$)} & \colhead{($\AU$)}}
\startdata
1    & 446 Aeternitas                 & P     & Minor body                & Sun          & 2.79 AU   & 0.126    & 10.62    & 2.79     & 1.45   & \nodata      & \\
2    & 52 Europa                      & P     & Minor body                & Sun          & 3.09 AU   & 0.110    & 7.48     & 3.09     & 1.77   & $0.2 \farcs$ & \\
3    & 624 Hektor                     & P     & Minor body                & Sun          & 5.26 AU   & 0.023    & 18.16    & 5.26     & 4.15   & \nodata      & \\
4    & 434 Hungaria                   & P     & Minor body                & Sun          & 1.94 AU   & 0.074    & 22.51    & 1.94     & 0.83   & \nodata      & \\
5    & 16 Psyche                      & P     & Minor body                & Sun          & 2.92 AU   & 0.134    & 3.10     & 2.92     & 1.54   & $0.2 \farcs$ & \\
6    & 3628 Bo\v{z}n\v{e}mcov\'a      & P     & Minor body                & Sun          & 2.54 AU   & 0.297    & 6.88     & 2.54     & 0.78   & \nodata      & \\
7    & 420 Bertholda                  & P     & Minor body                & Sun          & 3.41 AU   & 0.030    & 6.69     & 3.41     & 2.33   & \nodata      & \\
8    & 1862 Apollo                    & P     & Minor body                & Sun          & 1.47 AU   & 0.560    & 6.35     & 1.47     & 0.03   & \nodata      & \\
\enddata
\tablecomments{\textbf{Name} -- \editbfOne{TBD refers to a type of technological object type whose Prototype is yet to be selected. \\}
\textbf{Samples} -- All samples a target is in. P: Prototype, S: Superlative, A: non-SETI Anomaly, E: SETI Anomaly, C: Control.\\
\textbf{Primary} -- Name of body the target orbits.\\
\textbf{$a$, $e$, $i$} -- Semimajor axis, eccentricity, inclination of target's orbit around body's primary, respectively.\\
\textbf{$a_{\odot}$} -- Semimajor axis of target's or primary's orbit around Sun.\\
\textbf{$\textrm{MOID}_{\oplus}$} -- Minimum orbital intersection distance with Earth's orbit.\\
\textbf{$\Theta$} -- Maximum angular size of body, as calculated from radius and $\textrm{MOID}_{\oplus}$.}
\tablerefs{\editbfTwo{(1) \citet{Chapman95,Belton96}; (2) \citet{Brown20}; (3) \citet{Brown20}; (4) \citet{Buie20}; (5) \citet{Kwiatkowski09}; (6) \citet{Jewitt20}; (7) \citet{Meech17}; (8) \citet{Brown07-Eris}; (9) \citet{Pal12}; (10) \citet{Ragozzine09}}}
\end{deluxetable*}

\startlongtable
\begin{deluxetable*}{lp{2.5cm}cp{1.75cm}cccccccp{0.3cm}}
\tabletypesize{\scriptsize}
\tablecolumns{12}
\tablecaption{The \ExoticaCatalog{}: Sidereal targets\label{table:FullSidereal}}
\tablehead{\colhead{ID} & \colhead{Name} & \colhead{Samples} & \colhead{Phyla} & \colhead{RA} & \colhead{Dec} & \colhead{$D_L^{\rm eff}$} & \colhead{$\mu_{\alpha}$} & \colhead{$\mu_{\delta}$} &  \colhead{$\Theta$} & \colhead{I17?} & \colhead{Refs}}
\startdata
35   & \object[{bet Pic}]{$\beta$ Pic}          & P     & Minor body                &  5:47:17.1 & -51:03:59 &  19.8 pc  & 4.7     & 83.1    & \nodata                    & \checkmark & 1 \\
63   & \object{WD 1145+017}                     & P     & Minor body                & 11:48:33.6 &  +1:28:59 & 141.7 pc  & -43.7   & -4.1    & \nodata                    &  & 2 \\
69   & \object{TW Hya}                          & P     & Minor body, Star          & 11:01:51.9 & -34:42:17 &  60.1 pc  & -68.4   & -14.0   & \nodata                    &  & 3 \\
70   & \object[{DoAr 24}]{EPIC 203937317}       & P     & Minor body                & 16:26:17.1 & -24:20:22 & 134.3 pc  & -6.6    & -27.1   & \nodata                    &  & 4 \\
71   & \object{GM Aur}                          & P     & Minor body                &  4:55:11.0 & +30:21:59 & 159.6 pc  & 3.9     & -24.5   & \nodata                    &  & 5 \\
72   & \object[{tau Cet}]{$\tau$ Cet}           & P     & Minor body, Star          &  1:44:04.1 & -15:56:15 &   3.6 pc  & -1721.0 & 854.2   & \nodata                    & \checkmark & 6 \\
73   & \object[{kap Psc}]{$\kappa$ Psc}         & P     & Minor body                & 23:26:56.0 &  +1:15:20 &  48.9 pc  & 87.1    & -95.7   & \nodata                    & \checkmark & 7 \\
74   & \object{Altair}                          & P     & Minor body                & 19:50:47.0 &  +8:52:06 &   5.1 pc  & 536.2   & 385.3   & \nodata                    & \checkmark & 8 \\
\enddata
\end{deluxetable*}

\startlongtable
\begin{deluxetable*}{lccp{6.0cm}p{2.5cm}}
\tabletypesize{\scriptsize}
\tablecolumns{5}
\tablecaption{Relationships of \editbfTwo{sidereal} \ExoticaCatalog{} targets\label{table:Relationships}}
\tablehead{\colhead{Object} & \colhead{ID} & \colhead{Relationship} & \colhead{Partner} & \colhead{Partner IDs}}
\startdata
\cutinhead{Within Exotica catalog}
NGC 7293 central star          &   77 & IN                  & Helix Nebula                             & 365 \\
Proxima b                      &   84 & BOUND               & $\alpha$ Cen AB                          & 321 \\
Luhman 16B                     &  137 & IN                  & Luhman 16                                & 323 \\
                               &      & MUTUAL\_ORBIT       & Luhman 16B                               & 137 \\
Luhman 16A                     &  138 & IN                  & Luhman 16                                & 323 \\
                               &      & MUTUAL\_ORBIT       & Luhman 16A                               & 138 \\
61 Cyg B                       &  146 & MUTUAL\_ORBIT       & 61 Cyg A                                 & 147 \\
61 Cyg A                       &  147 & MUTUAL\_ORBIT       & 61 Cyg B                                 & 146 \\
\enddata
\tablecomments{\textbf{Relationships} -- ADJACENT: object's sky location is projected within same parent object as partner.\\
\editbfOne{ALTERNATE\_POSITION: additional possible sky location for source if poorly localized.}\\
BOUND: gravitationally bound with, orbital motion of either partner too slow to detect \editbfTwo{or period $> 10,000\ \yr$}.\\
CONTAINS: partner's sky location is projected within boundaries of object.\\
GROUP\_OF: object consists entirely of collection of listed objects.\\
HOSTS: partner orbits object, object's reflex motion too small to detect.\\
IN: object's sky location is within bounds of other object.\\
LENSES: object is gravitational lens of partner, implying the partner appears projected within the object.\\
LENSED\_BY: object is gravitationally lensed by partner, implying the object appears projected within its partner.\\
MUTUAL\_ORBIT: object and partner bound, both with detectable orbital motion.\\
ORBITS: object orbits partner, partner's reflex motion too small to detect.\\
}
\end{deluxetable*}

\clearpage

\bibliographystyle{aasjournal}
\bibliography{Exotica_20E_v4_arXiv}

\end{document}